\documentclass[preprint,11pt]{elsarticle}

\usepackage[margin=1in]{geometry}
\usepackage{amsmath,amssymb,amsfonts,amsthm,bm,mathtools,bbm}
\usepackage{graphicx}
\usepackage{xcolor}
\usepackage{subcaption}
\usepackage{upgreek}
\usepackage{hyperref}

\hypersetup{
  colorlinks=true,
  linkcolor=blue,
  citecolor=red,
  urlcolor=blue
}

\newtheorem{theorem}{Theorem}
\newtheorem{lemma}{Lemma}
\newtheorem{remark}{Remark}

\journal{Journal of Computational Physics}

\begin{document}

\begin{frontmatter}

\title{Diagnosing symplecticity in simulations of high-dimensional Hamiltonian systems}

\author[LANL]{William Barham\corref{cor1}}
\ead{wbarham@lanl.gov}

\author[UT]{J.~W.~Burby}

\address[LANL]{Los Alamos National Laboratory, Theoretical Division}
\address[UT]{The University of Texas at Austin, Institute for Fusion Studies and Department of Physics}

\cortext[cor1]{Corresponding author}

\begin{abstract}
Integrals of the Liouville $1$-form, known as the first Poincar{\'e} integral invariant, provide a computable figure of merit for monitoring the conservation of symplecticity in the numerical integration of Hamiltonian systems. For smooth loop data, these integrals may be approximated with spectral convergence in the number of sample points, with rates limited by regularity. We devise a numerical integral invariant diagnostic for checking preservation of symplecticity in particle-in-cell (PIC) kinetic plasma simulation codes. As a first application of this diagnostic tool, we check the preservation of symplecticity in symplectic electrostatic particle-in-cell (PIC) methods. Surprisingly, such PIC methods fail to have symplectic time-advance maps if the charge is interpolated to the grid using linear shape functions, as is commonly done in practice. It is found that at least quadratic interpolation is needed to avoid this failure of symplecticity preservation. 
\end{abstract}


\end{frontmatter}

\tableofcontents

\section{Introduction}

Symplectic integrators are a natural and popular method for the numerical integration of Hamiltonian ordinary differential equations. In addition to simulating Hamiltonian ordinary differential equations \cite{geo_num_int}, they are frequently used for the temporal integration of wave-like partial differential equations (e.g. symplectic Runge-Kutta methods \cite{sanz1988runge} or splitting methods \cite{lubich2008splitting}), and are commonly used in time-stepping schemes for structure-preserving particle-in-cell (PIC) kinetic plasma simulation methods \cite{squire2012geometric, qin2015canonical, kraus2017gempic, perse2021geometric, xiao2015explicit, evstatiev2013variational, he2015hamiltonian, campos2022variational, glasser2022gauge, glasser2023generalizing, qiang2018symplectic, kormann2024dual, shadwick2014variational, stamm2014variational, qin2015canonical, jianyuan2018structure, jianyuan2021explicit}. Likewise, symplectic integration has been used to time-advance the Vlasov equation when coupled with spatial discretizations other than particle-in-cell, although the semi-discrete system is not Hamiltonian in such cases \cite{bernier2178952splitting, casas2017high}. Symplectic integrators are designed such that the time-advance map is a canonical transformation, thus conserving a symplectic form exactly. A consequence of this defining property of symplectic integrators is the conservation of a numerical energy which remains close to the true energy for the duration of the simulation. This energy stability property is frequently cited as a justification for the use of symplectic integrators. However, there are non-symplectic, energy-conserving integrators, somewhat dulling this argument for their utility \cite{chen2011energy,chen2014energy,chen2015multi, chen2020semi, chacon2013charge, chacon2016curvilinear, ricketson2025explicit, kormann2021energy, ji2023asymptotic}. A better justification for the value of symplectic integration is the exact conservation of the symplectic form. This more fundamental property of symplectic integrators, while a clear theoretical advantage, is difficult to directly quantify in a simulation, and is sometimes overlooked in practice due to its abstractness and the difficulty of measuring its impact. 

This work proposes a diagnostic tool based on the Poincar{\'e} integral invariant which directly measures the conservation of the symplectic form. This integral invariant was previously used to monitor symplecticity conservation in variational integrators and related schemes for low-dimensional degenerate Lagrangian systems \cite{kraus2017projectedvariationalintegratorsdegenerate}. A Julia implementation of this diagnostic can be found in \cite{kraus_poincareinvariants_2024}. This loop-integral diagnostic is complementary to Jacobian-based tests of symplecticity, such as the diagnostic used in \cite{sachan2024conformal}: rather than verifying the symplectic condition on the derivative of the flow map, we test preservation of the Poincar{\'e} integral invariant along advected loops. In this work, the suitability of such a diagnostic for monitoring the symplecticity of PIC methods is a central concern, with challenges such as high-dimensionality and low-regularity being addressed in detail. Notably, we are forced to address two issues associated with integral invariants absent from previous work \cite{kraus2017projectedvariationalintegratorsdegenerate} in low dimensions. (1) Chaos reigns in high dimensions, leading to generic and rapid shearing of the phase space loops around which the Liouville 1-form is integrated. (2) The time advance map for a discretized PDE system generally only enjoys limited regularity, and therefore does not fall under the purview of the most elementary results establishing integral invariance. Our work provides a general purpose tool to aid in software-development and in analyzing the relative performance of symplectic and non-symplectic time-stepping methods. 

\section{Mathematical preliminaries}
This section briefly introduces some essential background knowledge of Hamiltonian systems.

\subsection{Hamiltonian systems and symplecticity}

The diagnostic tool of interest in this work is applicable to numerical approximations of Hamiltonian systems. Given a state vector $\bm{z} = (\bm{q}, \bm{p}) \in \mathbb{R}^{2d}$, where $\bm{q}, \bm{p} \in \mathbb{R}^d$, and a Hamiltonian, $\mathcal{H}: \mathbb{R}^{2d} \to \mathbb{R}$, a canonical Hamiltonian system evolves as
\begin{equation} \label{eq:hamiltons_eqns}
    \dot{\bm{z}} = J_c \nabla \mathcal{H} \,,
    \quad  \quad
    J_c
    =
    \begin{pmatrix}
        0 & I_d \\
        -I_d & 0
    \end{pmatrix}
    ,
\end{equation}
where $I_d \in \mathbb{R}^{d\times d}$ is the identity matrix. The matrix $J_c$ is variously called the canonical Poisson tensor, the Hamiltonian bivector, or the Poisson matrix. The corresponding canonical symplectic form is $\omega_c = \mathsf{d} \bm{q} \wedge \mathsf{d} \bm{p} = \sum_{i=1}^d \mathsf{d} q_i \wedge \mathsf{d} p^i$; under the canonical pairing, the Poisson tensor and symplectic form define inverse bundle maps, up to the sign convention used to identify two-forms with matrices. In more common notation, we write
\begin{equation}
    \dot{\bm{q}}
    =
    \frac{\partial \mathcal{H}}{\partial \bm{p}} \,,
    \quad \text{and} \quad
    \dot{\bm{p}}
    =
    - \frac{\partial \mathcal{H}}{\partial \bm{q}} \,.
\end{equation}
Such systems, and their complementary Lagrangian formulation, are the central object of study in classical mechanics.  

Systems of this form conserve energy and symplecticity. Let the time-advance map for solutions of equation \eqref{eq:hamiltons_eqns} be denoted by $\Phi_{h}: \bm{z}(t) \mapsto \bm{z}(t + h)$, and let $D_{\bm{z}} \Phi_{h}$ denote its Jacobian. Then these two conservation laws may be expressed as follows:
\begin{itemize}
    \item \textit{Energy is conserved}: $\mathcal{H}(\bm{z}(t)) = \text{const}$. 
    \item \textit{The time-advance map is symplectic}: $D_{\bm{z}} \Phi_{h}^T J_c D_{\bm{z}} \Phi_{h} = J_c$, $\forall t$.
\end{itemize}
The former is a direct consequence of the anti-symmetry of the Poisson matrix, while the conservation of symplecticity is a deeper result associated with the geometry of phase-space. 

A symplectic integrator is a computable approximation to the time-advance map, $\tilde{\Phi}_{h} \approx \Phi_{h}$, such that $D_{\bm{z}} \tilde{\Phi}_{h}^T J_c D_{\bm{z}} \tilde{\Phi}_{h} = J_c$. One may show \cite{geo_num_int} that if $\mathcal{H}$ is real analytic then energy is nearly conserved over exponentially long time intervals by a symplectic integrator: 
\begin{equation}
    | \mathcal{H}(\tilde{\Phi}_{h}^n \bm{z}_0) - \mathcal{H}(\bm{z}_0) | = O \left( h^p \right)
    +
    O( n h e^{-\gamma/(\omega h)} ) \quad \forall n \geq 0 \,,
\end{equation}
where $p$ is the order of discretization error, $\gamma$ depends on the method, and $\omega$ is related to the Lipschitz constant of the vector field. That is, the energy computed oscillates around a mean value in a band of width $O(h^p)$. 

While energy conservation is quite simple to verify, reliably monitoring the conservation of symplecticity in a simulation is difficult. The ease of monitoring energy conservation follows from the energy conservation law's dependence on individual trajectories. In contrast, symplecticity pertains to the organization of trajectories in phase space; there is no simple method for checking symplecticity preservation along an individual trajectory without knowledge of nearby trajectories. As we will see, probing symplecticity conservation can be accomplished using parameterized families of trajectories. Despite this additional challenge in monitoring the conservation of symplecticity, the endeavor is worthwhile, as energy conservation is not an ideal figure of merit to monitor symplecticity. Symplectic integrators do not identically conserve energy, and many non-symplectic methods conserve energy exactly, both in the context of PIC \cite{chen2011energy,chen2014energy,chen2015multi, chen2020semi, chacon2013charge, chacon2016curvilinear, ricketson2025explicit, kormann2021energy, ji2023asymptotic}, or for general Hamiltonian systems using discrete gradients and related methods \cite{gonzalez1996time, mclachlan1999geometric, quispel2008new, celledoni2009energy, hairer2010energy}. 

\subsection{A conserved loop integral}

To introduce a more appropriate figure of merit to monitor symplecticity, it is necessary to briefly establish some notation and terminology. The \textit{Liouville $1$-form} is $\vartheta = \bm{p} \cdot \mathsf{d} \bm{q} = \sum_{i=1}^d p_i \mathsf{d} q^i$. Minus the exterior derivative of the Liouville $1$-form is the canonical symplectic form: $\omega_c = -\mathsf{d} \vartheta$. For a given Hamiltonian, $\mathcal{H}$, we denote the corresponding Hamiltonian vector field by $\bm{X}_H$, and write $\iota_{\bm X_H}$ for contraction with this vector field. The Hamiltonian vector field is determined by $\mathcal{H}$ and $\omega_c$ according to
\begin{equation} \label{eq:abstract_hamiltons_eqn}
    \iota_{\bm{X}_H} \omega_c = \mathsf{d} \mathcal{H} \,,
\end{equation}
or, equivalently, $\bm{X}_H = J_c \nabla \mathcal{H}$, since this work considers the state space $\mathbb{R}^{2d}$ with its standard inner product. When it exists, the \textit{flow map} for $\bm{X}_H$, $\Phi_t^\mathcal{H}:\mathbb{R}^{2d}\rightarrow\mathbb{R}^{2d}$, is the unique $t$-dependent mapping such that
\begin{align}
\forall \bm{z}_0\in\mathbb{R}^{2d},\quad \frac{d}{dt}\Phi_t^\mathcal{H}(\bm{z}_0) = \bm{X}_H(\Phi_t^\mathcal{H}(\bm{z}_0))\text{ and }\Phi_0^\mathcal{H}(\bm{z}_0) = \bm{z}_0.
\end{align}

It is well-known that $\omega_c$ is conserved by a Hamiltonian flow. Here $(\Phi_t^\mathcal{H})^*$ denotes pullback and $\mathcal L_{\bm X_H}$ denotes the Lie derivative:
\begin{equation} \label{eq:conservation_of_symp_2}
     \frac{\mathsf{d}}{\mathsf{d} t}  (\Phi_t^\mathcal{H})^* \omega_c
    =
    (\Phi_t^\mathcal{H})^*(\mathcal{L}_{\bm{X}_H} \omega_c)
    =
    (\Phi_t^\mathcal{H})^*(\mathsf{d} \iota_{\bm{X}_H} \omega_c + \iota_{\bm{X}_H} \mathsf{d} \omega_c)
    =
    (\Phi_t^\mathcal{H})^*(\mathsf{d}^2 \mathcal{H} + 0)
    = 0 \,,
\end{equation}
where the first equality follows from the definition of the Lie derivative, the second from Cartan's formula, the third from equation \eqref{eq:abstract_hamiltons_eqn} and the fact that $\omega_c$ is closed, and finally the last equality follows from the fact that $\mathsf{d}^2 = 0$. Note, this result is equivalent to the previously stated criterion for the time-advance map being symplectic. 

For any oriented $2$-dimensional submanifold $D \subset \mathbb{R}^{2d}$, we have that
\begin{equation}
    \int_{\Phi_t^\mathcal{H}(D)} \omega_c = \int_{D} (\Phi_t^\mathcal{H})^* \omega_c = \int_D \omega_c \,. 
\end{equation}
In particular, if $\partial D = \gamma$ such that $\gamma: \mathbb{R}/\mathbb{Z} \to \mathbb{R}^{2d} : \theta \mapsto (\gamma^x(\theta), \gamma^p(\theta))$ is a parameterized loop in phase-space, one may use Stokes' theorem to deduce a second conserved integral:
\begin{equation} \label{eq:poincare_integral_inv}
    \mathcal{I}(t)
    =
    \int_{\Phi_t^\mathcal{H}\circ \gamma } \vartheta 
    =
    \int_{\gamma} \vartheta \,.
\end{equation}
This follows because Stokes' theorem implies
\begin{equation} \label{eq:conservation_of_integral_invariant}
    \int_{\Phi_t^\mathcal{H}\circ \gamma } \vartheta = \int_{\partial \Phi_t^\mathcal{H}(D)}\vartheta
    =
    -\int_{\Phi_t^\mathcal{H}(D)} \omega_c
    =
    -\int_{D} \omega_c
    =
    \int_{\gamma} \vartheta \,,
\end{equation}
as long as $\partial(\Phi_s^\mathcal{H}(D)) = \Phi_s^\mathcal{H} \circ \gamma$. This is ensured if the map $\Phi_s^\mathcal{H}$ is a bijective $C^0$ map. Stokes' theorem is being applied to a smooth spanning surface on which $\vartheta$ and $\omega_c$ are smooth and well-defined. The loop integral $\mathcal{I}(t)$ defined in equation \eqref{eq:poincare_integral_inv} is known as the (first) \emph{Poincar{\'e} integral invariant}. This quantity is the basis for the numerical diagnostic tool considered in this work. 

\subsection{Approximating the loop integral} \label{sec:approx_loop_int}

An arbitrary loop $\gamma$ in phase space cannot generally be computed
analytically. Hence, we must approximate the integral $\int_\gamma \vartheta$.
We wish to make this approximation only having access to a finite number of
samples of a parameterized loop. That is, given
$\gamma: \mathbb{R}/\mathbb{Z} \to \mathbb{R}^{2d}$, we have access to the
data
\begin{equation}
    \gamma(s_k)
    =
    (\bm{\mathsf{q}}_k, \bm{\mathsf{p}}_k)
    \in \mathbb{R}^d \times \mathbb{R}^d \,,
    \qquad
    s_k = (k-1)\Delta s \,,
    \quad
    k = 1,\ldots,N_s \,,
\end{equation}
where $\Delta s = 1/N_s$. We use periodic indexing throughout, so that
$\bm{\mathsf{q}}_{N_s+1}=\bm{\mathsf{q}}_1$ and
$\bm{\mathsf{p}}_{N_s+1}=\bm{\mathsf{p}}_1$. As a matter of notation, we
collect the loop-data in matrices
$\bm{\mathsf{q}}, \bm{\mathsf{p}} \in \mathbb{R}^{N_s \times d}$, such that
\begin{equation}
    \mathsf{q}_{ki} = q_i(s_k) \,,
    \qquad
    \mathsf{p}_{ki} = p_i(s_k) \,.
\end{equation}
This organization of the data makes it convenient to apply linear operators
in loop space as matrices. In particular, the derivative in loop space is
applied as a matrix. We consider two different approximations of the loop
integral whose convergence rates depend on the regularity of the loop data:
a Fourier-pseudospectral approximation and a finite difference approximation.
The proofs of convergence for these two methods are given in
\ref{appendix:conv_high_regularity} and
\ref{appendix:conv_low_regularity}, respectively.

\paragraph{Pseudospectral approximation:} Let $\mathbb{F}$ be the discrete
Fourier transform matrix, and let $\bm{\kappa}$ denote the vector of signed
Fourier modes associated with the $N_s$-point DFT. With the convention
$e^{2\pi i k s}$, the pseudospectral approximation to the derivative in loop
space is given by
\begin{equation}
    \mathbb{D} \bm{\mathsf{q}}
    =
    \mathbb{F}^{-1}
    \left(
        2\pi i\,\operatorname{diag}(\bm{\kappa})
        \mathbb{F}\bm{\mathsf{q}}
    \right) .
\end{equation}
To be absolutely clear,
\begin{equation}
    (\mathbb{D}\bm{\mathsf{q}})_{ki}
    =
    \sum_{j=1}^{N_s}
    \mathbb{D}_{kj}\mathsf{q}_{ji}.
\end{equation}
For smooth periodic loop data this approximation converges rapidly, and for
Sobolev data the algebraic convergence rate is quantified in
\ref{appendix:conv_high_regularity}. Using trapezoidal rule, the
loop integral may be approximated as follows:
\begin{equation}
    \mathcal{I}
    =
    \oint_{\gamma} \bm{p} \cdot \mathsf{d} \bm{q}
    =
    \int_0^1 \bm{p}(s) \cdot \frac{\partial \bm{q}}{\partial s} \mathsf{d} s
    \approx
    \bm{\mathsf{p}}
    :
    (\mathbb{D} \bm{\mathsf{q}})
    \Delta s
    =
    \sum_{i=1}^{d}
    \sum_{j,k=1}^{N_s}
    \mathsf{p}_{ki}
    \mathbb{D}_{kj}
    \mathsf{q}_{ji}
    \Delta s
    \eqcolon
    \mathcal{I}^{\mathrm{ps}}_{N_s} \,,
\end{equation}
where the colon notation contracts two matrices: for
$A,B \in \mathbb{R}^{n \times m}$,
$A:B = \sum_{i=1}^n\sum_{j=1}^m A_{ij} B_{ij}$.

\begin{remark}
    As a technical note, if $(\bm{q},\bm{p})$ represents the state of a
    discretized Hamiltonian PDE system with periodic boundary conditions, it
    is necessary to unwrap the data prior to computing the pseudospectral
    loop integral approximation to eliminate jumps at the edges of the
    spatial domain.
\end{remark}

\paragraph{Finite difference approximation:} For low-regularity or
discontinuous loop data, we use a one-sided quadrature rule based on
increments of $\bm{q}$. For each component $i=1,\ldots,d$, let
$\mathtt{K}_i$ denote the set of retained grid cells. Then
\begin{equation}
    \mathcal{I}^{\mathrm{fd}}_{N_s}
    \coloneq
    \sum_{i=1}^d
    \sum_{k \in \mathtt{K}_i}
    \mathsf{p}_{ki}
    \left(
        \mathsf{q}_{k+1,i} - \mathsf{q}_{ki}
    \right) \,.
\end{equation}
Here and below, sample indices are understood periodically modulo \(N_s\).
On cells containing no discontinuity, the increment
$q_i(s_{k+1})-q_i(s_k)$ is exactly the integral of $q_i'$ over that cell.
Thus, on retained cells, the quadrature error comes from approximating
$p_i(s)$ by its left endpoint value $p_i(s_k)$.

In practice, the retained set $\mathtt{K}_i$ may be estimated using
finite-difference cutoffs. Given dimensionless cutoff parameters
$\lambda_q,\lambda_p>0$, one may take
\begin{multline}
    \mathtt{K}_i(\lambda_q,\lambda_p)
    =
    \bigg\{
        k \in \{1,\hdots,N_s\}
        :
        \left|
        \frac{\mathsf{p}_{k+1,i} - \mathsf{p}_{ki}}{\Delta s}
        \right|
        \leq
        \lambda_p \| p_i \|_{L^\infty([0,1])}
        \\
        \text{and}
        \left|
        \frac{\mathsf{q}_{k+1,i} - \mathsf{q}_{ki}}{\Delta s}
        \right|
        \leq
        \lambda_q \| q_i \|_{L^\infty([0,1])}
    \bigg\} \,.
\end{multline}
This cutoff procedure is a heuristic for detecting cells containing jump
discontinuities. The convergence result in
\ref{appendix:conv_low_regularity} assumes that all cells
containing jumps have been omitted; the cutoff rule inherits this estimate
when it successfully removes these cells and introduces only a controlled
number of false positives.

\begin{remark}
    This approximation converges quite slowly; see
    \ref{appendix:conv_low_regularity}. Therefore, the finite
    difference approximation should only be used in cases where the
    pseudospectral approximation fails to converge, e.g.\ loops with
    discontinuities.
\end{remark}

\begin{remark}
    To our knowledge, there is no complete theory for the conservation of the
    Poincar{\'e} integral invariant by Hamiltonian flows with insufficient
    regularity to guarantee the spatial continuity of the flow map, e.g.\
    systems with discontinuous Hamiltonian vector fields. Indeed, the
    classical existence and uniqueness theory for ODEs assumes Lipschitz
    vector fields \cite{hartman2002ordinary}, and, for systems with less
    regularity, one must resort to a weaker interpretation of the solution
    map \cite{ambrosio2004transport}. Despite these difficulties, we
    consider an approximation of the loop integral which remains defined even
    for discontinuous data for reasons discussed in
    Section~\ref{sec:low_regularity_systems}.
\end{remark}

\subsection{Convergence of the loop integral approximations}

In \ref{appendix:conv_high_regularity}, we show that the scalar
Fourier-pseudospectral approximation satisfies
\begin{equation}
    |
    \mathcal{I}(p,q)
    -
    \mathcal{I}^{\mathrm{ps}}_N(p,q)
    |
    \leq
    C_r
    N^{-\beta(r)}
    \|p\|_{H^r}
    \|q\|_{H^r},
    \qquad
    \beta(r)=\min\{2r-1,r\},
\end{equation}
for $p,q \in H^r_{per}([0,1])$ and $r>1/2$. In
\ref{appendix:conv_low_regularity}, we show that the finite
difference approximation satisfies a first-order estimate for piecewise
smooth data with a finite number of jumps, provided the cells containing
jumps are omitted.

In these appendices, we demonstrate the convergence of approximations to
loop integrals of the form
\begin{equation}
    \mathcal{I}(p,q)
    =
    \int_0^1 p(s) \mathsf{d} q(s)
    =
    \int_0^1 p(s) q'(s) \mathsf{d} s \,.
\end{equation}
In the low-regularity case, this integral is understood in the piecewise
smooth sense described in \ref{appendix:conv_low_regularity}, where
the jump contribution to $\mathsf{d}q$ is omitted. Higher-dimensional phase
spaces are handled by adding up the individual one-dimensional loops:
\begin{equation}
    \mathcal{I}(\bm{p}, \bm{q})
    =
    \int_\gamma \bm{p} \cdot \mathsf{d} \bm{q}
    =
    \int_0^1 \bm{p}(s) \cdot
    \frac{\mathsf{d} \bm{q}}{\mathsf{d} s}
    \mathsf{d} s
    =
    \sum_{i=1}^{d}
    \int_0^1 p_i(s)
    \frac{\mathsf{d} q_i}{\mathsf{d} s}
    \mathsf{d} s \,.
\end{equation}
The triangle inequality allows us to reduce the convergence analysis of
approximations to loop integrals in higher-dimensional phase space to the
scalar case. This is because both the continuous and approximate loop
integrals split additively over each degree of freedom. Hence, the
pseudospectral approximation satisfies
\begin{equation}
    |
    \mathcal{I}(\bm{p}, \bm{q})
    -
    \mathcal{I}^{\mathrm{ps}}_{N_s}(\bm{p}, \bm{q})
    |
    \leq
    C_r N_s^{-\beta(r)}
    \sum_{i=1}^{d}
    \|p_i\|_{H^r}
    \|q_i\|_{H^r}.
\end{equation}
In particular, if the component Sobolev norms are bounded uniformly in $i$,
then the error scales like
\begin{equation}
    |
    \mathcal{I}(\bm{p}, \bm{q})
    -
    \mathcal{I}^{\mathrm{ps}}_{N_s}(\bm{p}, \bm{q})
    |
    =
    O(d N_s^{-\beta(r)}) \,.
\end{equation}
Similarly, the finite difference approximation satisfies a componentwise
estimate of the form
\begin{equation}
    |
    \mathcal{I}(\bm{p}, \bm{q})
    -
    \mathcal{I}^{\mathrm{fd}}_{N_s}(\bm{p}, \bm{q})
    |
    \leq
    \frac{1}{N_s}
    \sum_{i=1}^{d}
    L_{q_i}
    \left(
        \mathrm{TV}_{[0,1]}(p_i)
        +
        \#(\mathcal{O}_{N_s,i})
        \|p_i\|_{L^\infty([0,1])}
    \right),
\end{equation}
where $L_{q_i}$ is the piecewise $L^\infty$ bound on $q_i'$ and
$\mathcal{O}_{N_s,i}$ is the set of omitted cells for the $i$-th component.
Under uniform bounds, this gives
\begin{equation}
    |
    \mathcal{I}(\bm{p}, \bm{q})
    -
    \mathcal{I}^{\mathrm{fd}}_{N_s}(\bm{p}, \bm{q})
    |
    =
    O(d N_s^{-1}) \,.
\end{equation}
For high-dimensional systems with low regularity, a large number of sample
points may be needed.

For high-regularity loop data, we do not anticipate any other approximation
based on equispaced loop samples to converge faster than the
Fourier-collocation approach. In the low-regularity case, the finite
difference approximation of the loop integral converges slowly. For
discontinuous data, high-order convergence from equispaced samples is
generally not available without additional information about the locations
and sizes of the discontinuities. Endpoint-corrected or discontinuity-aware
quadrature rules can recover higher-order convergence when such information
is available, but this is not the setting of the present diagnostic
\cite{fornberg2019improved}. Generally, we will not know the location of the
discontinuities when applying these loop integral approximations in a
symplectic diagnostic tool.

To decide which approximation scheme to use, one requires some a priori
knowledge of the regularity of the loop parameterization. For smooth
periodic loops, the pseudospectral approximation is preferred. For
discontinuous or piecewise smooth loops where jump cells can be identified
and omitted, the finite difference approximation is more robust. When both
estimates are applicable, the pseudospectral rate
$N_s^{-\min\{2r-1,r\}}$ should be compared with the first-order finite
difference rate. In particular, for $r>1$, the pseudospectral estimate is
asymptotically faster than first order, while at $r=1$ the two estimates are
both first order.

\begin{remark}
    These results establish sufficient criteria for convergence. If the
    regularity of a given loop has been ascertained and an appropriate
    resolution has been used, then it is safe to conclude that the integral
    approximation may be trusted. However, because there may be uncertainty
    about the regularity of the loop data, and approximations to
    low-regularity loop integrals may converge slowly, it is frequently
    helpful to perform a parameter sweep to verify convergence.
\end{remark}

\subsection{Approximating the loop integral as a function of time}

In order to compute the approximate loop integral for multiple temporal
snapshots of a simulation, it is necessary to consider how the regularity of
an initially smooth loop in phase space deteriorates as the system evolves.

For a general autonomous ODE, $\dot{x} = F(x)$, over a compact
$n$-dimensional manifold without boundary, $M$, with solution map
$\Phi_t: M \to M$, if $F \in H^r(M)$, then one may show that
$\Phi_t \in H^r(M)$ as a function of space \cite{ebin1970groups}. Here
$\Phi_t$ denotes the true solution, not an approximation by some integration
scheme. After restricting attention to a compact region of phase space, this
suggests the following heuristic. Given a Hamiltonian
$\mathcal{H} \in H^r(T^*\mathbb{T})$ locally on such a region, where
$\mathbb{T}$ denotes the one-dimensional torus and $T^*\mathbb{T}$ denotes its
cotangent bundle, an initially smooth curve $\gamma$ may evolve to a loop with
only $H^{r-1}$ regularity. The
pseudospectral loop integral approximation is guaranteed to converge when
the evolved loop lies in $H^\rho$ for some $\rho>1/2$, in which case the
rate is $N_s^{-\min\{2\rho-1,\rho\}}$. Thus, in the heuristic situation
where the evolved loop has regularity $H^{r-1}$, the pseudospectral estimate
requires $r>3/2$.

The least regular Hamiltonians considered in this work have piecewise
linear potentials. Compactly supported piecewise linear functions on
$\mathbb{R}$ can be shown to be in $H^{3/2-\epsilon}(\mathbb{R})$ for every
$\epsilon > 0$; see \ref{appendix:piecewise_linear_regularity}.
The corresponding evolved loops fall below the $H^{1/2}$ threshold suggested
by this heuristic, and the pseudospectral convergence theorem does not
apply. For such examples, we use the finite difference approximation when
the loop data are piecewise smooth and the jump cells can be identified.

For sufficiently regular Hamiltonians, convergence of the pseudospectral
approximation is rapid enough that the loop integral can be approximated
with relatively small ensembles parameterizing the loop. This is crucial
since the approximation can be quite costly for high-dimensional systems:
$N_s$ independent simulations must be run to compute the loop integral. This
is somewhat mitigated by the fact that these simulations can be done in
parallel. Moreover, as we shall see subsequently, an effective way to
monitor symplecticity conservation is to check the error incurred in the
loop integral over a single time-step---likewise mitigating the cost.

As a final note, a distinction must be made between the exact flow map
$\Phi_t$ and its approximation via a symplectic integrator. As we will see
in Section~\ref{sec:low_regularity_systems}, certain low-regularity systems
with symplectic continuous flow maps yield non-symplectic approximate flow
maps when integrated with standard symplectic integrators like Strang
splitting. The conservation or non-conservation of the loop integral over
successive time-steps is entirely determined by the time-advance map used to
generate the data at successive snapshots in time.

\begin{remark}
    The regularity of the Hamiltonian generating the flow determines the
    regularity of the loop which in turn determines the convergence rate of
    the loop integral approximations. Some delicacy is required, however, as
    the discrete-time flow obtained from a time-stepping method may not have
    the same spatial regularity as the continuous-time flow.
\end{remark}

\begin{remark}
    It is worth briefly highlighting how the approximate loop integral
    differs from energy as a computable figure of merit. To compute the
    approximate loop integral for a temporal simulation of a Hamiltonian
    system, $N_s$ different simulations with initial data parameterizing the
    loop must be performed. This is in contrast with energy, which may be
    computed for a single trajectory. This is because symplecticity is a
    property associated with the geometry of phase-space and can only be
    measured with an ensemble of trajectories.
\end{remark}
 
\section{A computable diagnostic for symplecticity}
This section introduces the core contribution of this paper: a numerical diagnostic to monitor the conservation of symplecticity. Although applicable to any time-stepping method for Hamiltonian systems, the diagnostic is formulated as a tool for particle-in-cell methods. As such, special attention is given to difficulties arising when applying the tool to high-dimensional, low-regularity systems. 

\subsection{A diagnostic for symplecticity}

Suppose that we evolve a phase-space loop in time, keeping track only of a finite number of sub-samples determined by the initial parameterization of the loop:
\begin{equation}
    \{ (\bm{\mathsf{q}}_{k+1}(t), \bm{\mathsf{p}}_{k+1}(t)) \}_{k=0}^{N_s-1}
    \quad \text{such that} \quad
    (\bm{\mathsf{q}}_{k+1}(0), \bm{\mathsf{p}}_{k+1}(0)) = \gamma(k \Delta s) \,, 
\end{equation}
for $k \in \{0, 1, 2, \hdots, N_s-1 \}$, where $\Delta s = 1/N_s$. We may approximate the loop integral in time simply by applying one of the loop integral approximations from section \ref{sec:approx_loop_int} to the time-dependent loop. In the case of the pseudospectral approximation, we have
\begin{equation}
    \mathcal{I}_N^{\mathrm{ps}}(t)
    =
    \bm{\mathsf{p}}(t)
    :
    (\mathbb{D} \bm{\mathsf{q}}(t))
    \Delta s
    =
    \sum_{i=1}^{d}
    \sum_{j,k=1}^{N_s}
    \mathsf{p}_{ki}(t)
    \mathbb{D}_{kj}
    \mathsf{q}_{ji}(t)
    \Delta s \,.
\end{equation}
In the case of the finite difference approximation, we have
\begin{equation}
    \mathcal{I}_{N}^{\mathrm{fd}}(t)
    =
    \sum_{i=1}^d
    \sum_{k \in \mathtt{K}_i(\lambda_q,\lambda_p)} \mathsf{p}_{ki}(t) ( \mathsf{q}_{k+1,i}(t) - \mathsf{q}_{ki}(t)) \,.
\end{equation}
While $\mathcal{I}(t) = \int \bm{p} \cdot \mathsf{d} \bm{q}$ is conserved exactly by the flow, $\mathcal{I}_N(t)$ is conserved up to discretization error. The figure of merit considered in this work is the relative error incurred by the loop integral over a single time-step:
\begin{equation}
    \mathcal{E}(h, N_s)
    =
    \left|
    \frac{\mathcal{I}_N(h) - \mathcal{I}_N(0)}{\mathcal{I}_N(0)}
    \right|
    =
    \left|
    \exp
    \left(
        \log |\mathcal{I}_N(h)|
        -
        \log |\mathcal{I}_N(0)|
    \right)
    - 1
    \right| \,,
\end{equation}
with the second equality holding provided $\mathcal{I}_N(h)$ and $\mathcal{I}_N(0)$ are non-zero and have the same sign. The logarithmic form is algebraically equivalent in this case, but can be evaluated more stably for small relative errors. For a Hamiltonian system with $2d$ degrees of freedom, if the time advance map is symplectic, then, for the pseudospectral approximation, we find that $\mathcal{E}(h, N_s) = O(d N_s^{-\beta(r)})$ uniformly in $h$, where $r$ denotes the Sobolev regularity of the loop data entering the integral and $\beta(r) = \min\{2r-1,r\}$. For the finite difference approximation, $\mathcal{E}(h,N_s) = O(d N_s^{-1})$. Hence, assuming the loop is adequately resolved, this figure of merit is a proxy for symplecticity conservation. 

The diagnostic tool is defined as a parameter sweep in time-step size in which $\mathcal{E}(h, N_s)$ is computed for each value of $h$. To be specific:
\begin{itemize}
    \item To achieve $R$  digits of accuracy, initialize a smooth loop in phase-space with 
    \begin{itemize}
        \item at least $N_s = O((d \times 10^R)^{1/\beta(r)})$ equispaced points for the pseudospectral approximation if the integrand is $H^r$ with $r > 1/2$,
        \item or with at least $N_s = O(d \times 10^R)$ equispaced points for the finite difference approximation if the integrand has low regularity. 
    \end{itemize}  
    \item Compute $\mathcal{E}(h, N_s)$ for a range of time-steps $h = 2^{-m}$ for $m \in \{1, 2, \hdots, M\}$. 
    \item Plot the error versus time-step trend in log-log space.
\end{itemize}
A symplectic method should achieve $R$ digits of accuracy regardless of time-step size, while a non-symplectic method will recover the convergence trend of the time-stepping method (with some caveats discussed below). Deviation from this expected behavior signals either that the integral was not properly resolved, in which case one should refine and try again, or that the method is not conserving the loop integrals---a signal that the method is not symplectic. Note that some experimentation may be needed to determine a suitable number of points since other factors (e.g.\! the local Lyapunov exponent of the flow) can limit convergence in addition to regularity. 

It is possible that the integrator incurs errors in the loop integral diagnostic smaller than the number of digits a given simulation is capable of resolving. Integrators which conserve symplecticity up to $O(h^k)$, for some $k>0$ are more challenging to distinguish from true symplectic integrators. Further, in low-regularity cases, an unreasonable number of points may be required to resolve the approximate loop integral to machine precision. In such cases, it is still possible to convincingly distinguish discretization errors from symplecticity errors by additionally running a convergence sweep in the parameter $N_s$ with an extremely coarse time-step, e.g. $h = O(1)$. A true symplectic method exactly conserves the exact loop integral regardless of time-step size. The primary issue in these corner cases is distinguishing approximation error of the loop integral from errors in symplecticity conservation, and the large time-step amplifies potential errors in symplecticity preservation. If the expected error trend continues to the maximum resolution tested, either the method is symplectic, or the error in symplecticity is too small to be registered by the test. However, if the trend plateaus at a finite value above machine precision, then one has a source of error above and beyond discretization error, i.e.\! an error in symplecticity conservation for the time integration scheme itself. 

\subsection{Illustrative example: the shearing of phase-space loops}

\begin{figure}[t]
    \centering

\begin{subfigure}[t]{0.49\textwidth}
        \centering
        \includegraphics[width=\textwidth]{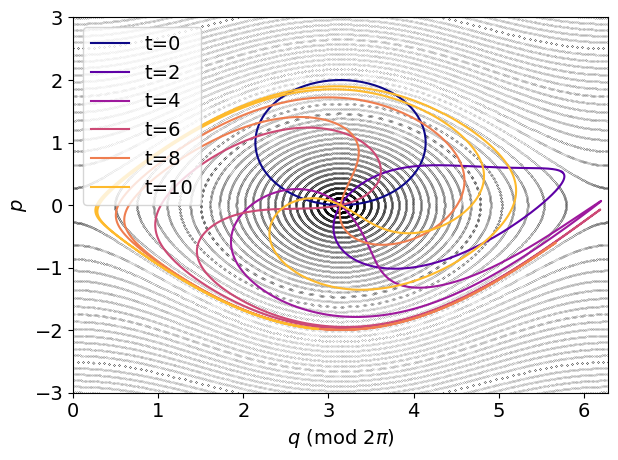}
        \caption{Visualization of the deformation of an initially smooth loop by the nonlinear pendulum.}
        \label{fig:nl_pend_phase_space}
    \end{subfigure}
    \hfill \begin{subfigure}[t]{0.49\textwidth}
        \centering
        \includegraphics[width=\textwidth]{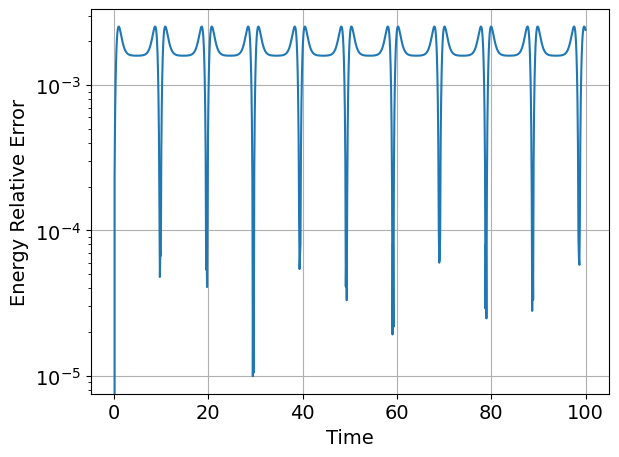}
        \caption{Energy conservation for trajectory starting at $(q_0,p_0) = (1,0)$.}
        \label{fig:nl_pend_energy}
    \end{subfigure}
    
\caption{Visualization of the behavior of the nonlinear pendulum.}
    \label{fig:nl_pend_vis}
\end{figure}

\begin{figure}[t]
    \centering
    
\begin{subfigure}[t]{0.49\textwidth}
        \centering
        \includegraphics[width=\textwidth]{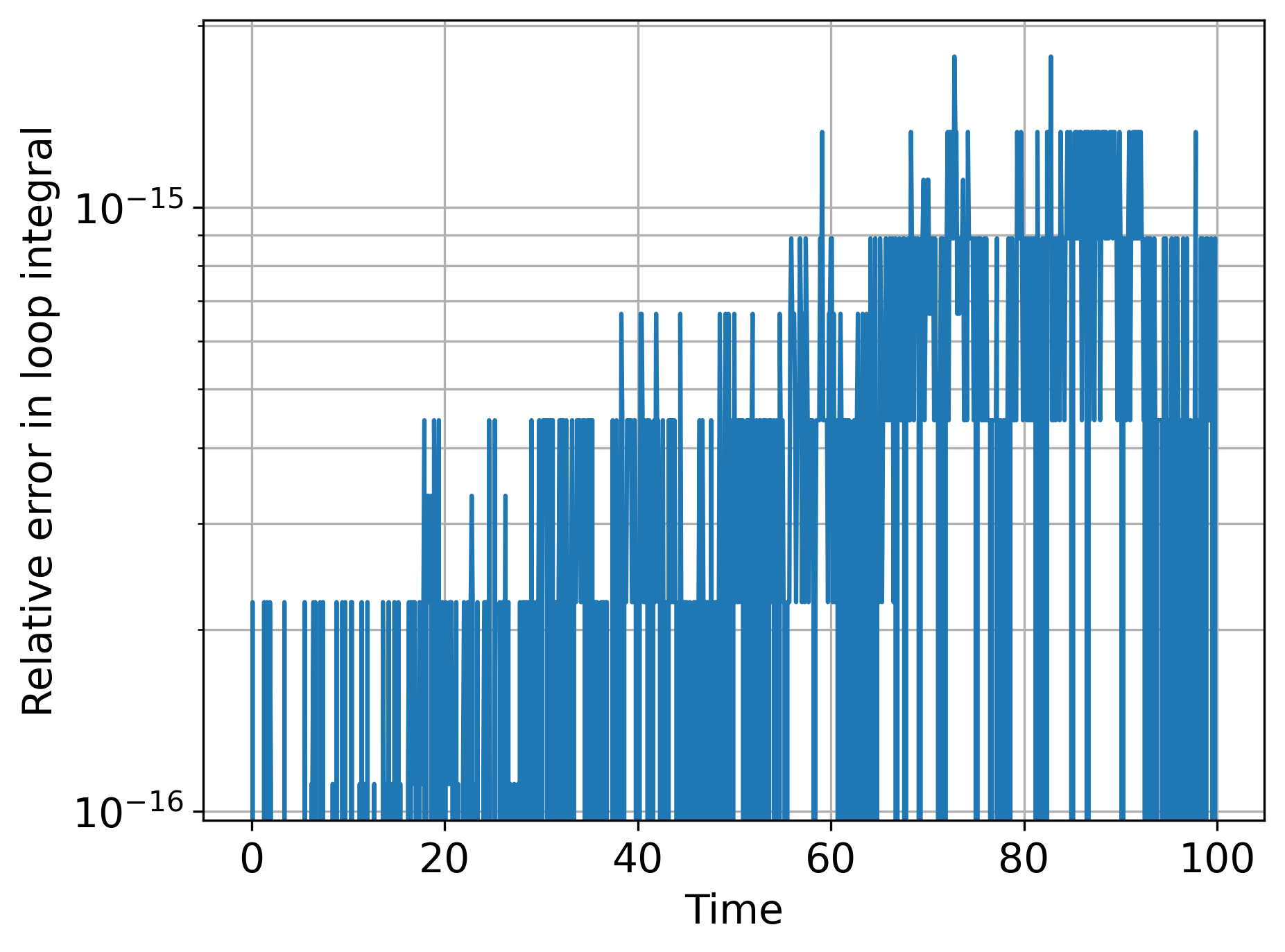}
        \caption{Approximate loop integral as a function of time. Initial loop of radius $1$ centered at $(q_0,p_0) = (\pi,1/2)$.}
        \label{fig:nl_pend_diag_good}
    \end{subfigure}
    \hfill \begin{subfigure}[t]{0.49\textwidth}
        \centering
        \includegraphics[width=\textwidth]{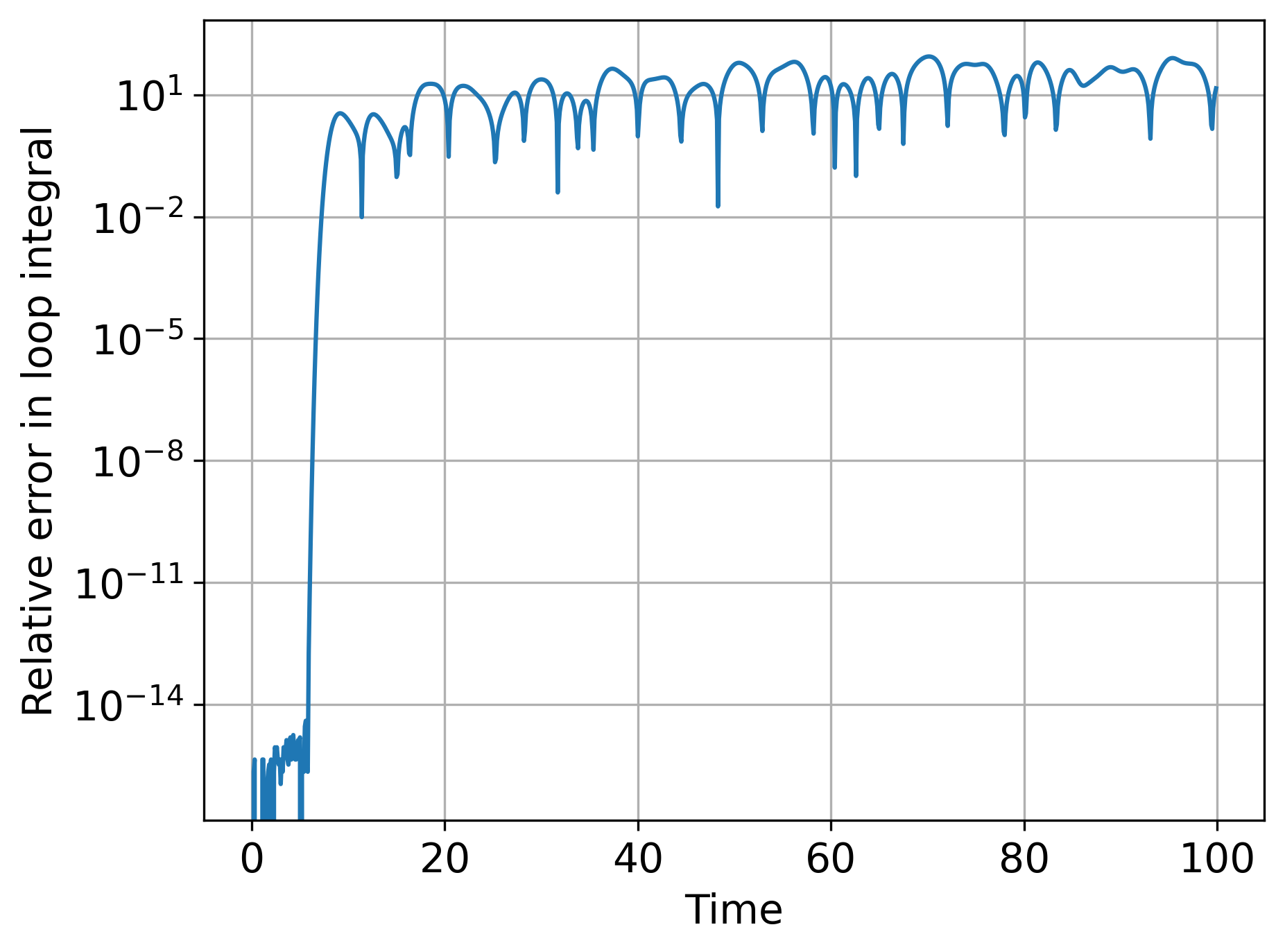}
        \caption{Approximate loop integral as a function of time. Initial loop of radius $1$ centered at $(q_0,p_0) = (0,0)$.}
        \label{fig:nl_pend_diag_xpoint}
    \end{subfigure}
    
\caption{Comparison of approximate loop integral conservation by the nonlinear pendulum for two different initial loops.}
    \label{fig:nl_pend_diag}
\end{figure}

It is illustrative to look at the nonlinear pendulum, i.e. $\mathcal{H}(q,p) = p^2/2 + \sin(q)$. This system is especially nice since its Hamiltonian is $C^\infty$ and the system is low dimensional. Therefore, this example serves to illustrate the difficulties the loop integral approximation can encounter in even the tamest systems. Because the flow is smooth, we use the pseudospectral method to approximate the loop integral. See Figure \ref{fig:nl_pend_phase_space} for a visualization of the evolution of an initial loop under the time-advance map and \ref{fig:nl_pend_energy} for the near conservation of energy by Strang splitting for a particular trajectory. See Figure \ref{fig:nl_pend_diag} for the relative error in the approximate loop integral as a function of time for two different initial loops. These figures demonstrate the severe shearing of the initial loop over time. Moreover, the number of points needed to resolve the approximate loop integral for a long simulation varies greatly depending on the location of the loop. However, in both cases considered in Figure \ref{fig:nl_pend_diag}, if the loop is well-resolved initially, it remains well-resolved for at least a few time-steps before the shearing of phase-space causes issues. The behavior seen in this example is quite generic and motivates our definition of the diagnostic tool as a calculation over a single time-step. Of course, there may be utility in considering errors in the loop integral over longer simulations, but such considerations are outside the scope of this work. 

\subsection{Conservative, nearly-symplectic systems}

As mentioned previously, nearly-symplectic systems are particularly difficult to distinguish from symplectic systems using the numerical diagnostic. As such, it is worth highlighting a particularly well-known case of near-symplecticity to alert the reader that extra care may be needed when studying such time-steppers. A family of non-symplectic time-stepping methods known as conjugate symplectic methods nearly conserve the Poincar{\'e} integral invariants over long simulations \cite{kraus2017projectedvariationalintegratorsdegenerate}. If $\Phi_{h}^{CS}$ is the flow of the conjugate symplectic method, there exists a symplectic map, $\Phi_{h}^{S}$, and a near-identity formal diffeomorphism, $\chi_{h}$, such that
\begin{equation}
    \Phi_{h}^{CS} = \chi_{h}^{-1} \circ \Phi_{h}^{S} \circ \chi_{h} \,.
\end{equation}
By near-identity, we mean that $\chi_{h}(\bm{z}) = \bm{z} + O(h^p)$ for some $p \geq 1$. Such methods can be shown to possess long-time near-symplecticity:
\begin{equation}
    \left| (D_{\bm{z}} \Phi_{h}^{CS})^T J_c D_{\bm{z}} \Phi_{h}^{CS} - J_c \right| = O(h^p) \,.
\end{equation}
An example of a conjugate symplectic method is the leapfrog method (which staggers position and momentum updates in time): this method differs from Strang splitting (or St{\"o}rmer-Verlet) by a half-timestep. Certain energy-conserving methods, including the popular averaged vector field discrete gradient method \cite{quispel2008new}, have been shown to possess a slightly weakened form of conjugate symplecticity \cite{hairer2010energy}. This work compares only Strang splitting (an explicit symplectic method) and second-order explicit Runge-Kutta (RK2) as a baseline for comparison. Conjugate symplectic and energy conserving methods are outside the scope of this work. However, the performance of the diagnostic on these systems is of substantial interest and worth pursuing in a future work. 
 
\section{On the symplecticity of low regularity Hamiltonian systems}
We now consider the symplecticity of flow maps of low regularity Hamiltonian
systems and the symplecticity of their discrete-time approximation. We find
that Hamiltonian systems with discontinuous vector fields, even when their
continuous-time flow is symplectic, can yield discrete-time flow maps which
are not symplectic when approximated using Hamiltonian splitting methods.
This is notable, because Hamiltonian systems with piecewise linear
potentials, commonly found in PIC methods, can fall into this category. In
particular, the classic and ubiquitous \textit{cloud-in-cell} scheme uses
linear shape functions \cite{birdsall2018plasma, hockney2021computer,
villasenor1992rigorous}. Many structure-preserving particle-in-cell methods,
while designed to be compatible with shape-functions of general polynomial
degree, do not explicitly disallow the use of linear shape functions
\cite{kraus2017gempic, glasser2022gauge, glasser2023generalizing,
burby2017finite, morrison2017structure}. Finally, it is not just piecewise
linear interpolation which can be problematic: interpolation schemes which do
not yield a globally $C^1$ interpolant, e.g.\ PIC methods with field
discretizations based on $C^0$ finite elements \cite{bettencourt2021empire,
glasser2022gauge, glasser2023generalizing, finn2023numerical} or Lagrange
interpolation \cite{kormann2024dual}, can exhibit the same issues described
in this section. The following discussion demonstrates this behavior in a
simple model problem.

\subsection{Interpretation of flow maps for low-regularity Hamiltonian systems} \label{sec:low_regularity_flow}

Given a Hamiltonian system with
\begin{equation}
    \mathcal{H}(q,p) = \frac{1}{2}p^2 + V(q) \,,
\end{equation}
where $V(q)$ is only piecewise smooth, the associated Hamiltonian vector
field may have jump discontinuities. Consequently, the classical existence
and uniqueness theory for ODEs does not apply directly, and one must adopt a
notion of weak solution. One possible interpretation is to approximate $V$
by smooth potentials and study limits of the corresponding smooth
Hamiltonian flows:
\begin{equation}
    \mathcal{H}_\epsilon(q,p)
    =
    \frac{1}{2}|p|^2 + V_\epsilon(q) \,,
\end{equation}
where $V_\epsilon \in C^\infty(\mathbb{R}^d)$ and $V_\epsilon \to V$ away
from the nonsmooth set, e.g.\ $V_\epsilon = V \ast \eta_\epsilon$ with a
standard mollifier $\eta_\epsilon$. Under suitable compactness and uniqueness
assumptions, such limits are related to the notion of regular Lagrangian
flow \cite{ambrosio2004transport}, which rigorously defines solution maps
for ODEs generated by vector fields of bounded variation.

To our knowledge, the conservation of the Poincar{\'e} integral invariant in
the regular Lagrangian flow framework has not been studied in the literature.
Intuitively, preservation of the Poincar{\'e} integral invariant is tied to
the preservation of phase space loop topology. The continuous-time flow of a
piecewise smooth Hamiltonian system should preserve the topology of loops,
provided the initial loop excludes points whose trajectories encounter the
nonsmooth set in a singular manner. In many practical cases, this set of
problematic initial conditions has zero Lebesgue measure. For example, if the
nonsmooth sets are $C^1$ hypersurfaces, then tangential approaches are
typically exceptional; for a mechanical Hamiltonian, such a tangential
approach corresponds to $p\cdot n(q)=0$ at the crossing, where $n(q)$ is the
normal to the hypersurface. Hence, we expect that the Poincar{\'e} integral
invariant may be conserved for a large class of low-regularity Hamiltonian
systems interpreted in this weak sense, even though a full rigorous proof is
not yet available.

\subsection{Hamiltonian splitting flow maps for piecewise linear potentials} \label{sec:low_regularity_systems}

Consider the canonical Hamiltonian system generated by the Hamiltonian
$\mathcal{H}(q,p) = p^2/2 + |q|$:
\begin{equation} \label{eq:piecewise_linear_ex}
\begin{aligned}
    \dot{q} &= p \,, \\
    \dot{p} &= - \operatorname{sign}(q) \,.
\end{aligned}
\end{equation}
The system is formally equivalent to the model of a particle falling in a
uniform gravitational potential which experiences a perfect elastic collision
with the boundary $q=0$, and parabolic trajectories for $q>0$. Such a system
is known as a hybrid system \cite{clark2023invariant}, so named because it
possesses qualities of both continuous and discrete dynamical systems, and
its flow may be shown to preserve the symplectic form. For this system, if
we consider trajectories which remain within a compact subset of
$\mathbb{R}^2$, we may cut off the potential in a smooth manner without
affecting the dynamics. After such a localization, the nonsmooth part of the
Hamiltonian has $H^{3/2-\epsilon}$ regularity in $q$, for every
$\epsilon>0$. Thus, this example lies outside the regime where the
pseudospectral convergence theorem is naturally applicable, while the
finite-difference loop integral approximation remains applicable to the
piecewise smooth loop data, provided jump cells are identified and omitted.
In what follows, we show that Hamiltonian splitting time-advance maps can
fail to conserve the loop integral over a single time-step.

Given a general Hamiltonian system with $\mathcal{H}(q,p) = K(p) + V(q)$,
the most basic Hamiltonian splitting method, the Lie-Trotter method, is
obtained from the partial flow maps
\begin{equation}
    \Phi_{h}^K(q,p)
    =
    (q + h \nabla K(p), p) \,,
    \quad \text{and} \quad
    \Phi_{h}^V(q,p)
    =
    (q, p - h \nabla V(q)) \,,
\end{equation}
which are obtained by exactly integrating the Hamiltonian system while
alternately setting the kinetic or potential energy to zero over a single
time-step, $h$. The two $O(h)$ Lie-Trotter update maps are defined to be
\begin{equation}
    \Phi^q_{h}(q,p)
    =
    \Phi_{h}^V
    \circ
    \Phi_{h}^K(q,p) \,,
    \quad \text{or} \quad
    \Phi^p_h(q,p)
    =
    \Phi_h^K
    \circ
    \Phi_h^V(q,p) \,,
\end{equation}
where $\Phi^q_{h}$ might be called the \textit{position-first update}, and
$\Phi^p_{h}$ the \textit{momentum-first update}. When applied to Hamiltonian
systems with sufficient regularity, these Lie-Trotter update rules are
symplectic maps, and higher-order splitting methods, such as
Strang-splitting, may be obtained as compositions of basic steps of this form
\cite{geo_num_int}.

Suppose that one wishes to integrate the original absolute value potential
system using Hamiltonian splitting. The partial flow-maps are given by
\begin{equation}
    \Phi_{h}^K(q,p)
    =
    (q + h p, p) \,,
    \quad \text{and} \quad
    \Phi_{h}^V(q,p)
    =
    (q, p - h \, \operatorname{sign}(q)) \,.
\end{equation}
We now examine whether the Lie-Trotter flow maps $\Phi^q_{h}$ and
$\Phi^p_{h}$ are symplectic when applied to this particular system. The
potential flow is discontinuous along $q=0$. Consequently, $\Phi_h^p$ is
discontinuous along $q=0$, while $\Phi_h^q$ is discontinuous along the line
$q+hp=0$. The formula
$D_{\bm{z}} \Phi_h^T J_c D_{\bm{z}} \Phi_h = J_c$ may therefore only be
verified, and indeed holds, away from these discontinuity sets. We can,
however, check what happens to conservation of loop integrals when such
discontinuities are encountered.

Define the phase-space loop
$\gamma: \mathbb{R}/2\pi\mathbb{Z} \to \mathbb{R}^2$ by
$\gamma(\theta) = (\sin(\theta), \cos(\theta))$. Note that
\begin{equation}
    \int_\gamma p \mathsf{d} q
    =
    \int_0^{2\pi} \cos^2(\theta) \mathsf{d} \theta
    =
    \pi \,.
\end{equation}
Composing this loop with the update maps, one finds
\begin{equation}
    \Phi_h^q \circ \gamma(\theta)
    =
    \left(
    \sin(\theta) + h \cos(\theta)
    \,,\,
    \cos(\theta) - h \, \operatorname{sign}( \sin(\theta) + h \cos(\theta))
    \right) \,,
\end{equation}
and
\begin{equation}
    \Phi_h^p \circ \gamma(\theta)
    =
    \left(
    \sin(\theta) + h ( \cos(\theta) - h \, \operatorname{sign}( \sin(\theta)) )
    \,,\,
    \cos(\theta) - h \, \operatorname{sign}( \sin(\theta))
    \right) \,.
\end{equation}
The curve $\Phi_h^q \circ \gamma(\theta)$ has discontinuities where
$\sin(\theta) + h \cos(\theta) = 0$, i.e.\ at
\begin{equation}
    \tan(\theta) = - h
    \implies
    \theta \in \{ \pi - \arctan(h), 2\pi - \arctan(h) \} \,.
\end{equation}
The curve $\Phi_h^p \circ \gamma(\theta)$ has discontinuities at
$\theta \in \{ 0, \pi, 2 \pi \}$. We compute integrals over these deformed
loops by only integrating over the continuous sub-intervals of each curve.
Equivalently, writing the distributional measure $\mathsf{d} q$ as an
absolutely continuous part with respect to Lebesgue measure in $\theta$ plus
possible atomic jump terms, the piecewise loop integral retains only the
absolutely continuous contribution.
Doing so, one finds that
\begin{equation}
    I_1(h) = \int_{\Phi_h^q \circ \gamma} p \mathsf{d} q
    =
    \pi \,,
    \quad \text{and} \quad
    I_2(h) = \int_{\Phi_h^p \circ \gamma} p \mathsf{d} q
    =
    \pi + 4 h^2 \,.
\end{equation}
Thus, the position-first Lie-Trotter update happens to preserve this
particular loop integral, while the momentum-first update fails to preserve
it for any nonzero time-step. Similarly, if one defines the Strang splitting
updates to be
\begin{equation}
    \Phi_h^{q,S}
    =
    \Phi_{h/2}^K \circ \Phi_h^V \circ \Phi_{h/2}^K \,,
    \quad \text{and} \quad
    \Phi_h^{p,S}
    =
    \Phi_{h/2}^V \circ \Phi_h^K \circ \Phi_{h/2}^V \,,
\end{equation}
then, for sufficiently small $h>0$, it may be shown that
\begin{equation}
    \int_{\Phi_h^{q,S} \circ \gamma} p \mathsf{d} q
    =
    \pi + \frac{4 h^2}{\sqrt{h^2 + 4}} \,,
    \quad \text{and} \quad
    \int_{\Phi_h^{p,S} \circ \gamma} p \mathsf{d} q
    =
    \pi - h^3 + 2 h^2 \,.
\end{equation}
The flow maps obtained from higher-order splitting methods can likewise fail
to conserve the loop integral.

The one-step defects above can accumulate under iteration. Indeed, if $\Psi_h$
denotes the time-advance map and $\gamma_n=\Psi_h^n\circ\gamma$, then
\begin{equation}
    \int_{\gamma_n}p\mathsf{d}q-\int_{\gamma}p\mathsf{d}q
    =
    \sum_{m=0}^{n-1}
    \left(
    \int_{\Psi_h\circ\gamma_m}p\mathsf{d}q
    -
    \int_{\gamma_m}p\mathsf{d}q
    \right).
\end{equation}
Thus an $O(h^2)$ one-step defect may lead to an $O(nh^2)$ accumulated
defect, absent cancellations. Over a fixed time interval $T=nh$, this is an
$O(Th)$ effect. The actual accumulation depends on how the advected loop
intersects the discontinuity set at each step.

\begin{remark}
The preceding convention should be distinguished from applying splitting to a
smooth regularization of the Hamiltonian. For example, if
\begin{equation}
    V_\epsilon(q)=\sqrt{q^2+\epsilon^2},
\end{equation}
then, for every fixed $\epsilon>0$, the corresponding Lie-Trotter and Strang
splitting maps are compositions of smooth Hamiltonian flows and are therefore
symplectic. In the singular limit $\epsilon\to0$ at fixed $h$, the force term
satisfies
\begin{equation}
    \frac{q}{\sqrt{q^2+\epsilon^2}}
    \to
    \operatorname{sign}(q)
    \qquad
    \text{for } q\neq0.
\end{equation}
One should therefore expect the loop integral for the mollified system to
converge to a quantity that may include jump contributions. The precise
weighting of those
jump contributions is determined by the chosen regularization and by how the
regularized loop deforms before the singular limit is taken. Therefore, a
sample-based diagnostic applied only to the discontinuous limiting loop does
not by itself determine the singular limit of the mollified loop integrals.
We instead use the absolutely continuous part of the limiting loop integral,
since this is sufficient to detect the breaking of loops.
To associate a conserved loop integral with a conserved symplectic form,
Stokes' theorem requires that the time-advanced disk have a continuous boundary.
A discontinuous time-advance map can violate this topological condition.
Thus, the diagnostic detects failure of symplecticity 
in the usual strong sense of a continuous, classically symplectic time-advance 
map. Whether a weaker,
measure-valued loop-integral invariant persists after the breaking of loops
is a subtler question that is not easily diagnosed from the limiting
discontinuous loop alone. The limits $h\to0$ and $\epsilon\to0$ need not
commute.
\end{remark}

As previously mentioned, we expect that the continuous-time flow conserves
symplecticity. For the Hamiltonian
$\mathcal{H}(q,p) = \frac{1}{2}p^2 + |q|$, the Hamiltonian vector field is
smooth except on the measure-zero set $q=0$. Away from the exceptional
equilibrium $(q,p)=(0,0)$, trajectories cross $q=0$ continuously in phase
space. Equivalently, one may view the flow as being obtained by gluing
together smooth Hamiltonian trajectories on $q>0$ and $q<0$. Thus, unlike
the discontinuous discrete-time splitting maps above, the continuous-time
flow does not break loops in phase space. This gives the mechanism by which
the continuous-time flow preserves the Poincar{\'e} integral invariant,
whereas the discontinuous discrete-time maps need not.

This simple example illustrates a problem which can occur in symplectic PIC
with piecewise linear interpolation and Strang-splitting time-stepping.
Indeed, one can manufacture a case in which a given particle effectively
experiences a signed absolute-value potential; see \ref{appendix:single_particle}.
Because the approximate time-advance map implied by Hamiltonian splitting can
be discontinuous in phase space, crossings of cell-boundaries can cause
topological loops to break, leading to a loss of conservation of the
Poincar{\'e} integral invariant. It is worth emphasizing that we expect the
continuous-time flow of the piecewise linear PIC system to be symplectic, as
with this simple example. However, the discontinuous character of the
discrete-time flow maps obtained from Hamiltonian splitting methods for such
systems can result in the breaking of topological loops. In general, one
should not expect PIC methods with insufficiently regular interpolation and
explicit Hamiltonian-splitting time-advance maps to preserve symplecticity
automatically. At minimum, sufficient spatial regularity of the interpolated
fields is needed, i.e. higher-order interpolation with smoothness guarantees.

\begin{remark}
    We saw in this example that spatial continuity of the flow map is
    essential to ensure the conservation of the Poincar{\'e} loop integral
    invariant in low regularity flows. It is reasonable to ask why we take
    such pains to design a loop integral approximation which is convergent
    for discontinuous loops. The reason is the non-equivalence of the spatial
    regularity of the exact and approximate flow maps, as seen in this
    example: while the continuous-time flow may preserve loop continuity, a
    discrete or approximate flow can still fail to do so. Having a convergent
    symplectic diagnostic is therefore important to detect conservation
    reliably, even when it is not obvious whether a given time-stepping
    scheme preserves the spatial continuity of the underlying continuous
    flow.

    More specifically, pseudospectral approximations of the loop integral
    converge for loop data $q,p \in H^r$ with $r>1/2$, but the guaranteed
    rate becomes very slow near the endpoint $r=1/2$. Moreover, the
    pseudospectral convergence theorem does not apply to discontinuous loop
    data. In contrast, a finite-difference loop integral approximation
    remains applicable and convergent for piecewise smooth discontinuous
    loops, provided the cells containing jumps are identified and omitted.
    This allows us to distinguish between cases where the flow truly
    preserves the loop integral and cases where a numerical approximation
    fails due to insufficient spatial regularity.
\end{remark}

\subsection{Example: an interpolated nonlinear pendulum} \label{sec:interp_pend}

This example considers a coarse-grained version of a potential which interpolates its values on a uniform grid with spacing, $\Delta q$. The polynomial basis used to approximate the potential are B-splines. This basis is used because it ensures differentiability up to the polynomial degree used: a degree-$p$ B-spline is $C^{p-1}$. This allows us to directly test the behavior of the loop integral diagnostic when applied to Hamiltonian systems of specified regularity. The piecewise character of potentials also mimics those encountered in particle-in-cell methods which will be considered subsequently. 

Let $q_i = i \Delta q $, $\forall i \in \mathbb{Z}$. In one spatial dimension, the constant and linear B-spline basis functions may be written as follows:
\begin{equation} \label{eq:linear_bspline}
    B_i^0(q)
    =
    \begin{cases}
        1 \,, & q \in [q_i, q_{i+1}] \\
        0 \,, & \text{else} \,,
    \end{cases}
    \quad \text{and} \quad
    B_i^1(q)
    =
    \begin{cases}
        \dfrac{\Delta q - |q - q_{i+1}|}{\Delta q} \,, & q \in [q_i, q_{i+2}] \\
        0 \,, & \text{else} \,,
    \end{cases}
\end{equation}
The higher-order B-splines are obtained via convolutions of the lower-order B-splines:
\begin{equation} \label{eq:higher_order_bspline}
    B_i^\alpha(q)
    =
    \underbrace{(B_i^0 * B_i^0 * \cdots * B_i^0)}_{\alpha + 1 \text{ times}}(q)
    =
    B_i^0 * B_i^{\alpha-1} (q)
    \,.
\end{equation}
Notice, this increases the width of the degree-$\alpha$ basis function to encompass $\alpha+1$ cells. The Hamiltonian with an interpolated potential is written as
\begin{equation}
    \mathcal{H}(q,p)
    =
    \frac{p^2}{2} + V_{\Delta q}^\alpha(q) \,,
\end{equation}
where 
\begin{equation}
    V_{\Delta q}^\alpha(q)
    =
    \sum_{j = i - \alpha + 1}^{i+1} \mathsf{V}_j B^\alpha_{j}(q) 
    \quad \text{for} \quad
    q \in [q_i, q_{i+1}] \,,
\end{equation}
where the coefficients $\mathsf{V}_i$ are selected so that $V_{\Delta q}^\alpha(q_i) = V(q_i)$. 

In this example, we use $V(q) = \sin(q) + \cos^2(2 q + 1)$, and $\Delta q = 2 \pi/25 \approx 0.25$. The large grid spacing helps to ensure that issues with symplecticity preservation due to low regularity, if they are present, are sufficiently exaggerated so as to readily appear in the numerical results. The symplecticity of the algorithm does not depend on the magnitude of $\Delta q$, so this choice does not influence the conclusions drawn from the numerical tests. Tests are performed using $\alpha = 1,2,3$, and for the original, non-interpolated potential. The loop is initialized as a circle of radius $1$ centered at $(q_0,p_0) = (\pi, 0)$. The cases with $\alpha = 2,3$, and the original, non-interpolated potential enjoy high enough regularity that the pseudospectral loop integral approximation converges. The case $\alpha = 1$ uses the finite difference loop integral approximation. 

\begin{figure}[t]
    \centering
\begin{subfigure}[t]{0.49\textwidth}
        \centering
        \includegraphics[width=\textwidth]{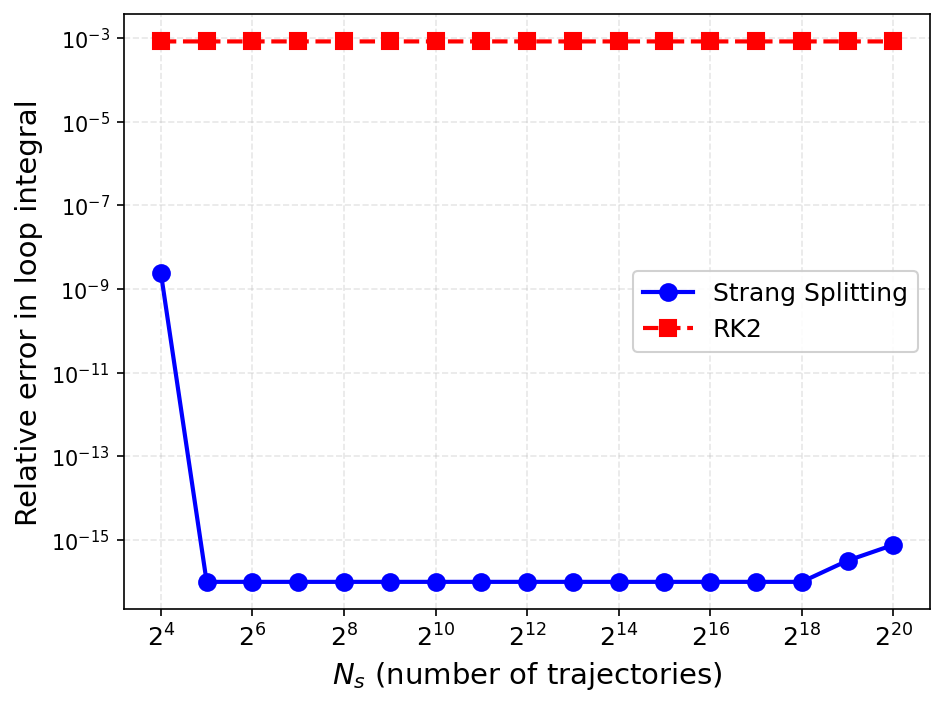}
        \caption{Non-interpolated}
        \label{fig:ip_M_0}
    \end{subfigure}
    \hfill \begin{subfigure}[t]{0.49\textwidth}
        \centering
        \includegraphics[width=\textwidth]{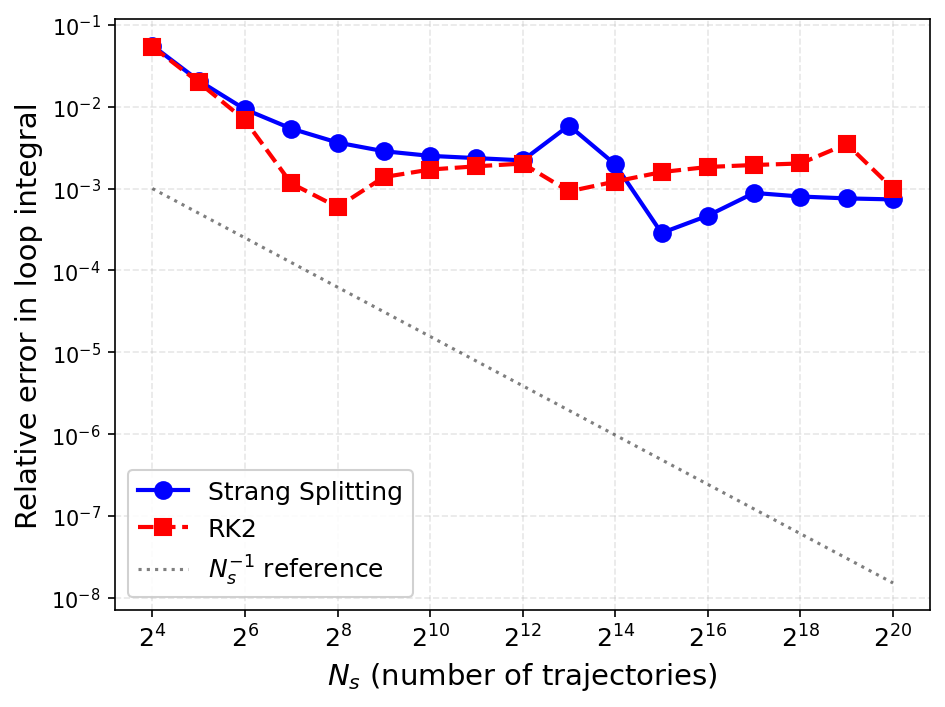}
        \caption{Piecewise linear}
        \label{fig:ipM_1}
    \end{subfigure}
\begin{subfigure}[t]{0.49\textwidth}
        \centering
        \includegraphics[width=\textwidth]{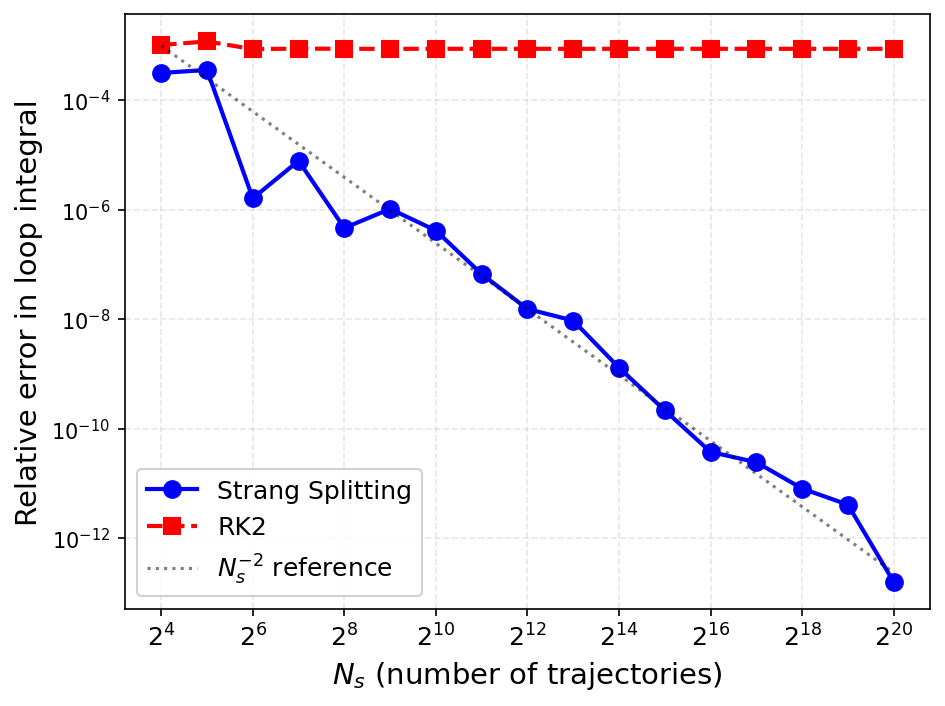}
        \caption{Piecewise quadratic}
        \label{fig:ip_M_2}
    \end{subfigure}
    \hfill \begin{subfigure}[t]{0.49\textwidth}
        \centering
        \includegraphics[width=\textwidth]{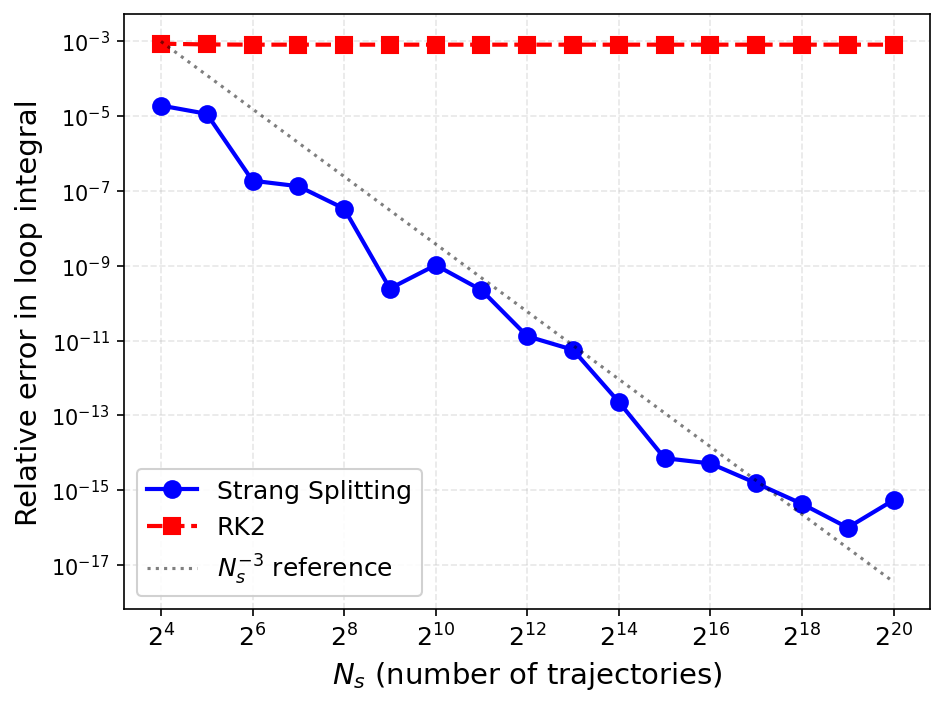}
        \caption{Piecewise cubic}
        \label{fig:ip_M_3}
    \end{subfigure}
\caption{Interpolated potential conservation results. Sweep in the number of points approximating the loop. The approximation of the diagnostic for the linear case saturates at an error of about $10^{-6}$, indicating a failure to converge the symplectic diagnostic.}
    \label{fig:interp_pend_M_sweep}
\end{figure}

See Figure \ref{fig:interp_pend_M_sweep} for the relative error in the symplectic diagnostic over a single time-step as a function of $N_s$, the number of points resolving the loop, using a fixed time-step size of $\Delta t = 0.1$. This parameter sweep serves to verify that the loop integral approximations are converging with the expected rate. Indeed, the convergence rates roughly correspond with the order of polynomial interpolation for the Strang splitting integrator. In all cases, RK2 achieves an $O(10^{-3})$ error. In the test where $\alpha = 1$, both Strang-splitting and RK2 saturate about $10^{-3}$. Strang splitting is not expected to be symplectic in the case $\alpha=1$ due to the low regularity of the Hamiltonian, so this result is expected. 

\begin{figure}[t]
    \centering
\begin{subfigure}[t]{0.49\textwidth}
        \centering
        \includegraphics[width=\textwidth]{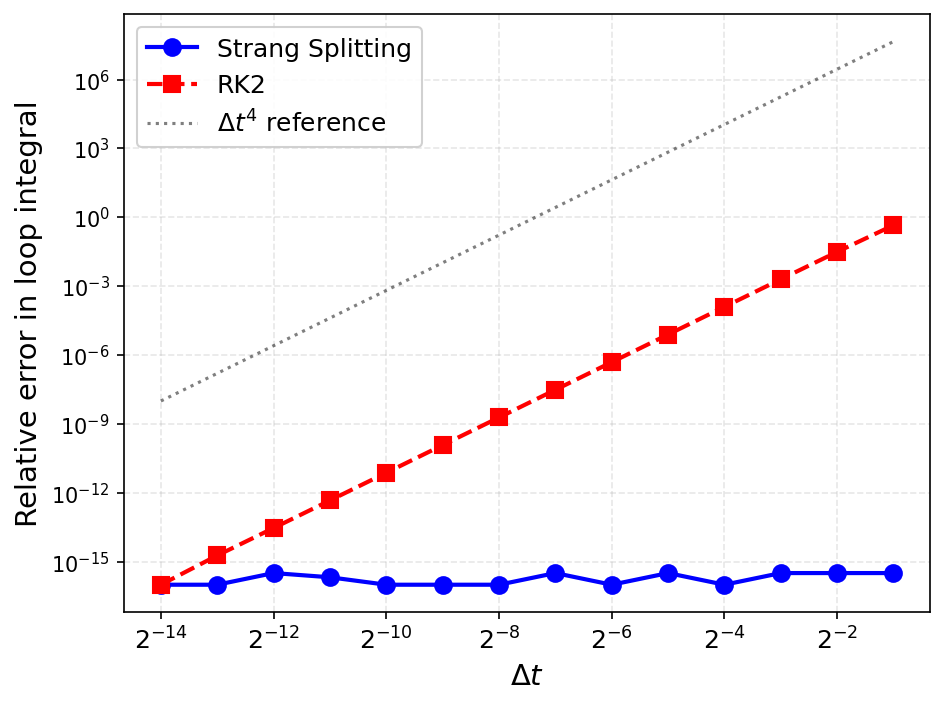}
        \caption{Non-interpolated}
        \label{fig:ip_dt_0}
    \end{subfigure}
    \hfill \begin{subfigure}[t]{0.49\textwidth}
        \centering
        \includegraphics[width=\textwidth]{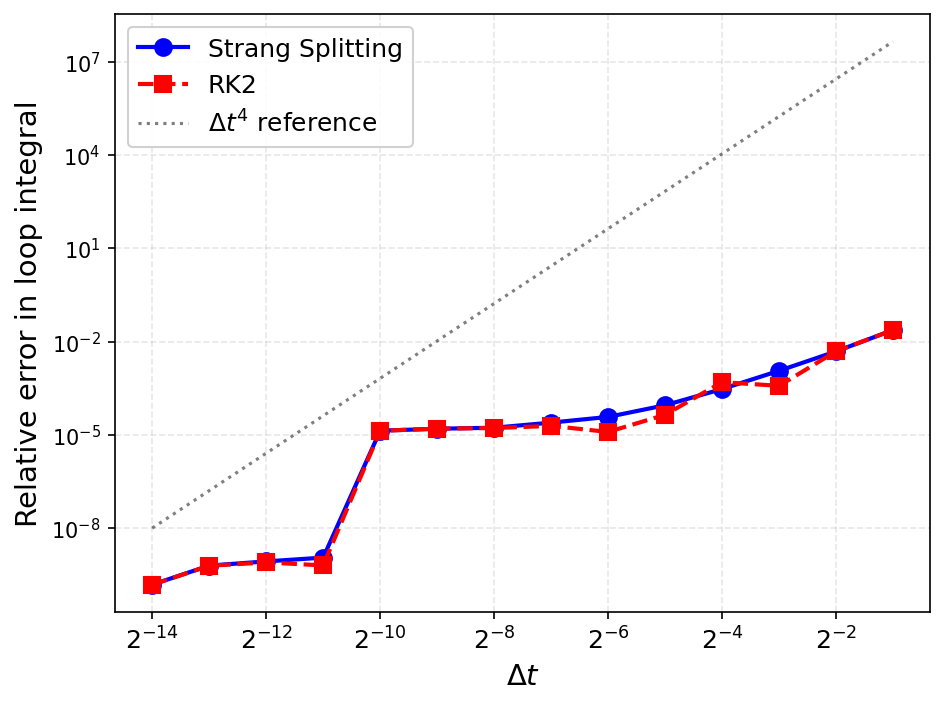}
        \caption{Piecewise Linear ($N_s = 2^{20}$)}
        \label{fig:ip_dt_1}
    \end{subfigure}
\begin{subfigure}[t]{0.49\textwidth}
        \centering
        \includegraphics[width=\textwidth]{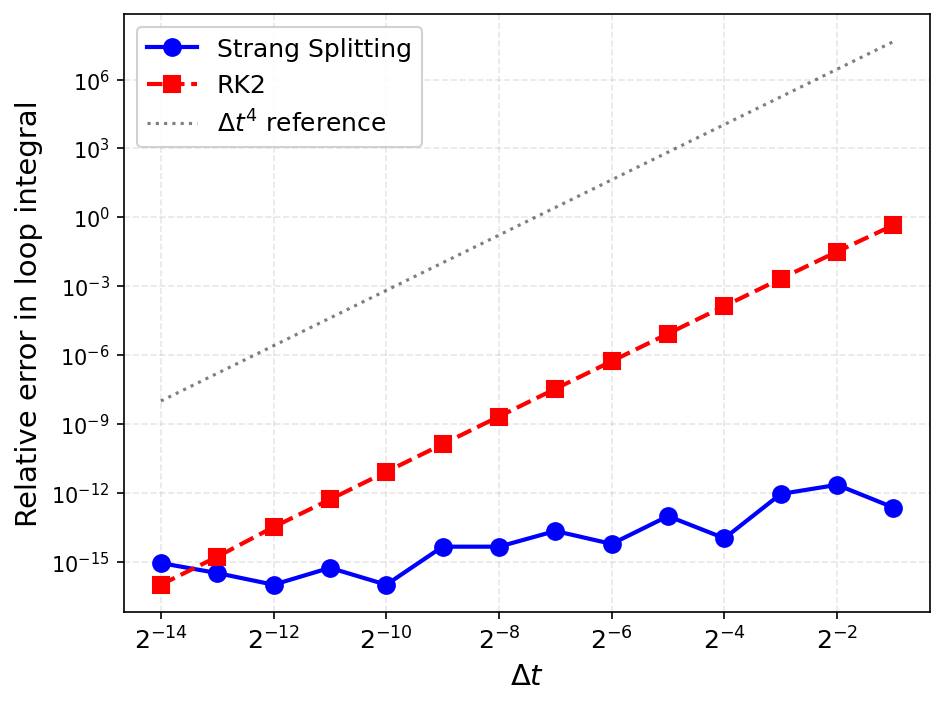}
        \caption{Piecewise quadratic ($N_s = 2^{20}$)}
        \label{fig:ip_dt_2}
    \end{subfigure}
    \hfill \begin{subfigure}[t]{0.49\textwidth}
        \centering
        \includegraphics[width=\textwidth]{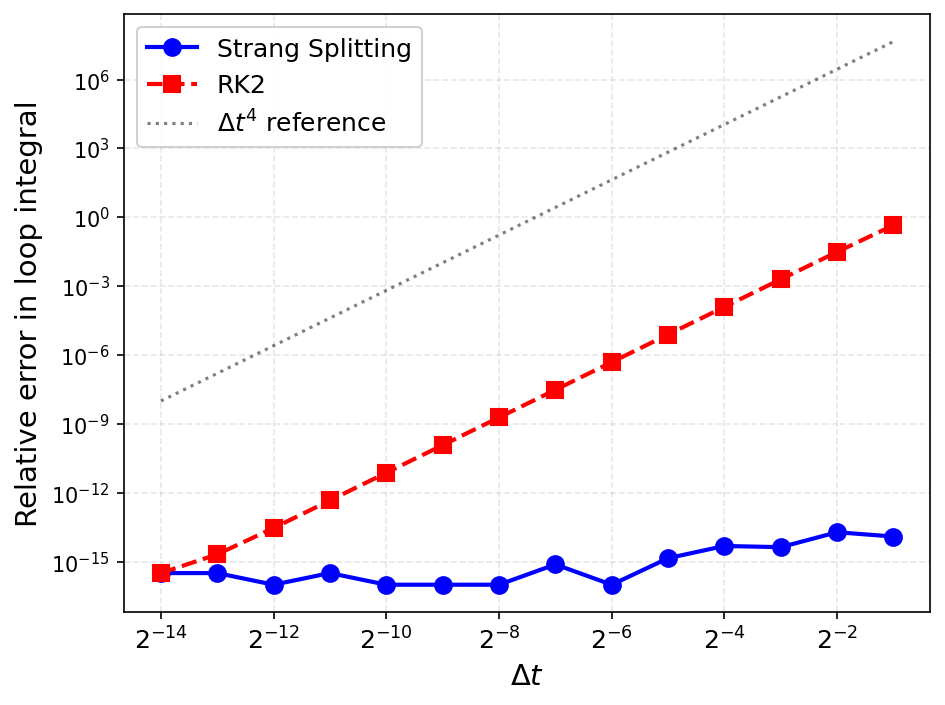}
        \caption{Piecewise cubic}
        \label{fig:ip_dt_3}
    \end{subfigure}
\caption{Interpolated potential conservation results. Sweep in the time-step size.}
    \label{fig:interp_pend_dt_sweep}
\end{figure}

See Figure \ref{fig:interp_pend_dt_sweep} for the relative error in the loop integral over a single time-step as a function of time-step size, $\Delta t$, with $N_s = 2^{16}$ points approximating the loop in the cases $\alpha = 3$ and the non-interpolated case, and $N_s = 2^{20}$ points in the cases $\alpha =1, 2$. In the tests which use the pseudospectral loop integral approximation, these resolutions are chosen to make loop-quadrature error smaller than the observed time-stepping trends, as verified by the corresponding sweep in $N_s$. The finite difference approximation in the case $\alpha = 1$ cannot be expected to converge to machine precision. Even with $2^{20} = 1,048,576$ points, the $O(N_s^{-1})$ convergence rate only suggests about six digits of accuracy. This is sufficient to see that the performance of Strang-splitting and RK2 is largely indistinguishable. For the finite difference loop integral approximation, the parameters detecting jumps are set to $\lambda_q = \lambda_p = 10$, because large discontinuities are not expected. One can see that, in each case, the error incurred in the loop integral by RK2 follows a convergence trend as the time-step is refined, whereas Strang splitting in the non-interpolated and $\alpha = 2,3$ test cases robustly conserve the diagnostic for all time-step sizes. Interestingly, in all cases except $\alpha = 1$, the convergence trend in $\Delta t$ for the error incurred by RK2 appears to have a slope of $\Delta t^4$, rather than the expected slope of $\Delta t^2$. The sweep in $\Delta t$ in the piecewise linear case indicates a failure to conserve symplecticity: Strang splitting and RK2 perform comparably in conservation of the loop integral. Note that we only trust the results in this test for time-step sizes coarse enough that discretization error is smaller than the error in symplecticity: i.e.\! for $\Delta t$ greater than about $2^{-6}$. The tests $\alpha = 2,3$ and the non-interpolated case all indicate symplecticity conservation.

\subsection{Example: a nonlinear pendulum array}

\begin{figure}[t]
    \centering
\begin{subfigure}[t]{0.49\textwidth}
        \centering
        \includegraphics[width=\textwidth]{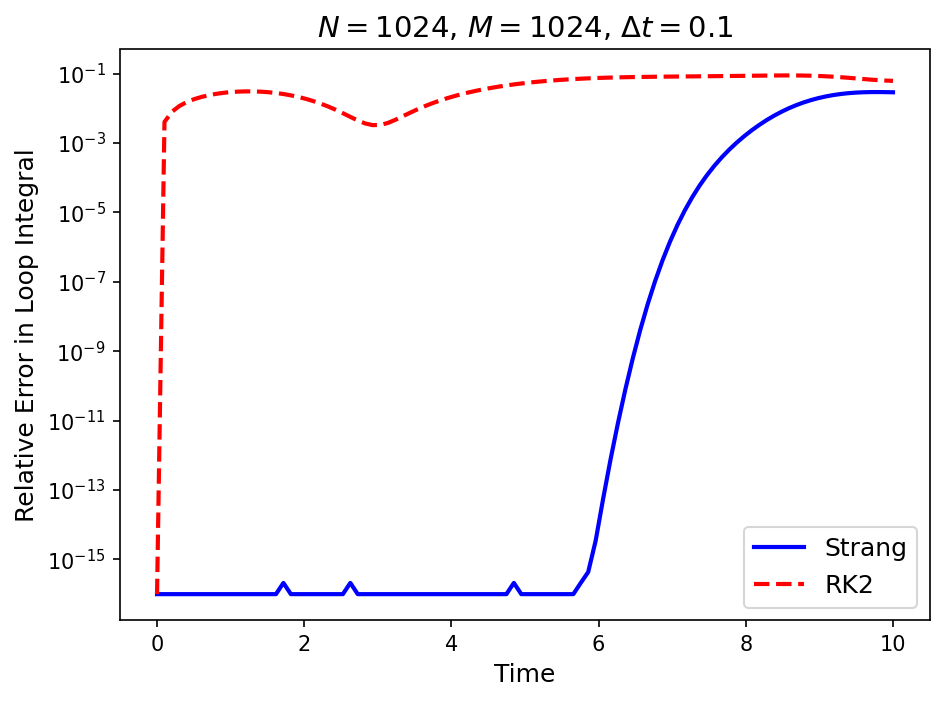}
        \caption{}
        \label{fig:pend_array_time}
    \end{subfigure}
    \hfill \begin{subfigure}[t]{0.49\textwidth}
        \centering
        \includegraphics[width=\textwidth]{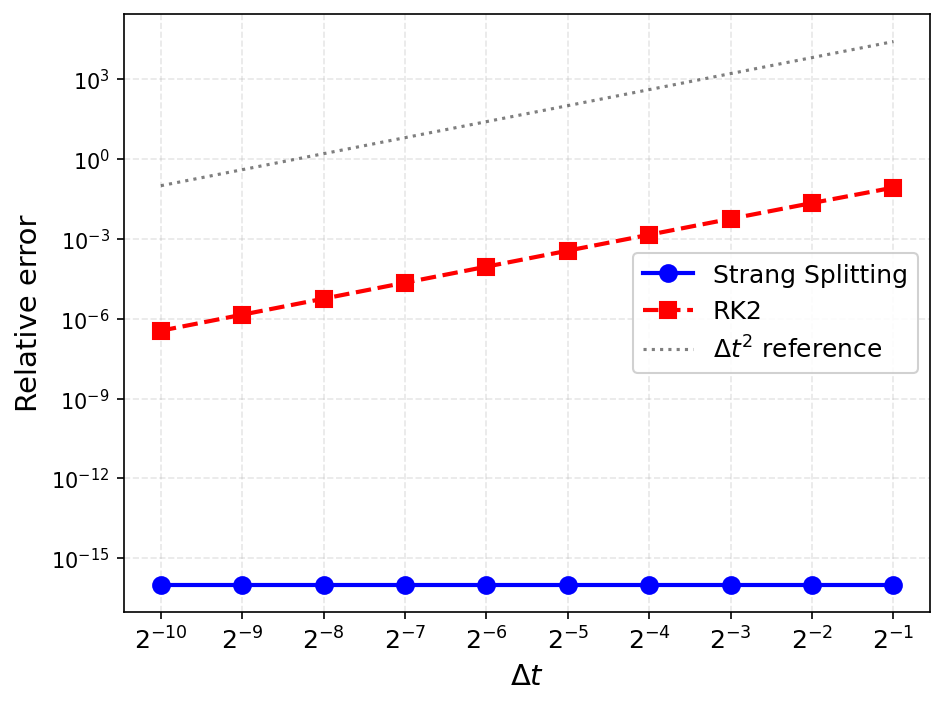}
        \caption{}
        \label{fig:pend_array_conv}
    \end{subfigure}
    \caption{Results for nonlinear pendulum array test.}
    \label{fig:pend_array}
\end{figure}

We next consider how the diagnostic performs when applied to a high dimensional system with a real analytic Hamiltonian. Consider a row of $N$ equispaced nonlinear pendula, each coupled to its neighbors by linear Hookean springs. The position coordinates $\{ q_i \}_{i=1}^{N}$ are the positions of the particles relative to their resting position. The evolution equations are then given by
\begin{equation}
    \ddot{q}_i
    =
    \alpha_i \sin(q_i)
    +
    k_i (q_{i-1} - q_i - \Delta q_{i-1})
    +
    k_{i+1} (q_{i+1} - q_i + \Delta q_i) \,, 
    \quad 
    i = 1, \hdots, N \,,
\end{equation}
where $\alpha_i$ is the spring constant for the $i^{th}$ pendulum, $k_i$ is the spring constant for the spring connecting the $(i-1)^{th}$ and $i^{th}$ pendulums, and $\Delta q_i$ is the difference between the rest positions of pendulum $i$ and $i+1$. For simplicity, we assume that $\alpha_i = \alpha > 0$, $k_i = k > 0$, and $\Delta q_i = \Delta q$ $\forall i$. Further, we let the first and last springs couple with each other so that $q_{N+1} = q_1$ and $q_0 = q_N$: i.e., the array is arranged in a ring. Hence, the evolution equations become
\begin{equation}
    \ddot{q}_i = \alpha \sin(q_i) + k (q_{i+1} - 2 q_i + q_{i-1}) \,, \quad
    i = 1, \hdots, N \,.
\end{equation}
This evolution equation comes from the Lagrangian
\begin{equation}
    L(q_i, \dot{q}_i)
    =
    \sum_{i=1}^N
    \left(
    \frac{1}{2}
    \dot{q}_i^2
    -
    \alpha
    \cos(q_i)
    -
    \frac{1}{2}
    k
    q_i
    (- q_{i+1} + 2 q_i - q_{i-1})
    \right) \,.
\end{equation}
The Lagrangian may be written in vector notation as follows:
\begin{equation}
    L(\bm{q}, \dot{\bm{q}})
    =
    \frac{1}{2} \dot{\bm{q}}^T \dot{\bm{q}}
    -
    \alpha
    \mathbbm{1}^T \cos(\bm{q})
    -
    \frac{1}{2}
    k
    \bm{q}^T \mathbb{L} \bm{q} \,,
\end{equation}
where $\mathbbm{1} = (1, 1, \hdots, 1)$, and $\mathbb{L}$ is the circulant matrix with stencil $(-1, 2, -1)$. Alternatively, this may be modeled as a canonical Hamiltonian system with Hamiltonian:
\begin{equation}
    \mathcal{H}(\bm{q}, \bm{p})
    =
    \frac{1}{2} \bm{p}^T \bm{p}
    +
    \alpha
    \mathbbm{1}^T \cos(\bm{q})
    +
    \frac{1}{2} k
    \bm{q}^T \mathbb{L} \bm{q} \,,
\end{equation}
which yields the evolution equations
\begin{equation}
    \dot{\bm{p}} = \alpha \sin(\bm{q}) - k \mathbb{L} \bm{q} \,,
    \quad
    \dot{\bm{q}} = \bm{p} \,.
\end{equation}
This system may be thought of as a finite difference approximation of the one-dimensional Sine-Gordon equation if we let $k = \Delta q^{-2}$. In the tests considered here, $\alpha = k = 1$, and the initial conditions are generated as uniform random numbers in the interval $[0,1]$. The loop for computing the diagnostic is generated by periodically perturbing these random initial conditions along a circle of radius $1$. As the Hamiltonian is $C^\infty$, the pseudospectral loop integral approximation is applicable, and converges rapidly. See Figure \ref{fig:pend_array_time} for the relative error in the loop integral as a function of time. See Figure \ref{fig:pend_array_conv} for the relative error in the symplectic diagnostic over a single time-step as a function of time-step size, $\Delta t$, using $N = 4096$ pendulums and $N_s = 4096$ points in the loop integral. The diagnostic is well-resolved across the parameter sweep with a modest number of points resolving the loop (relative to system size) due to the analyticity of the Hamiltonian. 

\subsection{Discussion}

The tests in this section confirm that it is regularity of the Hamiltonian, and not dimensionality, which limits the applicability of the loop integral diagnostic. The loop integral diagnostic is easy to resolve to machine precision with a modest number of sample points if the Hamiltonian is smooth, and, conversely, a large number of points are needed in cases of low regularity. The flow maps arising from Hamiltonian splitting methods applied to piecewise linear Hamiltonians fail to conserve the loop integral diagnostic, indicating that these flow maps are not symplectic. 

These observations lead to the conclusion that Hamiltonian splitting methods should not be used to temporally integrate Hamiltonian systems with piecewise linear interpolated fields, these, respectively, being staple temporal \cite{kraus2017gempic, he2015hamiltonian} and spatial \cite{birdsall2018plasma, hockney2021computer, villasenor1992rigorous} approximation methods in PIC. Note that structure-preserving PIC methods are generally not limited to linear shape functions, and indeed frequently use high-order shape functions in their practical numerical tests. However, they do offer the option of using linear shape functions as a viable choice in the algorithm. That the use of linear shape functions leads to a violation of symplecticity preservation appears to be absent in the structure-preserving PIC literature. Perhaps more importantly, it further follows that PIC methods which use higher-order interpolation but do not yield a globally $C^1$ interpolant, such as standard $C^0$ finite element PIC methods \cite{bettencourt2021empire, glasser2022gauge, glasser2023generalizing, finn2023numerical} or methods based on Lagrange interpolation/histopolation \cite{kormann2024dual}, can likewise suffer from this issue. Regarding symplecticity preservation, structure-preserving PIC methods based on higher-order B-splines \cite{kraus2017gempic,campos2022variational} or Fourier interpolation \cite{evstatiev2013variational, campos2022variational, shen2024particle, muralikrishnan2024parapif} are highly favorable since these yield globally smoother interpolants.

Section \ref{sec:symp_pic} will further explore these claims by applying the diagnostic to a structure-preserving particle-in-cell method. 
 
\section{Symplectic electrostatic particle-in-cell methods} \label{sec:symp_pic}
The use of the symplectic diagnostic to study the preservation of symplecticity in electrostatic particle-in-cell (PIC) methods presents two difficulties. First, these systems are high-dimensional and expensive to simulate. Therefore, it is desirable to use as few trajectories to approximate the loop integral as possible. Second, the regularity of the Hamiltonian in a symplectic PIC method is limited by the degree of polynomial interpolation used when depositing charge to the grid. This forces one to use more points to approximate the loop integral. 

The system of interest in this section is the Vlasov-Poisson system (with physical constants set to unity):
\begin{equation}
    \partial_t f + \bm{v} \cdot \nabla_{\bm{x}} f - \nabla \phi \cdot \nabla_{\bm{v}} f = 0 \,,
    \quad  \quad
    - \Delta \phi
    =
    \int f \mathsf{d} \bm{v} - \rho_0 \,,
\end{equation}
where $\rho_0$ is a neutralizing background charge density. This system is Hamiltonian \cite{marsden1982hamiltonian, morrison1980maxwell} with Poisson bracket 
\begin{equation}
    \{F, G\}
    =
    \int f \left[ \frac{\delta F}{\delta f}, \frac{\delta G}{\delta f} \right] \mathsf{d} \bm{x} \mathsf{d} \bm{v} \,,
    \quad  \quad
    [u_1,u_2] = \nabla_{\bm{x}} u_1 \cdot\nabla_{\bm{v}} u_2 - \nabla_{\bm{x}} u_2 \cdot\nabla_{\bm{v}} u_1 \,,
\end{equation}
and Hamiltonian
\begin{equation}
\begin{aligned}
    \mathcal{H}[f]
    &=
    \int \int \frac{1}{2} | \bm{v} |^2 f(\bm{x}, \bm{v}) \mathsf{d} \bm{v}\mathsf{d} \bm{x}
    +
    \frac{1}{2}\int \rho(\bm{x}) \phi(\bm{x}) \mathsf{d} \bm{x} \,,
    \\
    \rho(\bm{x})
    &=
    \int f(\bm{x},\bm{v})\mathsf{d}\bm{v} - \rho_0
    \quad \text{and} \quad
    - \Delta \phi
    =
    \rho(\bm{x}) \,.
\end{aligned}
\end{equation}
We consider only periodic boundary conditions in space. 

\subsection{Symplectic PIC methods} \label{sec:symp_pic_method}

Many structure-preserving PIC methods have been derived based on the Hamiltonian or variational structure of the Vlasov equation \cite{squire2012geometric, qin2015canonical, kraus2017gempic, perse2021geometric, xiao2015explicit, evstatiev2013variational, he2015hamiltonian, campos2022variational, glasser2022gauge, glasser2023generalizing, qiang2018symplectic, kormann2024dual, shadwick2014variational, stamm2014variational, qin2015canonical, jianyuan2018structure, jianyuan2021explicit, burby2017finite, morrison2017structure}. Following the general approach described in these prior works, one may obtain an electrostatic PIC method for the Vlasov-Poisson system written as a nearly-canonical Hamiltonian system which captures many salient features common to these structure-preserving PIC methods. The Hamiltonian is given by
\begin{equation}
    \mathcal{H}(\bm{x}_1, \bm{v}_1, \hdots, \bm{x}_{N_p}, \bm{v}_{N_p})
    =
    \frac{1}{2} \left( \sum_{a=1}^{N_p} w_a | \bm{v}_a |^2
    +
    \bm{\uprho}^T \mathbb{L}^{-1} \bm{\uprho} \right) \,,
    \quad \text{where} \quad
    \uprho_i \label{eq:pic_charge}
    =
    \sum_a w_a B_i^p(\bm{x}_a) - \rho_0,
\end{equation}
where $\bm{\uprho}$ is the vector of nodal charge residuals, $\{ B_i^p \}$ is a piecewise polynomial basis, and $\mathbb{L}$ is the discrete Laplacian matrix associated with the basis. Because we use periodic boundary conditions, $\mathbb{L}^{-1}$ denotes the inverse on the mean-zero subspace, with $\rho_0$ chosen so that $\bm{\uprho}$ lies in this subspace. Here $(\bm{x}_a, \bm{v}_a) \in \mathbb{R}^{2d}$ are the position and velocity of the $a$-th particle and $w_a$ is its weight. The Poisson bracket is given by
\begin{equation}
    \{F, G\}
    =
    \sum_{a=1}^{N_p} \frac{1}{w_a} \left( \frac{\partial F}{\partial \bm{x}_a} \cdot \frac{\partial G}{\partial \bm{v}_a}
    -
    \frac{\partial G}{\partial \bm{x}_a} \cdot \frac{\partial F}{\partial \bm{v}_a}
    \right) \,.
\end{equation}
Only the constant scaling factor in the bracket prevents this system from being canonical. The evolution equations are given by
\begin{equation}
\begin{aligned}
    \dot{\bm{x}}_a &= \{ \bm{x}_a, \mathcal{H} \} = \bm{v}_a \,, \\
    \dot{\bm{v}}_a &= \{ \bm{v}_a, \mathcal{H} \} = \bm{E}_h(\bm{x}_a) = - \nabla \phi_h(\bm{x}_a) \,,
\end{aligned}
\end{equation}
where
\begin{equation}
    (\mathbb{L} \bm{\upphi})_i
    =
    \uprho_i
    \quad \text{and} \quad
    \phi_h(\bm{x})
    =
    \sum_i \upphi_i B_i^p(\bm{x}) \,.
\end{equation}
The discrete Laplacian matrix is computed as the finite element stiffness matrix associated with the basis: $\mathbb{L}_{ij} = ( \nabla B_i^p, \nabla B_j^p)_{L^2}$. We interpolate with $p^{th}$ order B-splines over a uniform grid, hence the notation for the basis functions. In one spatial dimension, the domain is $\Omega = [0,L]$, and with $N$ grid-points, the grid-spacing is $\Delta x = L/N$. The B-spline basis functions are constructed as described in Equations \eqref{eq:linear_bspline} and \eqref{eq:higher_order_bspline}. A basis in higher dimensions may be obtained as a tensor product of the one-dimensional basis.

As in the rest of this paper, we temporally integrate using Strang splitting:
\begin{equation}
\begin{aligned}
    \bm{x}_a^{n+1/2}
    &=
    \bm{x}_a^n
    +
    \frac{\Delta t}{2} \bm{v}_a^n \,, \\
    \bm{v}_a^{n+1}
    &=
    \bm{v}_a^n
    -
    \Delta t
    \nabla \phi_h(\bm{x}_a^{n+1/2}) \,, \\
    \bm{x}_a^{n+1}
    &=
    \bm{x}_a^{n+1/2}
    +
    \frac{\Delta t}{2} \bm{v}_a^{n+1} \,.
\end{aligned}
\end{equation}
By doing the velocity update second, only a single field solve is needed per time-step. As in the rest of the paper, for comparison, we consider the non-symplectic second-order explicit Runge-Kutta (RK2) method. This section considers only the 1D1V PIC method for simplicity since the behavior of the diagnostic may be readily seen in this simplified setting. 

\subsection{Applying the symplectic diagnostic to PIC}

The Landau damping and two-stream instability test cases are used as initial conditions for the simulations. For Landau damping, the initial positions are sampled from the distribution:
\begin{equation}
    f_x(x)
    \propto
    1 + A \cos(2 \pi k x/L) \,, 
\end{equation}
where $A \in (0,1)$, and $k \in \{1, 2, \hdots\}$. The initial velocities are sampled from a normal distribution:
\begin{equation}
    f(x,v)
    \propto
    \exp \left( - \frac{v^2}{v_{th}^2} \right) \,.
\end{equation}
For the two-stream instability, the initial conditions are $x_a \sim U([0,L])$, a uniform distribution on $[0,L]$, and $v_a$ are sampled with the density:
\begin{equation}
    f_v(v)
    \propto
    \exp \left( - \frac{(v - v_b)^2}{v_{th}^2} \right)
    + 
    \exp \left( - \frac{(v + v_b)^2}{v_{th}^2} \right)
\end{equation}
where $v_b$ is the velocity of the counter-propagating beams, and $v_{th}$ is the width of the beams (the thermal speed). We seed the instability by periodically perturbing each particle velocity:
\begin{equation}
    \tilde{v}_a
    =
    (1 + A \sin(2 \pi x_a/L)) v_a \,. 
\end{equation}
For our tests, we consider a domain of length $L=50$, $v_{th} = 1$, $v_b = 3$, $A=0.5$, and $k=1$. The particle weights are made to be uniform such that the charge density is unity. 

\begin{figure}[!htbp]
    \centering

    \begin{subfigure}[t]{0.49\textwidth}
        \centering
        \includegraphics[width=\textwidth]{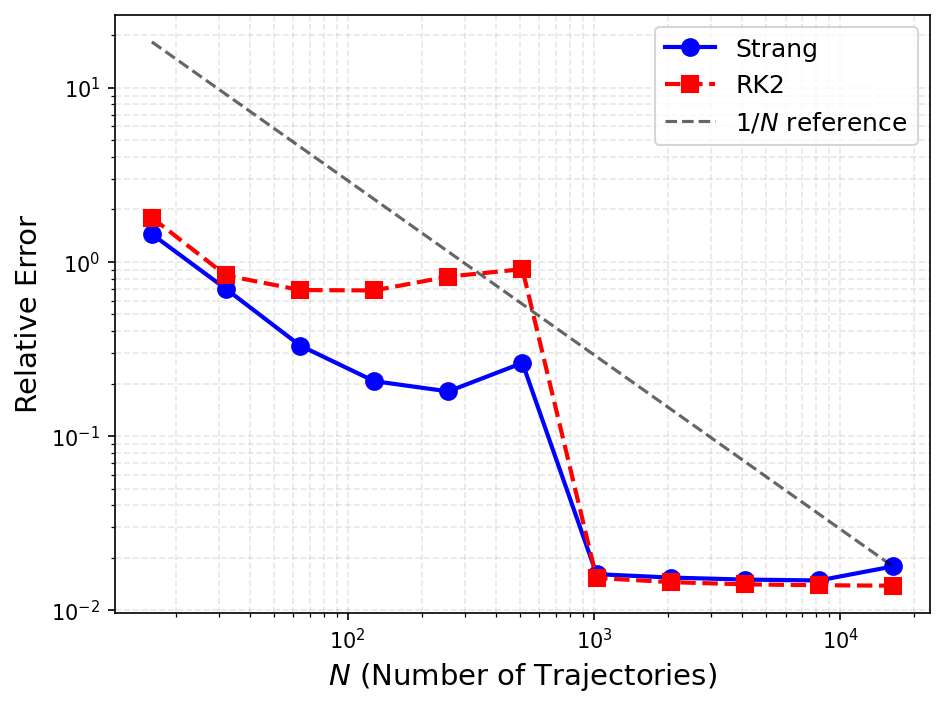}
        \caption{Landau with linear interpolation}
        \label{fig:pic_landau_1_N}
    \end{subfigure}
    \hfill \begin{subfigure}[t]{0.49\textwidth}
        \centering
        \includegraphics[width=\textwidth]{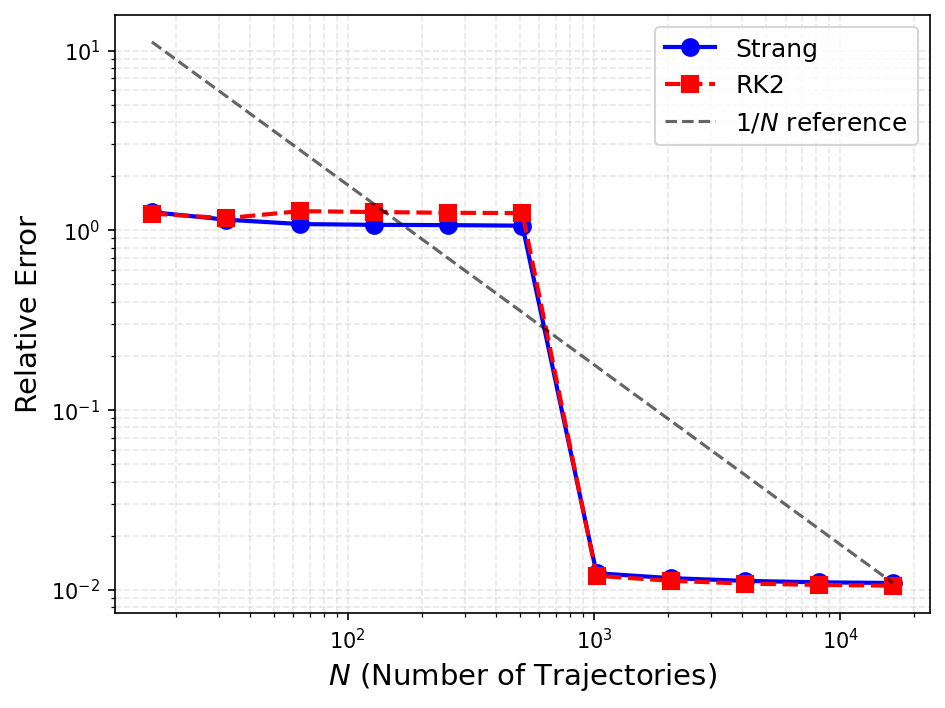}
        \caption{Two-stream with linear interpolation}
        \label{fig:pic_two_stream_1_N}
    \end{subfigure}

    \begin{subfigure}[t]{0.49\textwidth}
        \centering
        \includegraphics[width=\textwidth]{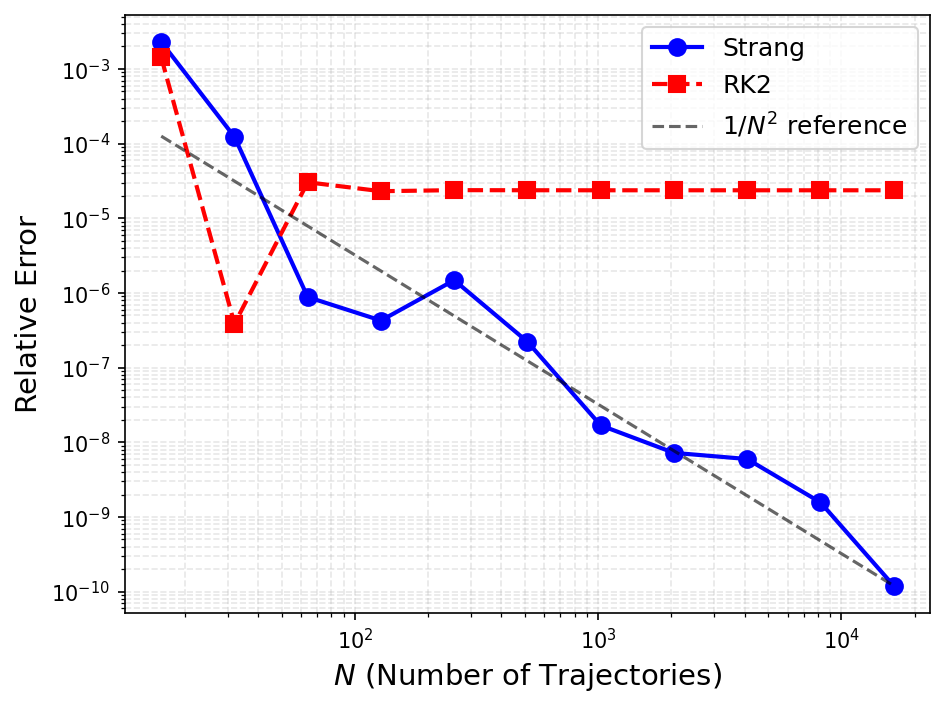}
        \caption{Landau with quadratic interpolation}
        \label{fig:pic_landau_2_N}
    \end{subfigure}
    \hfill \begin{subfigure}[t]{0.49\textwidth}
        \centering
        \includegraphics[width=\textwidth]{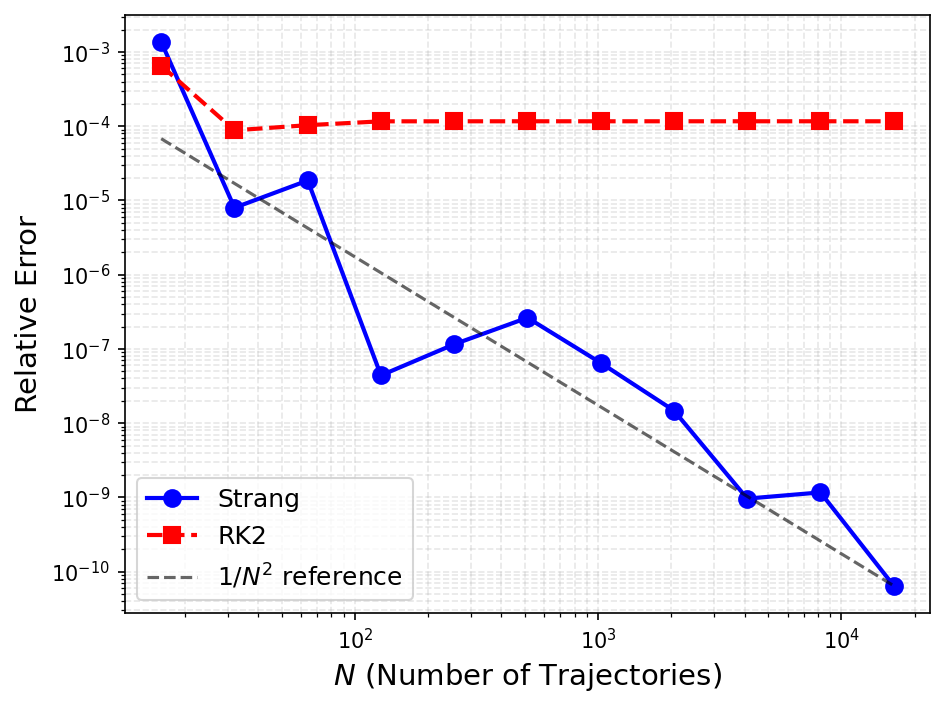}
        \caption{Two-stream with quadratic interpolation}
        \label{fig:pic_two_stream_2_N}
    \end{subfigure}

    \begin{subfigure}[t]{0.49\textwidth}
        \centering
        \includegraphics[width=\textwidth]{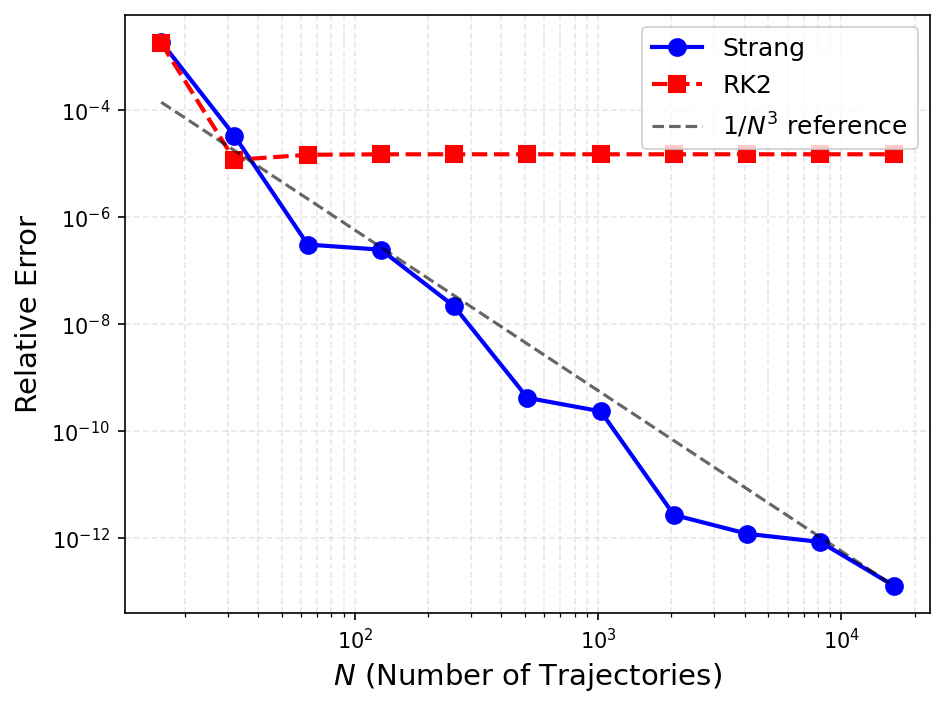}
        \caption{Landau with cubic interpolation}
        \label{fig:pic_landau_3_N}
    \end{subfigure}
    \hfill \begin{subfigure}[t]{0.49\textwidth}
        \centering
        \includegraphics[width=\textwidth]{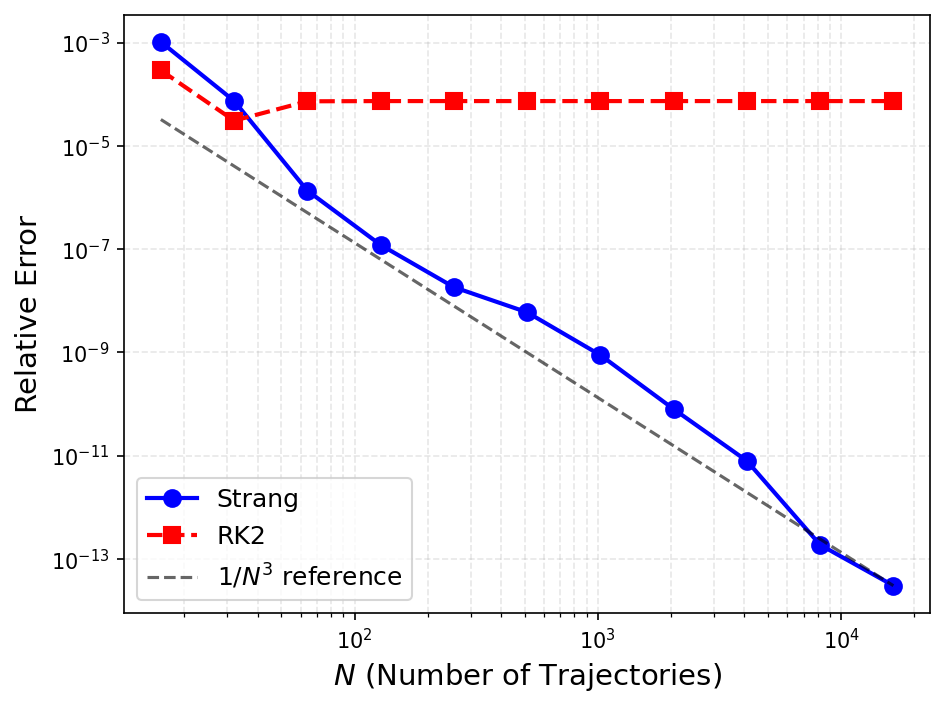}
        \caption{Two-stream with cubic interpolation}
        \label{fig:pic_two_stream_3_N}
    \end{subfigure}
    
    \caption{Convergence trends as a function of $N_s$ with $\Delta t = 0.1$, and $N_p = 128$.}
    \label{fig:pic_conv_N}
\end{figure}

Let $N_s$ be the number of samples to resolve the loop, and let $N_p$ be the number of particles in each simulation. For $p = 1$, linear interpolation, we must use the finite difference loop integral approximation, while for $p=2,3$, we use the pseudospectral loop integral approximation. To allow for possibly large gradients in the loop, we set $\lambda_q = \lambda_p = 1000$ in the finite difference loop integral approximation (and hence only converge for $N_s > 1000$). See Figure \ref{fig:pic_conv_N} for the convergence of the single-step loop integral error as a function of $N_s$ with fixed $\Delta t = 0.1$ and $N_p = 128$. For $p=1$, we do not observe the convergence trend, because $\lambda_q=\lambda_p = 1000$, and the quadrature error has already saturated for $N_s > 1000$. For $p=2$, we observe a convergence rate trend of $N_s^{-2}$. For $p=3$, we observe a convergence rate trend of $N_s^{-3}$. These empirical rates are used below to select $N_s$ for the $\Delta t$ sweep.

These results inform the parameters to use for the $\Delta t$ sweep. We expect an error of about $O(N_p/N_s)$ in the case $p=1$. If we use $N_p=2$ and $N_s = 2^{15}$, we can expect about four to five decimal places of accuracy. In the case $p=2$, we can expect nine digits of accuracy with $N_p=2$ and $N_s = 2^{15}$. In the case $p=3$, we can expect thirteen to fourteen digits of accuracy with $N_p=2$ and $N_s = 2^{15}$. Such small number of particles (while entirely inappropriate in a PIC simulation) is beneficial in testing symplecticity. The conservation/non-conservation of symplecticity does not depend on the physical realism of the simulation. 

\begin{figure}[!htbp]
    \centering

    \begin{subfigure}[t]{0.49\textwidth}
        \centering
        \includegraphics[width=\textwidth]{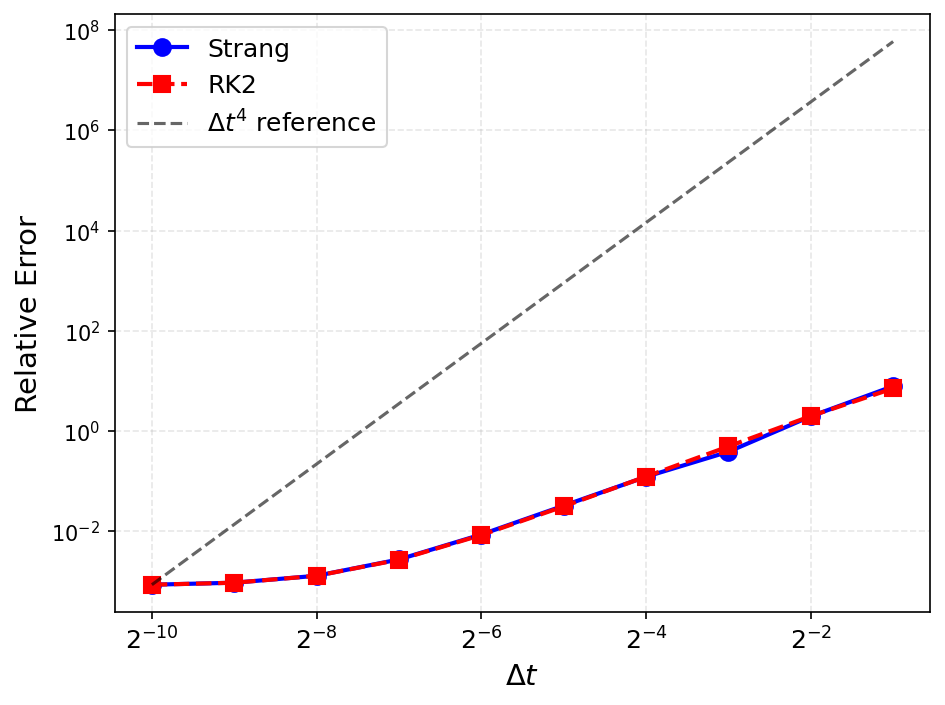}
        \caption{Landau with linear interpolation}
        \label{fig:pic_landau_1}
    \end{subfigure}
    \hfill \begin{subfigure}[t]{0.49\textwidth}
        \centering
        \includegraphics[width=\textwidth]{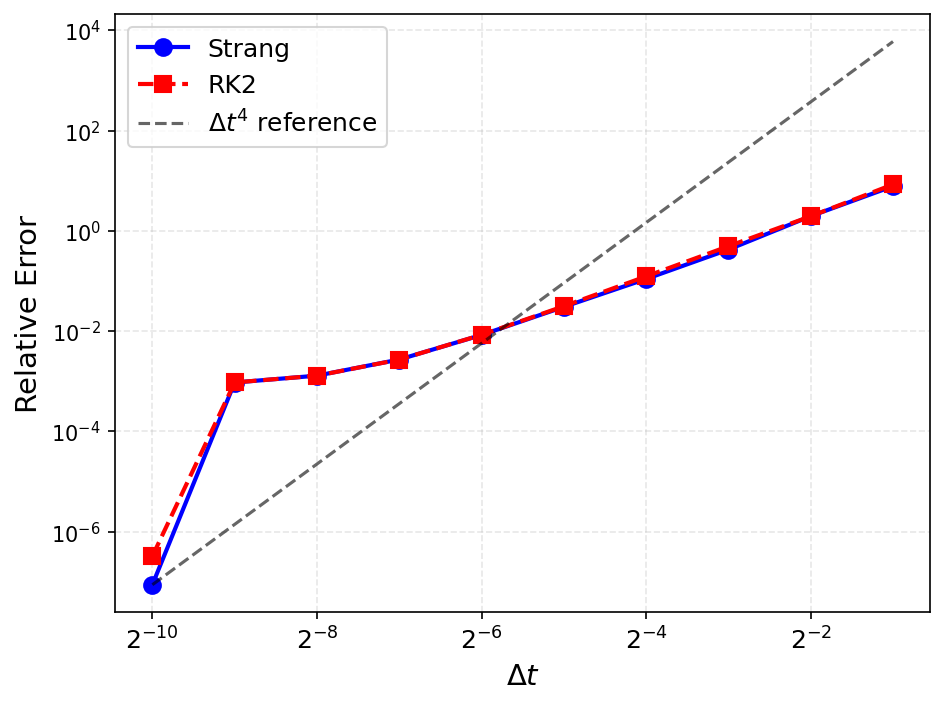}
        \caption{Two-stream with linear interpolation}
        \label{fig:pic_two_stream_1}
    \end{subfigure}

    \begin{subfigure}[t]{0.49\textwidth}
        \centering
        \includegraphics[width=\textwidth]{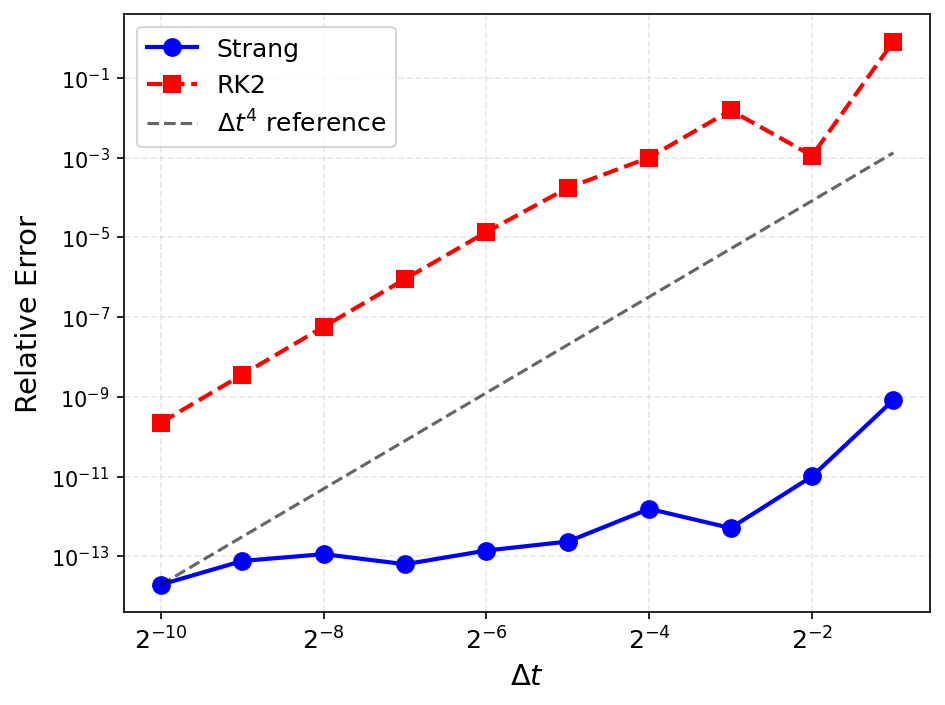}
        \caption{Landau with quadratic interpolation}
        \label{fig:pic_landau_2}
    \end{subfigure}
    \hfill \begin{subfigure}[t]{0.49\textwidth}
        \centering
        \includegraphics[width=\textwidth]{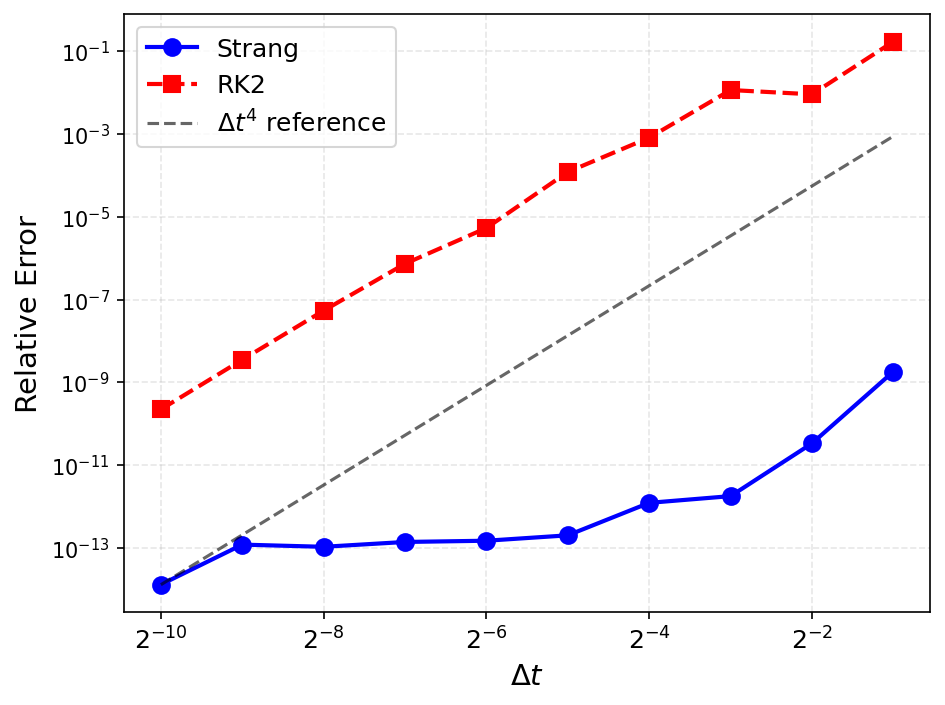}
        \caption{Two-stream with quadratic interpolation}
        \label{fig:pic_two_stream_2}
    \end{subfigure}

    \begin{subfigure}[t]{0.49\textwidth}
        \centering
        \includegraphics[width=\textwidth]{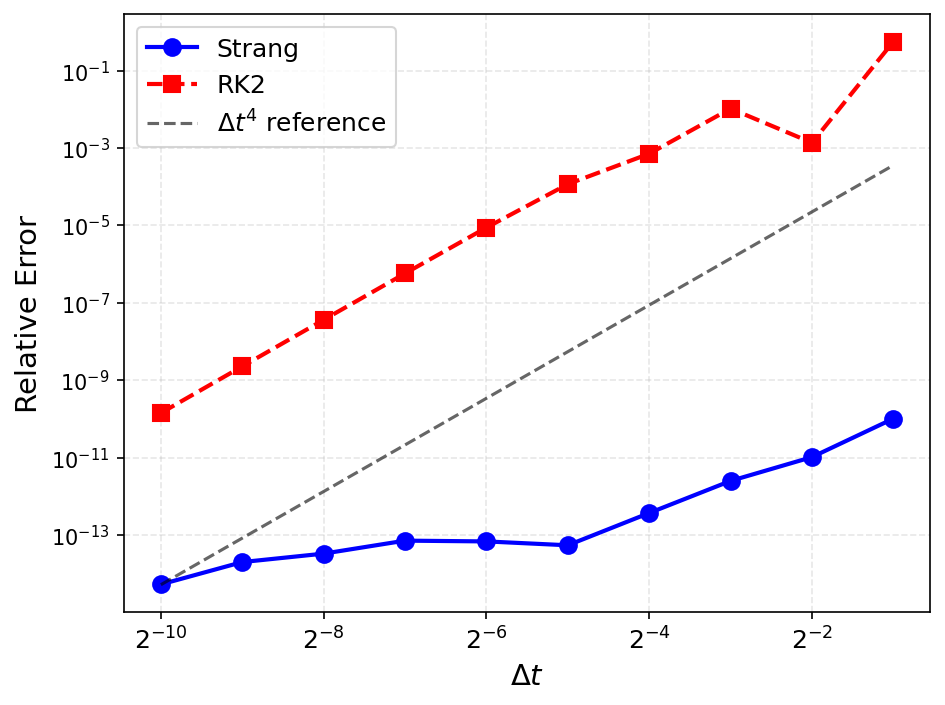}
        \caption{Landau with cubic interpolation}
        \label{fig:pic_landau_3}
    \end{subfigure}
    \hfill \begin{subfigure}[t]{0.49\textwidth}
        \centering
        \includegraphics[width=\textwidth]{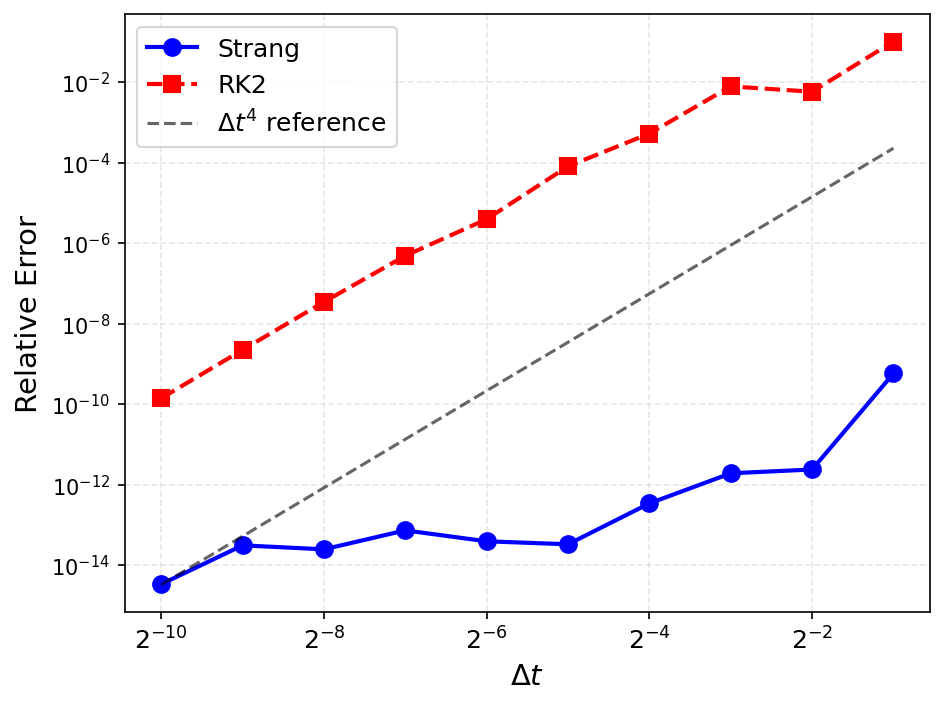}
        \caption{Two-stream with cubic interpolation}
        \label{fig:pic_two_stream_3}
    \end{subfigure}
    
    \caption{Convergence trends of single-step loop integral errors as a function of $\Delta t$. These tests used $128$ grid points on a domain of length $L=50$, with $N_p = 2$ particles, and $N_s = 32,768$ points to resolve the loop integrals.}
    \label{fig:pic_conv}
\end{figure}

See Figure \ref{fig:pic_conv} for the relative error in the symplectic diagnostic over a single time-step as a function of $\Delta t$, with $N_s$ and $N_p$ selected as described above. Due to a lack of spatial regularity, the approximate flow map from the piecewise linear, $p=1$, PIC method fails to be symplectic, and no discernible difference in loop integral conservation can be found between Strang splitting and RK2. In the piecewise quadratic and cubic test cases, we expect about $9$ and $13$--$14$ digits of accuracy, respectively. This is roughly observed in the numerical results. In both tests, Strang splitting conserves the loop integral to within about two or three orders of magnitude of machine precision, with only a modest reduction in error as the time-step is refined. This is in contrast with RK2, which clearly follows a convergence trend as the time-step decreases. This, combined with the results reported in Figure \ref{fig:pic_conv_N}, strongly supports the conclusion that these Strang-splitting algorithms are symplectic, as expected.

\begin{figure}[!htbp]
    \centering
\begin{subfigure}[t]{\textwidth}
        \centering
        \includegraphics[width=\textwidth]{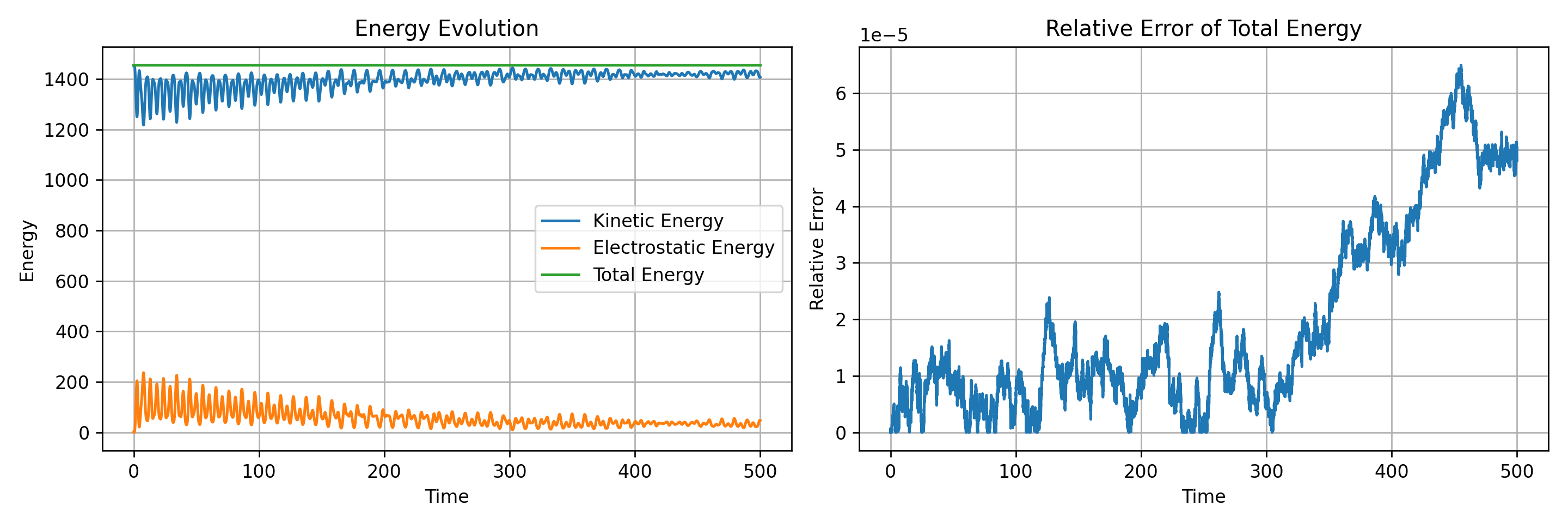}
        \caption{Linear interpolation}
        \label{fig:pic_nrg_1}
    \end{subfigure}

    \begin{subfigure}[t]{\textwidth}
        \centering
        \includegraphics[width=\textwidth]{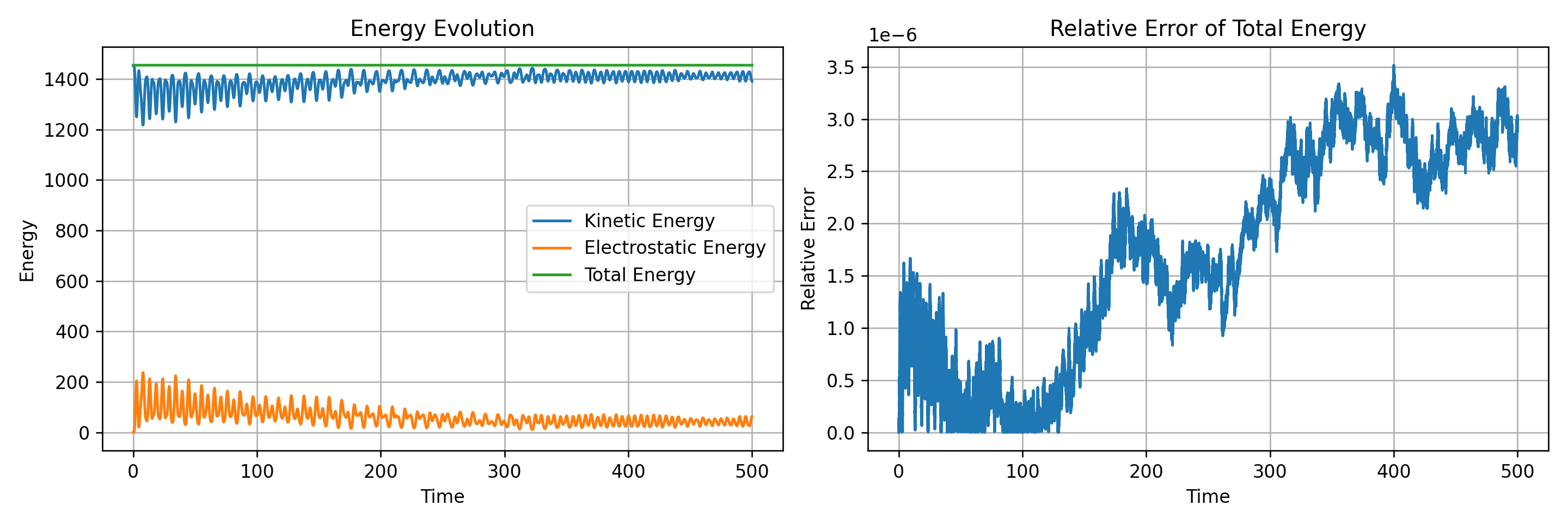}
        \caption{Quadratic interpolation}
        \label{fig:pic_nrg_2}
    \end{subfigure}

    \begin{subfigure}[t]{\textwidth}
        \centering
        \includegraphics[width=\textwidth]{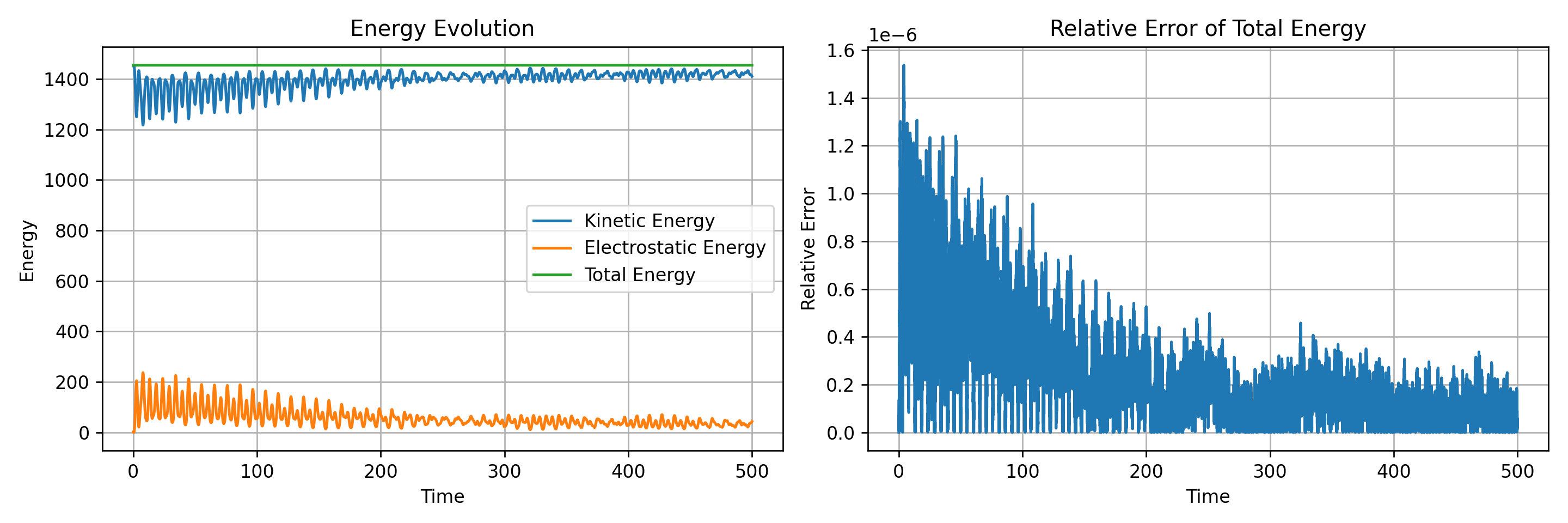}
        \caption{Cubic interpolation}
        \label{fig:pic_nrg_3}
    \end{subfigure}
    
    \caption{Energy conservation results for two-stream instability PIC simulations. These used $N_x = 200$ grid points, domain length $L=50$, time-step $\Delta t = 0.01$, and $N_p = 10,000$ particles.}
    \label{fig:pic_nrg}
\end{figure}

See Figure \ref{fig:pic_nrg} for a visualization of the relative error in the energy as a function of time. In all cases, energy is conserved well on average despite the linear case failing to conserve the loop integral diagnostic. 

\subsection{The effects of smoothing in PIC methods}

A common feature of PIC methods is smoothing via some kind of filtering operation \cite{birdsall2018plasma, hockney2021computer, chen2011energy}. These filters are introduced to reduce noise in the simulation by smoothing out the charge distribution prior to solving Poisson's equation. Binary filtering takes local averages in a moving window. Such filtering may be incorporated into the Hamiltonian structure as follows: given the circulant binary filtering matrix
\begin{equation}
    \mathbb{B}_{ij}
    =
    \begin{cases}
        1/2 \,, & i = j \\
        1/4 \,, & |i-j| = 1 \\
        0 \,, & \text{else} \,,
    \end{cases}
\end{equation}
the Hamiltonian is modified as follows:
\begin{equation}
    \mathcal{H}_s(\bm{x}_1, \bm{v}_1, \hdots, \bm{x}_{N_p}, \bm{v}_{N_p})
    =
    \frac{1}{2} \left( \sum_{a=1}^{N_p} w_a | \bm{v}_a |^2
    +
    \bm{\uprho}^T \mathbb{L}^{-1} \mathbb{B}^s \bm{\uprho} \right) \,,
\end{equation}
where $\bm{\uprho}$ is defined in equation \eqref{eq:pic_charge} and $\mathbb{B}^s$ denotes repeated filtering:
\begin{equation}
    \mathbb{B}^s
    =
    \underbrace{\mathbb{B} \mathbb{B} \hdots \mathbb{B}}_{s \text{ times}} \,.
\end{equation}
Hence, the filter is simply a post-processing step when depositing charge to the grid, applied prior to computing the electrostatic potential. With periodic boundary conditions and a uniform grid, both $\mathbb{B}^s$ and $\mathbb{L}^{-1}$ are circulant matrices. Hence, the two matrices commute. This is because circulant matrices commute with one another \cite{horn2012matrix}, indeed they are diagonalized by the discrete Fourier transform matrix. Therefore, although the notation does not make the fact explicit, the potential energy is still a symmetric positive semi-definite quadratic form in $\bm{\uprho}$. This filtering operation does not fundamentally alter the continuity of the Hamiltonian since the potential is still just a $p^{th}$-degree B-spline. However, smoothing should decrease the size of the discontinuities in the $p^{th}$ derivative as it effectively widens the interpolation stencil. 

Increasing $s$ while refining the grid so that $\sigma=\sqrt{s}\Delta x$ remains fixed effectively approximates smoothing by a Gaussian kernel. In Fourier space, the filter is given by:
\begin{equation}
    \widehat{\mathbb{B}}(k) = \frac{1}{2} + \frac{1}{2} \cos(k \Delta x) 
    \implies
    (\widehat{\mathbb{B}}(k))^s = \left( \frac{1}{2} + \frac{1}{2} \cos(k \Delta x) \right)^s \,.
\end{equation}
For $k \Delta x \ll 1$, that is for the low wave-numbers, we obtain:
\begin{equation}
    (\widehat{\mathbb{B}}(k))^s \approx \left( 1 - \frac{1}{4} (k \Delta x)^2 \right)^s \,.
\end{equation}
For large $s$, using the limit $(1 - x/s)^s \approx e^{-x}$, this expression approaches:
\begin{equation}
    \widehat{\mathbb{B}}_\infty(k)
    = \exp\left(-\left(\frac{\sigma k}{2}\right)^2 \right) \,,
    \quad \text{where} \quad
    \sigma = \sqrt{s} \Delta x \,.
\end{equation}
Transforming back to physical space, the filter remains Gaussian. Hence, even though smoothing does not eliminate the problem of limited regularity from low order polynomial interpolation in PIC, one might reasonably hope that sufficiently aggressive smoothing effectively mitigates the problem in an asymptotic sense. The following tests examine this supposition. 

\begin{figure}[!htbp]
    \centering
\begin{subfigure}[t]{0.49\textwidth}
        \centering
        \includegraphics[width=\textwidth]{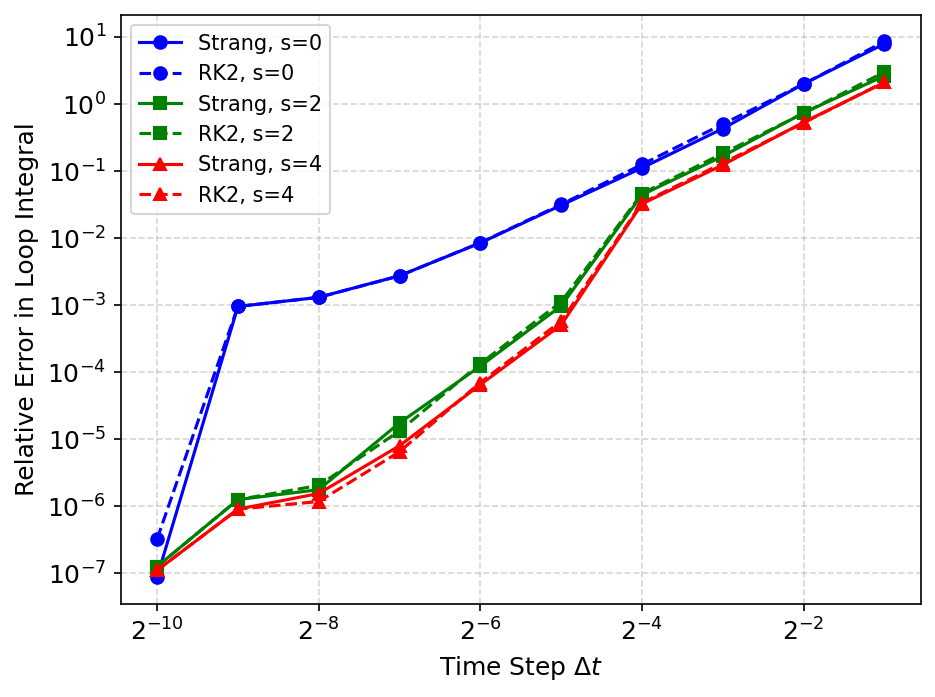}
        \caption{Linear interpolation, two-stream test}
        \label{fig:pic_filter_1}
    \end{subfigure}
    \hfill \begin{subfigure}[t]{0.49\textwidth}
        \centering
        \includegraphics[width=\textwidth]{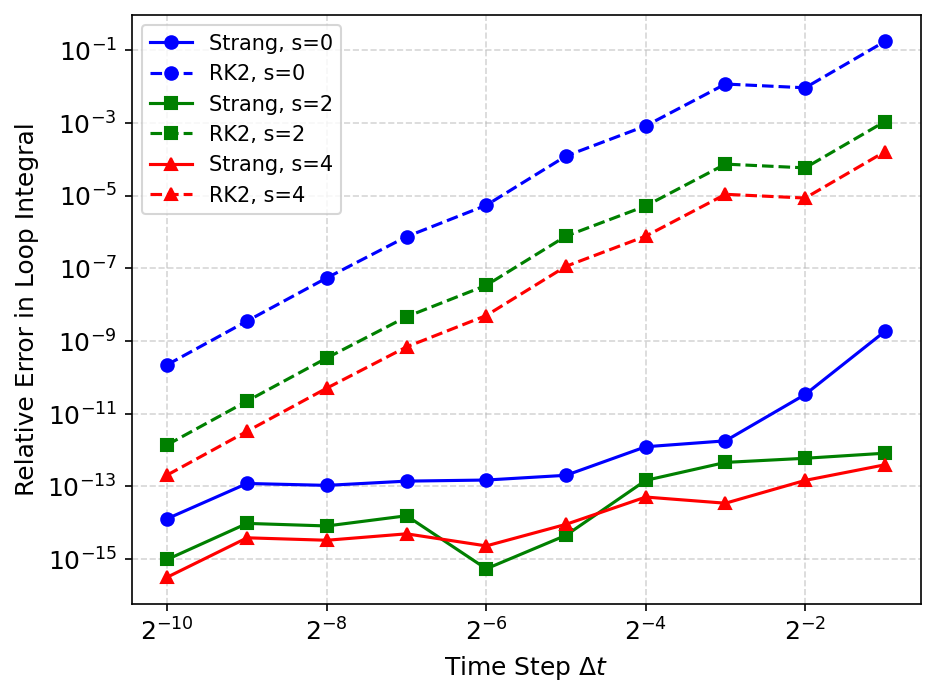}
        \caption{Quadratic interpolation, two-stream test}
        \label{fig:pic_filter_2}
    \end{subfigure}

    \caption{Convergence trends of single-step loop integral errors as a function of $\Delta t$ with binary filtering. These tests used $128$ grid points on a domain of length $L=50$, $\Delta t = 0.1$, with $N_p = 2$ particles, and $N_s = 32,768$ points to resolve the loop integrals.}
    \label{fig:pic_filter}

\end{figure}

See Figure \ref{fig:pic_filter} for the relative error in the symplectic diagnostic over a single time-step as a function of $\Delta t$ for $s=0,2,$ and $4$ applications of filtering for the linear and quadratic interpolation cases. The cubic interpolation test case is omitted since the results are entirely analogous to the quadratic case. We find that filtering reduces errors in the loop integral diagnostic slightly in the quadratic case, but has almost no effect in the linear case. Filtering most likely achieves two things: it modifies the convergence trend in $N_s$ allowing the loop integral approximation to converge to its true value more quickly, and also slightly reduces the magnitude of the error in the loop integral diagnostic incurred over a single time-step. The convergence rate of the approximation of the loop integral improves if the data is more regular (e.g.\! if the loop data has a smaller Lipschitz constant). However, filtering does not change the differentiability of the function, and therefore has limited impact as far as symplecticity conservation is concerned. 

\section{Conclusion}

This paper proposes a diagnostic tool for symplectic integration, with a particular emphasis on its use with symplectic PIC methods. While a diagnostic of this kind previously appeared in \cite{kraus2017projectedvariationalintegratorsdegenerate}, this work specifically considered how it works in relation with Hamiltonians of limited regularity and applied the diagnostic to study the symplecticity of structure-preserving particle-in-cell methods. The diagnostic converges spectrally for smooth data, with rates limited by the regularity of the Hamiltonian vector field. Due to this limitation, the diagnostic is most useful when applied to Hamiltonian systems with a high degree of regularity, but may nonetheless be used for low regularity systems if a sufficiently high resolution is used. 

The key finding of this work is that ``symplectic'' PIC methods which use piecewise linear interpolation do not, in fact, preserve symplecticity once particles cross cell interfaces. More generally, for explicit Hamiltonian-splitting PIC methods, any interpolation whose force field is discontinuous across cell interfaces generically produces discontinuous time-advance maps at such crossings, and therefore generically fails to preserve symplecticity. Such crossings are unavoidable in high-dimensional simulations except for specially prepared initial data over a short time horizon. This problem is not ameliorated by standard filtering techniques. Rather, to rule out this mechanism, one must ensure continuous differentiability of the Hamiltonian. This condition rules out the particular loop-breaking mechanism caused by discontinuous force fields, ensuring that the splitting substeps are spatially continuous. Classical Jacobian-based symplecticity requires more, namely differentiability of the time-advance map at the points where the symplecticity criterion is invoked. The loop-integral diagnostic indicates that a weaker sense of symplecticity can hold: we find preservation of the Poincar{\'e} integral invariant along advected loops, even when the flow map is merely continuous but not differentiable. 

In electrostatic PIC, this regularity criterion corresponds to requiring that the electrostatic potential be interpolated in a globally $C^1$ basis. One may reasonably expect a similar principle to apply for electromagnetic PIC methods with time-stepping based on Hamiltonian splitting \cite{he2015hamiltonian, kraus2017gempic}. For the purposes of efficiently resolving the loop integral diagnostic, cubic or higher interpolation is preferable to quadratic interpolation, as the diagnostic converges with a smaller ensemble of simulations. Furthermore, while B-splines are conventional in PIC, and this work used B-splines, it is not unknown to use Lagrange or other forms of polynomial interpolation \cite{kormann2024dual}, especially in finite element PIC methods \cite{bettencourt2021empire, glasser2022gauge, glasser2023generalizing, finn2023numerical}. One should be careful in such cases, as other forms of polynomial interpolation do not link interpolation degree with global regularity in such a direct manner as B-splines: e.g.\! degree-$(p+1)$ Lagrange interpolation on a non-uniform grid does not automatically yield a $C^p$ interpolant. Therefore, true symplectic PIC methods require more than a compatible choice of basis functions for the fields and a symplectic time-stepper. Regularity of the basis across cell interfaces is also crucial. 

This work exclusively considered explicit symplectic integration based on Strang splitting, as this form of time-stepping is conventional in many structure-preserving PIC algorithms. Implicit time-stepping methods are beyond the scope of this work, and may or may not admit symplectic algorithms with low-order spatial regularity. Such implicit symplectic integrators are a topic of particular interest for future work. The sensitivity of symplecticity preservation to solver tolerance is worth investigating. The conservation of the Poincar{\'e} integral invariant by conjugate symplectic methods is likewise of interest. Finally, the performance of the diagnostic in energy conserving PIC methods \cite{chen2011energy,chen2014energy,chen2015multi, chen2020semi, chacon2013charge, chacon2016curvilinear, ricketson2025explicit, kormann2021energy, ji2023asymptotic}, which are a significant thrust in contemporary PIC literature, is of interest. The desire to distinguish symplectic and conservative, non-symplectic PIC methods was a core motivation for developing this diagnostic tool, and will be considered in subsequent work. 

\section*{Declaration of Competing Interest}
The authors declare that they have no known competing financial interests or personal relationships that could have appeared to influence the work reported in this paper.

\section*{Data and Code Availability} 
The source code to reproduce all the results found in this paper is publicly available on GitHub at:
\url{https://github.com/wbarham/symplectic_diagnostic}. This repository also serves as a template for using the diagnostic in other contexts. 

\section*{Acknowledgements}
This material is based on work supported by the U.S. Department of Energy, Office of Science, Office of Advanced Scientific Computing Research, as a part of the Mathematical Multifaceted Integrated Capability Centers program, under Award Number DE-SC0023164. It was also supported by U.S. Department of Energy grant \# DE-FG02-04ER54742. WB was supported by the Laboratory Directed Research and Development program of Los Alamos National Laboratory under project number 20251151PRD1. Los Alamos Laboratory Report LA-UR-25-31632. 

\bibliographystyle{elsarticle-num} \bibliography{references}          

\appendix
\section{The convergence rate of the high-regularity loop integral approximation} 
\label{appendix:conv_high_regularity}

This appendix leverages results found in standard references on spectral and pseudospectral methods \cite{gottlieb1977numerical, boyd2001chebyshev}. We use the Fourier convention
\begin{equation}
    \hat f_k = \int_0^1 f(x)e^{-2\pi i kx}\,\mathsf d x,
    \qquad
    f(x) = \sum_{k\in\mathbb Z} \hat f_k e^{2\pi i kx}.
\end{equation}
Let
\begin{equation}
    \langle k\rangle = (1+|k|^2)^{1/2}.
\end{equation}
For $r\ge 0$, the periodic Sobolev space $H^r_{\mathrm{per}}([0,1])$ is defined by
\begin{equation}
    H^r_{\mathrm{per}}([0,1])
    =
    \left\{
    f\in L^2([0,1]) :
    \|f\|_{H^r([0,1])}^2
    :=
    \sum_{k\in\mathbb Z}\langle k\rangle^{2r}|\hat f_k|^2
    <\infty
    \right\}.
\end{equation}
This is the usual Sobolev space on the one-dimensional torus. If $r>1/2$, then $H^r_{\mathrm{per}}([0,1])$ embeds continuously into $C^0_{\mathrm{per}}([0,1])$, and point values such as $f(j/N)$ are well-defined.

For $p,q\in H^r_{\mathrm{per}}([0,1])$ with $r>1/2$, define the loop integral, or Sobolev-duality extension of the loop integral, by
\begin{equation}
    \mathcal I(p,q)
    =
    2\pi i\sum_{k\in\mathbb Z} k\,\hat p_{-k}\hat q_k .
\end{equation}
This series is absolutely convergent. Indeed,
\begin{equation}
    \sum_{k\in\mathbb Z}|k|\,|\hat p_{-k}|\,|\hat q_k|
    \le
    \sum_{k\in\mathbb Z}|k|^{1-2r}
    \left(|k|^r|\hat p_{-k}|\right)
    \left(|k|^r|\hat q_k|\right)
    \le
    \sum_{k\in\mathbb Z}
    \left(|k|^r|\hat p_{-k}|\right)
    \left(|k|^r|\hat q_k|\right)
    \le
    \|p\|_{H^r}\|q\|_{H^r},
\end{equation}
where the $k=0$ term is zero and $|k|^{1-2r}\le 1$ for $|k|\ge 1$ when $r>1/2$. When $q$ is sufficiently regular for $q'$ to be an ordinary $L^2$ function, this definition agrees with
\begin{equation}
    \mathcal I(p,q)
    =
    \int_0^1 p(s)q'(s)\,\mathsf d s.
\end{equation}
For $1/2<r<1$, however, $q'$ need not be an $L^2$ function, so the Fourier series above should be interpreted as the natural Sobolev-duality pairing rather than as a classical integral.

\subsection{The pseudospectral loop integral approximation}

We seek an approximation of
\begin{equation}
    \mathcal I(p,q)
    =
    2\pi i\sum_{k\in\mathbb Z} k\,\hat p_{-k}\hat q_k
\end{equation}
using only finitely many equispaced samples of the phase-space loop. To avoid an inessential Nyquist-frequency convention, assume throughout this subsection that $N=2M+1$ is odd and set
\begin{equation}
    \Lambda_N = \{-M,-M+1,\ldots,M-1,M\}.
\end{equation}
The even-$N$ case is handled similarly after fixing a convention for the Nyquist mode, for instance by discarding it or setting its derivative multiplier to zero.

Let
\begin{equation}
    \mathsf p_j = p(j/N),
    \qquad
    \mathsf q_j = q(j/N),
    \qquad
    j=0,\ldots,N-1.
\end{equation}
For $k\in\Lambda_N$, define the discrete Fourier transform coefficients
\begin{equation}
    \tilde{\mathsf f}_k
    =
    \frac{1}{N}\sum_{j=0}^{N-1} f(j/N)e^{-2\pi i k j/N}.
\end{equation}
The trigonometric interpolant is
\begin{equation}
    I_N f(x)
    =
    \sum_{k\in\Lambda_N}\tilde{\mathsf f}_k e^{2\pi i kx},
\end{equation}
and the pseudospectral derivative of $q$ at the grid points is
\begin{equation}
    (D_N I_Nq)(j/N)
    =
    \sum_{k\in\Lambda_N}2\pi i k\,\tilde{\mathsf q}_k e^{2\pi i k j/N}.
\end{equation}
The collocation approximation is
\begin{equation}
    \mathcal I_N(p,q)
    :=
    \frac{1}{N}\sum_{j=0}^{N-1}p(j/N)(D_N I_Nq)(j/N).
\end{equation}
Equivalently, by the discrete orthogonality relation on the grid,
\begin{equation}
    \mathcal I_N(p,q)
    =
    2\pi i\sum_{k\in\Lambda_N} k\,\tilde{\mathsf p}_{-k}\tilde{\mathsf q}_k .
\end{equation}
This is the same bilinear form as the Fourier expression for $\mathcal I(p,q)$, but with the exact Fourier coefficients replaced by aliased DFT coefficients and the sum restricted to the resolved modes.

The objective of this section is to bound
\begin{equation}
    \mathcal E_N(p,q)
    =
    \left|\mathcal I(p,q)-\mathcal I_N(p,q)\right|.
\end{equation}
There are two sources of error: Fourier truncation and aliasing of the Fourier coefficients induced by sampling.

\begin{lemma}[Fourier--Galerkin error estimate]\label{lem:fourier_galerkin}
Let $f,g\in H^r_{\mathrm{per}}([0,1])$ for some $r>1/2$, and let
\begin{equation}
    P_N f(x)=\sum_{k\in\Lambda_N}\hat f_k e^{2\pi i kx},
    \qquad
    P_N g(x)=\sum_{k\in\Lambda_N}\hat g_k e^{2\pi i kx}.
\end{equation}
Then
\begin{equation}
    \left|
    \mathcal I(f,g)
    -
    2\pi i\sum_{k\in\Lambda_N}k\,\hat f_{-k}\hat g_k
    \right|
    \le
    C_r N^{-(2r-1)}\|f\|_{H^r}\|g\|_{H^r}.
\end{equation}
Equivalently, whenever the classical integrals are defined,
\begin{equation}
    \left|
    \int_0^1 f(x)g'(x)\,\mathsf d x
    -
    \int_0^1 P_N f(x)(P_N g)'(x)\,\mathsf d x
    \right|
    \le
    C_r N^{-(2r-1)}\|f\|_{H^r}\|g\|_{H^r}.
\end{equation}
\end{lemma}

\begin{proof}
The error is the tail
\begin{equation}
    E_N
    =
    \left|
    2\pi i\sum_{|k|>M}k\,\hat f_{-k}\hat g_k
    \right|.
\end{equation}
Using Cauchy--Schwarz,
\begin{equation}
\begin{aligned}
    E_N
    &\le
    2\pi\sum_{|k|>M}|k|^{1-2r}
    \left(|k|^r|\hat f_{-k}|\right)
    \left(|k|^r|\hat g_k|\right) \\
    &\le
    2\pi M^{-(2r-1)}
    \left(\sum_{|k|>M}|k|^{2r}|\hat f_{-k}|^2\right)^{1/2}
    \left(\sum_{|k|>M}|k|^{2r}|\hat g_k|^2\right)^{1/2} \\
    &\le
    C_r N^{-(2r-1)}\|f\|_{H^r}\|g\|_{H^r},
\end{aligned}
\end{equation}
since $M=(N-1)/2$.
\end{proof}

\begin{lemma}[Aliasing formula for DFT coefficients]\label{lem:aliasing_lemma}
Let $f\in H^r_{\mathrm{per}}([0,1])$ for some $r>1/2$. For $k\in\Lambda_N$, define
\begin{equation}
    \tilde{\mathsf f}_k
    =
    \frac{1}{N}\sum_{j=0}^{N-1}f(j/N)e^{-2\pi i k j/N}.
\end{equation}
Then
\begin{equation}
    \tilde{\mathsf f}_k
    =
    \sum_{\ell\in\mathbb Z}\hat f_{k+\ell N}.
\end{equation}
\end{lemma}

\begin{proof}
Since $r>1/2$, the Fourier coefficients of $f$ are absolutely summable, and the Fourier series of $f$ converges uniformly to its continuous periodic representative. Thus
\begin{equation}
\begin{aligned}
    \tilde{\mathsf f}_k
    &=
    \frac{1}{N}\sum_{j=0}^{N-1}
    \sum_{m\in\mathbb Z}\hat f_m e^{2\pi i m j/N}e^{-2\pi i k j/N} \\
    &=
    \sum_{m\in\mathbb Z}\hat f_m
    \left[\frac{1}{N}\sum_{j=0}^{N-1}e^{2\pi i(m-k)j/N}\right].
\end{aligned}
\end{equation}
The bracketed sum is equal to one when $m-k$ is divisible by $N$ and zero otherwise. Therefore only indices $m=k+\ell N$ contribute.
\end{proof}

\begin{lemma}[Aliasing error in Sobolev norms]\label{lem:aliasing_sobolev}
Let $f\in H^r_{\mathrm{per}}([0,1])$ for some $r>1/2$. Then, for every $s$ with $0\le s\le r$,
\begin{equation}
    \|I_N f-P_N f\|_{H^s}
    \le
    C_{r,s}N^{s-r}\|f\|_{H^r}.
\end{equation}
\end{lemma}

\begin{proof}
Write
\[
    I_N f-P_N f
    =
    \sum_{k\in\Lambda_N} e_k e^{2\pi i kx},
    \qquad
    e_k=\tilde{\mathsf f}_k-\hat f_k
    =
    \sum_{\ell\ne0}\hat f_{k+\ell N}.
\]
Choose \(\alpha\) such that
\[
    1<\alpha\le 2r.
\]
Such a choice is possible because \(r>1/2\). By Cauchy--Schwarz,
\[
\begin{aligned}
    |e_k|^2
    &=
    \left|
    \sum_{\ell\ne0}
    (1+|\ell|)^{-\alpha/2}
    (1+|\ell|)^{\alpha/2}
    \hat f_{k+\ell N}
    \right|^2  \\
    &\le
    \left(
    \sum_{\ell\ne0}(1+|\ell|)^{-\alpha}
    \right)
    \left(
    \sum_{\ell\ne0}(1+|\ell|)^\alpha
    |\hat f_{k+\ell N}|^2
    \right)  \\
    &\le
    C_\alpha
    \sum_{\ell\ne0}(1+|\ell|)^\alpha
    |\hat f_{k+\ell N}|^2,
\end{aligned}
\]
where \(C_\alpha<\infty\) because \(\alpha>1\). Hence
\[
\begin{aligned}
    \|I_N f-P_N f\|_{H^s}^2
    &=
    \sum_{k\in\Lambda_N}
    \langle k\rangle^{2s}|e_k|^2  \\
    &\le
    C_\alpha
    \sum_{k\in\Lambda_N}
    \sum_{\ell\ne0}
    \langle k\rangle^{2s}
    (1+|\ell|)^\alpha
    |\hat f_{k+\ell N}|^2 .
\end{aligned}
\]

We now estimate the weight multiplying
\(|\hat f_{k+\ell N}|^2\). Since \(k\in\Lambda_N\), we have
\[
    \langle k\rangle \le C N.
\]
Therefore
\[
    \langle k\rangle^{2s}
    \le C_s N^{2s}.
\]
On the other hand, for \(k\in\Lambda_N\) and \(\ell\ne0\), the aliased
frequency \(k+\ell N\) satisfies
\[
    \langle k+\ell N\rangle
    \ge c N(1+|\ell|),
\]
for some constant \(c>0\). Consequently,
\[
    1+|\ell|
    \le C N^{-1}\langle k+\ell N\rangle,
\]
and so
\[
    (1+|\ell|)^\alpha
    \le C_\alpha N^{-\alpha}
    \langle k+\ell N\rangle^\alpha .
\]
Combining these two bounds gives
\[
\begin{aligned}
    \langle k\rangle^{2s}(1+|\ell|)^\alpha
    &\le
    C_{s,\alpha}
    N^{2s-\alpha}
    \langle k+\ell N\rangle^\alpha  \\
    &=
    C_{s,\alpha}
    N^{2(s-r)}
    N^{2r-\alpha}
    \langle k+\ell N\rangle^\alpha .
\end{aligned}
\]
Since \(\alpha\le 2r\) and \(\langle k+\ell N\rangle\ge cN\), we also have
\[
    N^{2r-\alpha}
    \le C_{r,\alpha}
    \langle k+\ell N\rangle^{2r-\alpha}.
\]
Thus
\[
\begin{aligned}
    \langle k\rangle^{2s}(1+|\ell|)^\alpha
    &\le
    C_{r,s}
    N^{2(s-r)}
    \langle k+\ell N\rangle^{2r-\alpha}
    \langle k+\ell N\rangle^\alpha  \\
    &=
    C_{r,s}
    N^{2(s-r)}
    \langle k+\ell N\rangle^{2r}.
\end{aligned}
\]
Therefore
\[
\begin{aligned}
    \|I_N f-P_N f\|_{H^s}^2
    &\le
    C_{r,s}N^{2(s-r)}
    \sum_{k\in\Lambda_N}\sum_{\ell\ne0}
    \langle k+\ell N\rangle^{2r}
    |\hat f_{k+\ell N}|^2 .
\end{aligned}
\]
Finally, the indices \(k+\ell N\), with \(k\in\Lambda_N\) and
\(\ell\ne0\), are a subset of the Fourier modes of \(f\). Hence
\[
    \sum_{k\in\Lambda_N}\sum_{\ell\ne0}
    \langle k+\ell N\rangle^{2r}
    |\hat f_{k+\ell N}|^2
    \le
    \sum_{n\in\mathbb Z}
    \langle n\rangle^{2r}|\hat f_n|^2
    =
    \|f\|_{H^r}^2.
\]
We conclude that
\[
    \|I_N f-P_N f\|_{H^s}^2
    \le
    C_{r,s}N^{2(s-r)}\|f\|_{H^r}^2.
\]
Taking square roots gives
\[
    \|I_N f-P_N f\|_{H^s}
    \le
    C_{r,s}N^{s-r}\|f\|_{H^r}.
\]
\end{proof}

We now state the full convergence estimate. The important point is that the Fourier--Galerkin truncation error controls the rate for $1/2<r\leq1$, while sampling and aliasing control the rate for $r\geq1$. Thus the overall rate is $N^{-\min\{2r-1,r\}}$.

\begin{theorem}[Error in the collocation approximation]\label{thm:collocation_error}
Let $f,g\in H^r_{\mathrm{per}}([0,1])$ for some $r>1/2$. Let $N=2M+1$, let $I_N f$ and $I_N g$ be the trigonometric interpolants from $N$ equispaced samples, and define
\begin{equation}
    \mathcal I_N(f,g)
    =
    \frac{1}{N}\sum_{j=0}^{N-1}f(j/N)(D_N I_Ng)(j/N)
    =
    2\pi i\sum_{k\in\Lambda_N}k\,\tilde{\mathsf f}_{-k}\tilde{\mathsf g}_k.
\end{equation}
Then
\begin{equation}
    \left|\mathcal I(f,g)-\mathcal I_N(f,g)\right|
    \le
    \begin{cases}
    C_r N^{-(2r-1)}\|f\|_{H^r}\|g\|_{H^r},
        & 1/2<r\le 1,\\[4pt]
    C_r N^{-r}\|f\|_{H^r}\|g\|_{H^r},
        & r\ge 1.
    \end{cases}
\end{equation}
\end{theorem}

\begin{proof}
Let
\begin{equation}
    \mathcal B_N(u,v)
    =
    2\pi i\sum_{k\in\Lambda_N}k\,\hat u_{-k}\hat v_k
\end{equation}
for trigonometric polynomials with frequencies in $\Lambda_N$. Then
\begin{equation}
    \mathcal I_N(f,g)=\mathcal B_N(I_N f,I_N g),
    \qquad
    \mathcal B_N(P_N f,P_N g)
    =
    2\pi i\sum_{k\in\Lambda_N}k\,\hat f_{-k}\hat g_k.
\end{equation}
By Lemma~\ref{lem:fourier_galerkin},
\begin{equation}
    \left|\mathcal I(f,g)-\mathcal B_N(P_N f,P_N g)\right|
    \le
    C_r N^{-(2r-1)}\|f\|_{H^r}\|g\|_{H^r}.
\end{equation}
It remains to bound the aliasing contribution. We first record the elementary estimate
\begin{equation}\label{eq:BN_bound}
    |\mathcal B_N(u,v)|
    \le
    C_{a,b}\|u\|_{H^a}\|v\|_{H^b},
    \qquad
    a,b\ge0,\quad a+b\ge1.
\end{equation}
Indeed,
\begin{equation}
\begin{aligned}
    |\mathcal B_N(u,v)|
    &\le
    2\pi\sum_{k\in\Lambda_N}|k|\,|\hat u_{-k}|\,|\hat v_k| \\
    &\le
    C_{a,b}
    \left(\sum_{k\in\Lambda_N}\langle k\rangle^{2a}|\hat u_{-k}|^2\right)^{1/2}
    \left(\sum_{k\in\Lambda_N}\langle k\rangle^{2b}|\hat v_k|^2\right)^{1/2},
\end{aligned}
\end{equation}
because $|k|\le \langle k\rangle^{a+b}$ when $a+b\ge1$.

Set
\begin{equation}
    e_f=I_N f-P_N f,
    \qquad
    e_g=I_N g-P_N g.
\end{equation}
Then
\begin{equation}
\begin{aligned}
    \mathcal B_N(I_N f,I_N g)-\mathcal B_N(P_N f,P_N g)
    &=
    \mathcal B_N(e_f,P_N g)
    +
    \mathcal B_N(P_N f,e_g)
    +
    \mathcal B_N(e_f,e_g).
\end{aligned}
\end{equation}
Let
\begin{equation}
    a_r=\max\{0,1-r\}.
\end{equation}
Then $a_r+r\ge1$ and $0\le a_r\le r$. Using \eqref{eq:BN_bound} and Lemma~\ref{lem:aliasing_sobolev},
\begin{equation}
\begin{aligned}
    |\mathcal B_N(e_f,P_N g)|
    &\le
    C_r\|e_f\|_{H^{a_r}}\|P_N g\|_{H^r}
    \le
    C_r N^{a_r-r}\|f\|_{H^r}\|g\|_{H^r}, \\
    |\mathcal B_N(P_N f,e_g)|
    &\le
    C_r\|P_N f\|_{H^r}\|e_g\|_{H^{a_r}}
    \le
    C_r N^{a_r-r}\|f\|_{H^r}\|g\|_{H^r}.
\end{aligned}
\end{equation}
If $1/2<r\le1$, then $a_r-r=1-2r=-(2r-1)$. If $r\ge1$, then $a_r-r=-r$.

For the remaining quadratic aliasing term, use \eqref{eq:BN_bound} with $a=b=1/2$ and Lemma~\ref{lem:aliasing_sobolev}:
\begin{equation}
    |\mathcal B_N(e_f,e_g)|
    \le
    C\|e_f\|_{H^{1/2}}\|e_g\|_{H^{1/2}}
    \le
    C_r N^{1-2r}\|f\|_{H^r}\|g\|_{H^r}.
\end{equation}
Combining these estimates with the Fourier--Galerkin truncation estimate gives
\begin{equation}
    |\mathcal I(f,g)-\mathcal I_N(f,g)|
    \le
    C_r N^{-\min\{2r-1,r\}}\|f\|_{H^r}\|g\|_{H^r},
\end{equation}
as claimed.
\end{proof}
 
\section{The convergence rate of the low-regularity loop integral approximation} \label{appendix:conv_low_regularity}

The appropriate space to study the convergence of numerical methods for
discontinuous functions is the space of functions of bounded variation
\cite{evans2015measure}. Recall that \(BV([0,1])\subset L^1([0,1])\)
consists of functions whose total variation is finite:
\begin{equation}
    \|f\|_{BV([0,1])}
    :=
    \|f\|_{L^1([0,1])}
    +
    \mathrm{TV}_{[0,1]}(f)
    <
    \infty ,
\end{equation}
where
\begin{equation}
    \mathrm{TV}_{[0,1]}(f)
    :=
    \sup
    \left\{
    \sum_{i=1}^n |f(x_i)-f(x_{i-1})|:
    0=x_0<x_1<\cdots<x_n=1
    \right\}.
\end{equation}

For a function \(f\) with a finite number of jump discontinuities, we denote
its jump set by \(\mathcal J_f\). At a jump point \(a\in\mathcal J_f\), we
write
\begin{equation}
    [f]_a := f(a^+)-f(a^-)
\end{equation}
for the jump size, where \(f(a^\pm)\) denote the one-sided limits. Given two
functions \(p\) and \(q\), we write
\begin{equation}
    \mathcal J := \mathcal J_p\cup \mathcal J_q
\end{equation}
for the union of their jump sets, and we let \(\mathcal S\) denote the
collection of connected components of \([0,1]\setminus\mathcal J\).

In this section, the loop integral is understood in a piecewise smooth
sense. Namely, if \(q\in W^{1,\infty}(I)\) for each \(I\in\mathcal S\), we
define
\begin{equation}
    \mathcal I(q,p)
    :=
    \sum_{I\in\mathcal S}
    \int_I p(s)q'(s)\,\mathsf{d}s .
\end{equation}
This definition retains only the absolutely continuous part of
\(\mathsf{d}q\) on the smooth pieces and excludes the jump contribution.

\begin{theorem}[First-order error estimate for the piecewise smooth loop integral]
Let \(p,q:[0,1]\to\mathbb R\) be bounded functions with finite sets of
interior jump discontinuities. Suppose that \(p\in BV([0,1])\) and that
\(q\in W^{1,\infty}(I)\) for each \(I\in\mathcal S\). Define
\begin{equation}
    L_q
    :=
    \max_{I\in\mathcal S}
    \|q'\|_{L^\infty(I)}
    < \infty .
\end{equation}

Define the piecewise smooth loop integral by
\begin{equation}
    \mathcal I(q,p)
    :=
    \sum_{I\in\mathcal S}
    \int_I p(s)q'(s)\,\mathsf{d} s .
\end{equation}
Let \(s_k=k/N\), \(k=0,\ldots,N\), let \(\Delta s=1/N\), and define
\begin{equation}
    G_k := [s_k,s_{k+1}],
    \qquad
    k=0,\ldots,N-1.
\end{equation}
Assume that no grid point \(s_k\) lies in \(\mathcal J\). Define the retained
and omitted cell sets by
\begin{equation}
    \mathcal K_N
    :=
    \{k\in\{0,\ldots,N-1\}:G_k\cap\mathcal J=\emptyset\},
\end{equation}
and
\begin{equation}
    \mathcal O_N
    :=
    \{k\in\{0,\ldots,N-1\}:G_k\cap\mathcal J\neq\emptyset\}.
\end{equation}
The discrete approximation is
\begin{equation}
    \mathcal I_N(q,p)
    :=
    \sum_{k\in\mathcal K_N}
    p(s_k)\bigl(q(s_{k+1})-q(s_k)\bigr).
\end{equation}
Then
\begin{equation}
    |\mathcal I(q,p)-\mathcal I_N(q,p)|
    \le
    \frac{L_q}{N}
    \left(
        \mathrm{TV}_{[0,1]}(p)
        +
        \#(\mathcal O_N)\,\|p\|_{L^\infty([0,1])}
    \right).
\end{equation}
In particular, since \(\#(\mathcal O_N)\le \#(\mathcal J)\), one obtains
\begin{equation}
    |\mathcal I(q,p)-\mathcal I_N(q,p)|
    \le
    \frac{L_q}{N}
    \left(
        \mathrm{TV}_{[0,1]}(p)
        +
        \#(\mathcal J)\,\|p\|_{L^\infty([0,1])}
    \right).
\end{equation}
\end{theorem}

\begin{proof}
We split the error into contributions from cells that do not contain a jump
and cells that do. For \(k\in\mathcal K_N\), define
\begin{equation}
    E_k
    :=
    \int_{G_k} p(s)q'(s)\,\mathsf{d} s
    -
    p(s_k)\bigl(q(s_{k+1})-q(s_k)\bigr).
\end{equation}
For \(k\in\mathcal O_N\), define
\begin{equation}
    R_k
    :=
    \sum_{I\in\mathcal S}
    \int_{G_k\cap I} p(s)q'(s)\,\mathsf{d} s .
\end{equation}
That is, \(R_k\) is the contribution of the exact integral over the smooth
portions of an omitted cell. Since the discrete approximation assigns no
contribution to omitted cells, we have
\begin{equation}
    \mathcal I(q,p)-\mathcal I_N(q,p)
    =
    \sum_{k\in\mathcal K_N} E_k
    +
    \sum_{k\in\mathcal O_N} R_k .
\end{equation}

First consider a retained cell \(G_k\), so that
\(G_k\cap\mathcal J=\emptyset\). On this cell,
\(q\in W^{1,\infty}(G_k)\), and hence
\begin{equation}
    q(s_{k+1})-q(s_k)
    =
    \int_{G_k} q'(s)\,\mathsf{d} s .
\end{equation}
Therefore
\begin{equation}
\begin{aligned}
    E_k
    &=
    \int_{G_k} p(s)q'(s)\,\mathsf{d} s
    -
    p(s_k)\int_{G_k}q'(s)\,\mathsf{d} s  \\
    &=
    \int_{G_k} \bigl(p(s)-p(s_k)\bigr)q'(s)\,\mathsf{d} s .
\end{aligned}
\end{equation}
Using the definition of \(L_q\), we obtain
\begin{equation}
    |E_k|
    \le
    L_q
    \int_{G_k}|p(s)-p(s_k)|\,\mathsf{d} s .
\end{equation}
Since \(p\in BV([0,1])\), for every \(s\in G_k\),
\begin{equation}
    |p(s)-p(s_k)|
    \le
    \mathrm{TV}_{G_k}(p).
\end{equation}
Hence
\begin{equation}
    \int_{G_k}|p(s)-p(s_k)|\,\mathsf{d} s
    \le
    \Delta s\,\mathrm{TV}_{G_k}(p),
\end{equation}
and so
\begin{equation}
    |E_k|
    \le
    \Delta s\,L_q\,\mathrm{TV}_{G_k}(p).
\end{equation}
Summing over retained cells gives
\begin{equation}
    \sum_{k\in\mathcal K_N}|E_k|
    \le
    \Delta s\,L_q
    \sum_{k\in\mathcal K_N}\mathrm{TV}_{G_k}(p)
    \le
    \Delta s\,L_q\,\mathrm{TV}_{[0,1]}(p).
\end{equation}

Now consider an omitted cell \(G_k\), with \(k\in\mathcal O_N\). The exact
integral contributes only over the smooth portions of \(G_k\), since
\(\mathcal I(q,p)\) excludes the jump part of \(\mathsf{d}q\). Thus
\begin{equation}
    |R_k|
    \le
    \sum_{I\in\mathcal S}
    \int_{G_k\cap I}|p(s)|\,|q'(s)|\,\mathsf{d} s
    \le
    \Delta s\,\|p\|_{L^\infty([0,1])}\,L_q .
\end{equation}
Summing over omitted cells,
\begin{equation}
    \sum_{k\in\mathcal O_N}|R_k|
    \le
    \Delta s\,\#(\mathcal O_N)\,
    \|p\|_{L^\infty([0,1])}\,L_q .
\end{equation}
Combining the retained-cell and omitted-cell estimates yields
\begin{equation}
    |\mathcal I(q,p)-\mathcal I_N(q,p)|
    \le
    \Delta s\,L_q
    \left(
        \mathrm{TV}_{[0,1]}(p)
        +
        \#(\mathcal O_N)\,\|p\|_{L^\infty([0,1])}
    \right).
\end{equation}
Since \(\Delta s=1/N\), this proves the first estimate. Finally, because no
grid point lies in \(\mathcal J\), each jump lies in a unique grid cell.
Consequently, each omitted cell contains at least one jump, and hence
\begin{equation}
    \#(\mathcal O_N)\le \#(\mathcal J).
\end{equation}
The stated bound follows.
\end{proof}

\subsection{Practical detection of jump discontinuities}

The quadrature rule above assumes that the grid cells containing jump
discontinuities are known and omitted. In practice, these cells must be
estimated from the sampled data. A simple approach is to flag cells for
which the observed finite difference is too large to be explained by the
expected smooth variation of the data.

Let
\begin{equation}
    D^+_N f(k)
    :=
    \frac{f((k+1)\Delta s)-f(k\Delta s)}{\Delta s},
    \qquad
    \Delta s = \frac{1}{N}.
\end{equation}
Given user-specified cutoffs \(\tau_q,\tau_p>0\), define the set of retained
cells by
\begin{equation}
    \mathtt K_N(\tau_q,\tau_p)
    :=
    \bigg\{
        k\in\{0,\ldots,N-1\}:
        |D^+_N q(k)|\le \tau_q
        \text{ and }
        |D^+_N p(k)|\le \tau_p
    \bigg\}.
\end{equation}
The corresponding computable approximation is
\begin{equation}
    \mathcal I_N(q,p;\tau_q,\tau_p)
    :=
    \sum_{k\in\mathtt K_N(\tau_q,\tau_p)}
    p(k\Delta s)
    \left(q((k+1)\Delta s)-q(k\Delta s)\right).
\end{equation}

This criterion should be interpreted as a jump-detection heuristic. The
cutoffs should be chosen larger than the largest finite difference expected
from smooth portions of the data, but smaller than the finite difference
created by a jump. To see the relevant scaling, suppose that
\(f\in\{p,q\}\) is \(W^{1,\infty}\) away from its jumps, and define
\begin{equation}
    M_f
    :=
    \max_{I\subset [0,1]\setminus\mathcal J_f}
    \|f'\|_{L^\infty(I)}.
\end{equation}
For a jump point \(a\in\mathcal J_f\), write
\begin{equation}
    [f]_a := f(a^+)-f(a^-).
\end{equation}
If \(\mathcal J_f\neq\emptyset\), define the minimum jump size by
\begin{equation}
    \delta_f
    :=
    \min_{a\in\mathcal J_f}|[f]_a|.
\end{equation}
On a smooth cell, one has
\begin{equation}
    |D^+_N f(k)| \le M_f.
\end{equation}
On a cell containing a single jump of magnitude \(|[f]_a|\), the smooth
variation can partially cancel the jump, but the lower bound
\begin{equation}
    |D^+_N f(k)|
    \ge
    N|[f]_a| - M_f
\end{equation}
still holds. Hence, if
\begin{equation}
    M_f < \tau_f < N\delta_f - M_f,
\end{equation}
then smooth cells are not flagged, while cells containing jumps of \(f\) are
flagged. In particular, for fixed \(\tau_f>M_f\), jump detection becomes
reliable once
\begin{equation}
    N > \frac{\tau_f+M_f}{\delta_f},
\end{equation}
provided \(N\) is also large enough that each grid cell contains at most one
jump.

In the numerical experiments in this work, we use fixed normalized thresholds of
the form
\begin{equation}
    \tau_q = \lambda_q \|q\|_{L^\infty([0,1])},
    \qquad
    \tau_p = \lambda_p \|p\|_{L^\infty([0,1])},
\end{equation}
with large dimensionless constants \(\lambda_q,\lambda_p\). This choice is
problem dependent. The constants must be large enough that steep but smooth
regions are not spuriously removed, but small enough that cells crossing
true jumps are excluded.

\begin{remark}
The estimate in the preceding theorem applies directly when the detected
set of omitted cells contains the true jump cells. False positives merely
omit additional smooth cells and contribute an error proportional to the
number of such omitted cells times \(\Delta s\). False negatives are more
serious: retaining a cell that crosses a jump of \(q\) generally introduces
an \(O(1)\) error, because the discrete increment then includes the jump
contribution whereas the piecewise smooth integral excludes it. Retaining a
cell that crosses only a jump of \(p\) is less severe, since such an error
is still controlled by the bounded variation of \(p\).
\end{remark}
 
\section{On the regularity of B-splines in 1D} \label{appendix:piecewise_linear_regularity}

Consider the constant cardinal B-spline,
\begin{equation}
    B_0(x)
    =
    \begin{cases}
        1 \,, & | x | \leq 1/2 \\
        0 \,, & \text{else} \,.
    \end{cases}
\end{equation}
Its Fourier transform is
\begin{equation}
    \hat{B}_0(k) 
    =
    \frac{1}{\sqrt{2 \pi}} \text{sinc} \left( \frac{k}{2} \right) \,.
\end{equation}
It is possible to show that
\begin{equation}
    \int_{\mathbb{R}} (1 + k^2)^s |\hat{B}_0(k) |^2 \mathsf{d} k < \infty 
\end{equation}
for $s \in [0, 1/2)$, since $\text{sinc}(k) \sim 1/k$ as $|k| \to \infty$. The degree-$p$ cardinal B-spline is 
\begin{equation}
    B_p
    =
    \underbrace{(B_0 * B_0 * \cdots * B_0)}_{p + 1 \text{ times}}(x) \,.
\end{equation}
Proceeding similarly to the case of linear B-splines, we find that
\begin{equation}
    \hat{B}_p(k)
    =
    \left[
    \frac{1}{\sqrt{2 \pi}} \text{sinc} \left( \frac{k}{2} \right)
    \right]^{p+1} \,,
\end{equation}
and therefore, we conclude that
\begin{equation}
    \int_{\mathbb{R}} (1 + k^2)^s |\hat{B}_p(k) |^2 \mathsf{d} k 
    < \infty \,,
\end{equation}
for $s \in [0,p + 1/2)$, since
\begin{equation}
    (1 + k^2)^s |\hat{B}_p(k) |^2
    \sim \frac{1}{k^{2(p+1-s)}}
    \quad \text{as} \quad
    k \to \infty \,.
\end{equation}
Thus, it follows that $B_p \in H^{p + 1/2 -\epsilon}(\mathbb{R})$ for all $\epsilon > 0$. Because $H^s(\mathbb{R})$ is closed under scaling, translation, and finite linear combinations, any compactly supported function obtained as a finite linear combination of degree-$p$ B-splines likewise lies in $H^{p + 1/2-\epsilon}(\mathbb{R})$.
 
\section{On the symplecticity of Strang splitting}

As the numerical tests in this work utilize Strang splitting for their time-stepping, and we operate under the assumption that this time-advance map is symplectic, it is helpful to briefly demonstrate the symplecticity of the method to keep this work self-contained. The symplecticity of this method is well known and may be demonstrated by several different approaches \cite{geo_num_int}, e.g.\! by deriving the method via Hamiltonian splitting, as a variational integrator, or as a symplectic partitioned Runge-Kutta method \cite{sanz1988runge}. 

Consider a Lagrangian system of the form
\begin{equation}
    L(\bm{q}, \dot{\bm{q}})
    =
    \frac{1}{2} |\dot{\bm{q}}| ^2
    -
    V(\bm{q}) \,.
\end{equation}
The conjugate momentum is given by $\bm{p} = \partial L/\partial \dot{\bm{q}} = \dot{\bm{q}}$. The type-I generating function mapping $\bm{q}_0 = \bm{q}(t_0)$ to $\bm{q}_1 = \bm{q}(t_1)$ is given by Hamilton's principle function:
\begin{equation}
    S_1(\bm{q}_0, \bm{q}_1; t_0, t_1)
    =
    \int_{t_0}^{t_1} 
    \left(
    \frac{1}{2} |\dot{\bm{q}}|^2 - V(\bm{q}) 
    \right)
    \mathsf{d} t \,.
\end{equation}
The symplectic map from $(\bm{q}_0, \bm{p}_0)$ to $(\bm{q}_1, \bm{p}_1)$ is implicitly defined by the system
\begin{equation} \label{eq:symp_map_typeI}
    \partial_{\bm{q}_0} S_1(\bm{q}_0, \bm{q}_1; t_0, t_1)
    =
    - \bm{p}_0 \,,
    \quad \text{and} \quad
    \partial_{\bm{q}_1} S_1(\bm{q}_0, \bm{q}_1; t_0, t_1)
    =
    \bm{p}_1 \,.
\end{equation}
Suppose we approximate the generating function using trapezoidal rule with the time derivative approximated by forward and backward differences at the endpoints $t = t_0$ and $t = t_1$, respectively. This yields the generating function
\begin{equation}
    S_d(\bm{q}_0, \bm{q}_1, t_0, t_1)
    =
    \frac{h}{2}
    \left(
    \left| \frac{\bm{q}_1 - \bm{q}_0}{h} \right|^2
    -
    V(\bm{q}_0)
    -
    V(\bm{q}_1)
    \right) \,.
\end{equation}
The symplectic map implied by equation \eqref{eq:symp_map_typeI} may be solved explicitly yielding
\begin{equation}
\begin{aligned}
    \bm{q}_1
    &=
    \bm{q}_0
    +
    h
    \left(
   \bm{p}_0
    -
    \frac{h}{2} \nabla V(\bm{q}_0)
    \right) \\
    \bm{p}_1
    &=
    \bm{p}_0
    -
    \frac{h}{2}
    \left[
    \nabla V( \bm{q}_0)
    +
    \nabla V
    \left(
    \bm{q}_0
    +
    h
    \left(
    \bm{p}_0
    -
    \frac{h}{2} \nabla V(\bm{q}_0)
    \right)
    \right)
    \right] \,,
\end{aligned}
\end{equation}
which may be alternatively written as
\begin{equation}
\begin{aligned}
    \bm{p}_{1/2}
    &= \bm{p}_0 - \frac{h}{2} \nabla V(\bm{q}_0) \\
    \bm{q}_1
    &=
    \bm{q}_0
    +
    h \bm{p}_{1/2}
    \\
    \bm{p}_1
    &=
    \bm{p}_{1/2}
    -
    \frac{h}{2}
    \nabla V( \bm{q}_1) \,.
\end{aligned}
\end{equation}
This is the \textit{momentum first} Strang splitting method. This derivation as a variational integrator is instructive, but unfortunately the corresponding \textit{position first} version of the algorithm does not admit a similarly simple and intuitive derivation.

However, we can derive both forms of Strang splitting as symplectic partitioned Runge-Kutta methods, which may be shown to be variational integrators in general \cite{geo_num_int}. Consider a differential equation of the form
\begin{equation}
    \dot{\bm{q}} = \bm{f}(\bm{q}, \bm{p}) \,,
    \quad 
    \dot{\bm{p}} = \bm{g}(\bm{q}, \bm{p}) \,.
\end{equation}
If $(a_{ij}, b_i)$ and $(\hat{a}_{ij}, \hat{b}_i)$ are the Butcher tableau for two Runge-Kutta methods, a partitioned Runge-Kutta method for the system is specified by the system:
\begin{equation}
\begin{aligned}
    \bm{k}_i &= \bm{f} \left( \bm{q}_0 + h \sum_{j=1}^s a_{ij} \bm{k}_j, \bm{p}_0 + h \sum_{j=1}^s \hat{a}_{ij} \bm{\ell}_j \right) \,, \\
    \bm{\ell}_i &= \bm{g} \left( \bm{q}_0 + h \sum_{j=1}^s a_{ij} \bm{k}_j, \bm{p}_0 + h \sum_{j=1}^s \hat{a}_{ij} \bm{\ell}_j \right) \,, \\
    \bm{q}_1 &= \bm{q}_0 + h \sum_{i=1}^s b_i \bm{k}_i \,, 
    \qquad
    \bm{p}_1 = \bm{p}_0 + h \sum_{i=1}^s \hat{b}_i \bm{\ell}_i \,.
\end{aligned}
\end{equation}
We can approximate a canonical Hamiltonian system with Hamiltonian $\mathcal{H}(\bm{q}, \bm{p})$ by letting
\begin{equation}
    \bm{f}(\bm{q}, \bm{p})
    =
    \nabla_{\bm{p}} \mathcal{H}(\bm{q}, \bm{p}) \,,
    \quad \text{and} \quad
    \bm{g}(\bm{q}, \bm{p})
    =
    -\nabla_{\bm{q}} \mathcal{H}(\bm{q}, \bm{p}) \,.
\end{equation}
In this case, the map $(\bm{q}_0, \bm{p}_0) \mapsto (\bm{q}_1, \bm{p}_1)$ is symplectic if $b_i \hat{a}_{ij} + \hat{b}_j a_{ji} = b_i \hat{b}_j$ \cite{geo_num_int, sanz1988runge}. Further, if the Hamiltonian is additively separable, that is if $\mathcal{H}(\bm{q},\bm{p}) = T(\bm{p}) + V(\bm{q})$, the condition $b_i = \hat{b}_i$ is no longer needed to ensure symplecticity. 

Now, consider a separable Hamiltonian $\mathcal{H}(\bm{q}, \bm{p}) =  \frac{1}{2} |\bm{p}|^2 + V(\bm{q})$, yielding:
\begin{equation}
    \dot{\bm{q}} = \bm{p}, \quad \dot{\bm{p}} = -\nabla V(\bm{q}) \,.
\end{equation}
As we previously saw, the momentum-first Strang splitting scheme is:
\begin{align*}
\bm{p}_{1/2} &= \bm{p}_0 - \frac{h}{2} \nabla V(\bm{q}_0), \\
\bm{q}_1 &= \bm{q}_0 + h \bm{p}_{1/2}, \\
\bm{p}_1 &= \bm{p}_{1/2} - \frac{h}{2} \nabla V(\bm{q}_1).
\end{align*}
This scheme has the Butcher tableau:
\begin{equation}
    \begin{array}{c|cc}
         & a  \\
         \hline
         & b
    \end{array}
    =
    \begin{array}{c|cc}
    & 0 & 0 \\
    & 1 & 0 \\
    \hline
     & 1 & 0
    \end{array}, 
    \quad \text{and} \quad
    \renewcommand{\arraystretch}{1.25}
    \begin{array}{c|cc}
         & \hat{a}  \\
         \hline
         & \hat{b}
    \end{array}
    =
    \renewcommand{\arraystretch}{1.25}
    \begin{array}{c|cc}
    & \frac{1}{2} & 0 \\
    & \frac{1}{2} & 0 \\
    \hline
     & \frac{1}{2} & \frac{1}{2}
    \end{array} \,.
\end{equation}
Symplecticity holds: $b_1 \hat{a}_{11} + \hat{b}_1 a_{11} = b_1 \hat{a}_{12} + \hat{b}_2 a_{21} = 1/2 = b_1 \hat{b}_1 = b_1 \hat{b}_2$ (the other terms vanish). On the other hand, the position-first Strang splitting scheme is:
\begin{align*}
\bm{q}_{1/2} &= \bm{q}_0 + \frac{h}{2} \bm{p}_0, \\
\bm{p}_1 &= \bm{p}_0 - h \nabla V(\bm{q}_{1/2}), \\
\bm{q}_1 &= \bm{q}_{1/2} + \frac{h}{2} \bm{p}_1.
\end{align*}
This scheme has the Butcher tableau:
\begin{equation}
    \begin{array}{c|cc}
         & a  \\
         \hline
         & b
    \end{array}
    =
    \renewcommand{\arraystretch}{1.25}
    \begin{array}{c|cc}
    & \frac{1}{2} & 0 \\
    & \frac{1}{2} & 0 \\
    \hline
     & \frac{1}{2} & \frac{1}{2}
    \end{array}, 
    \quad \text{and} \quad
    \renewcommand{\arraystretch}{1.25}
    \begin{array}{c|cc}
         & \hat{a}  \\
         \hline
         & \hat{b}
    \end{array}
    =
    \begin{array}{c|cc}
    & 0 & 0 \\
    & 1 & 0 \\
    \hline
     & 0 & 1
    \end{array} \,.
\end{equation}
As with the prior case, the symplecticity condition is satisfied. 
 
\section{Two-cell, single-particle linear PIC evolution equations} \label{appendix:single_particle}

We wish to demonstrate that it is possible to obtain a single particle in an absolute value potential as a special case of the linear PIC method described in section \ref{sec:symp_pic_method}. This can arise in the minimal example of a single particle influenced by its own field in a domain with only two cells. Let the domain $\Omega = [-L,L]$ be divided into two cells: $\Omega_- = [-L,0]$ and $\Omega_+ = [0,L]$ and identify the endpoints. The grid has only two nodes, $x_0 = 0$ and $x_L = \pm L$. Hence, there are only two shape functions:
\begin{equation}
    B_0(x)
    =
    1 - \left| \frac{x}{L} \right| \,,
    \quad \text{and} \quad
    B_L(x)
    =
    \left| \frac{x}{L} \right| \,.
\end{equation}
The discrete Laplacian associated with this basis is
\begin{equation}
    \mathbb{L} = \frac{2}{L}
    \begin{bmatrix}
        1 & -1 \\
        -1 & 1
    \end{bmatrix} \,.
\end{equation}
We now wish to discern the evolution equation for a single particle $(x_p, v_p)$ influenced by its own potential interpolated to this grid. The values of the potential at the two grid-points are therefore found to be
\begin{equation}
    \frac{2}{L}
    \begin{bmatrix}
        1 & -1 \\
        -1 & 1
    \end{bmatrix}
    \begin{bmatrix}
        \upphi_0 \\
        \upphi_L
    \end{bmatrix}
    =
    \frac{1}{L}
    \begin{bmatrix}
        B_0(x_p) - 1/2 \\
        B_L(x_p) - 1/2
    \end{bmatrix} \,,
\end{equation}
where we let $\rho_0 = 1/(2L)$ and $w_a = 1/L$ to achieve overall charge neutrality. Before proceeding, note that
\begin{equation}
    E_h(x)
    =
    - \frac{d \phi_h}{dx}
    =
    - ( \phi_0 B_0'(x) + \phi_L B_L'(x))
    =
    (\phi_0 - \phi_L) B_L'(x)
    =
    \frac{\phi_0 - \phi_L}{L} \text{sign}(x) \,,
\end{equation}
since
\begin{equation}
    B_0'(x) = - B_L'(x) = - \frac{1}{L} \text{sign}(x) \,.
\end{equation}
Finally, note that 
\begin{equation}
    \phi_0 - \phi_L 
    = \frac{1}{2} \left(B_0(x_p) - \frac{1}{2}\right) 
    = \frac{1}{2} \left( \frac{1}{2} - \left| \frac{x_p}{L} \right| \right) \,.
\end{equation}
Therefore, the evolution equations are given by
\begin{equation}
    \dot{x}_p = v_p \,,
    \quad 
    \dot{v}_p = \frac{1}{2 L} \left( \frac{1}{2} - \left| \frac{x_p}{L} \right| \right) \text{sign}(x_p) \,.
\end{equation}
These are just Hamilton's equations generated by the Hamiltonian
\begin{equation}
    \mathcal{H}(x_p,v_p)
    =
    \frac{v_p^2}{2}
    -
    \frac{1}{4} \left( \left| \frac{x_p}{L} \right| - \frac{x_p^2}{L^2} \right) \,.
\end{equation}
For $|x_p| \ll L$, we recover trajectories which resemble those from a signed absolute value potential.
  
\end{document}